\documentclass[
 reprint,
 amsfonts,amssymb,amsmath,
 showkeys,
 pra,
 superscriptaddress,
 twocolumn,
longbibliography,
nofootinbib,
floatfix,
]{revtex4-2}

\usepackage[usenames,dvipsnames]{xcolor}
\definecolor{shadecolor}{gray}{0.9}

\usepackage[caption=false]{subfig}
\usepackage{graphicx}
\usepackage{dcolumn}
\usepackage{bm}
\usepackage{bbm}
\usepackage{cancel}

\usepackage{comment}
\usepackage{amsmath,amssymb,amsthm,amssymb}
\usepackage{physics}
\usepackage{mathrsfs}  
\usepackage{cancel}
\usepackage{enumitem}
\usepackage{hyperref}
\hypersetup{
    colorlinks=false,
    linkcolor=MidnightBlue,
    filecolor=BrickRed,      
    urlcolor=RoyalBlue,
    citecolor=OliveGreen,
    pdftitle={Quantum-Probabilistic Information Geometry},
    pdfpagemode=FullScreen,
}
\usepackage{cleveref}

\usepackage{algorithm} 
\usepackage[noend]{algpseudocode} 
\usepackage{dsfont}
\usepackage[colorinlistoftodos]{todonotes}
\usepackage[toc,acronym]{glossaries}
\usepackage{varwidth}

\DeclareMathOperator{\Cov}{Cov}  
  
\DeclareMathOperator{\BKM}{BKM}

\newcommand{\spliteq}[1]{\begin{equation}
\begin{split}
#1
\end{split}
\end{equation}
}

\newcommand{\eps}{\varepsilon}
\newcommand{\floor}[1]{\left\lfloor #1 \right\rfloor}

\newtheorem{theorem}{Theorem}[section]
\newtheorem{corollary}{Corollary}[theorem]
\crefname{corollary}{corollary}{corollaries}
\newtheorem{lemma}[theorem]{Lemma}
\crefname{lemma}{lemma}{lemmas}

\crefname{exercise}{exercise}{exercises}
\newtheorem{definition}[theorem]{Definition}
\crefname{definition}{definition}{definitions}

\crefname{fact}{fact}{facts}
\newtheorem{example}[theorem]{Example}
\crefname{example}{example}{examples}
\newtheorem{proposition}[theorem]{Proposition}
\crefname{proposition}{proposition}{propositions}

\crefname{result}{result}{results}

\crefname{notation}{notation}{notation}
\newtheorem{assumption}[theorem]{Assumption}
\crefname{assumption}{assumption}{assumptions}
\newtheorem{remark}[theorem]{Remark}
\crefname{remark}{remark}{remarks}

\crefname{paragraph}{paragraph}{paragraphs}

\makeglossaries

\newacronym{qfi}{QFI}{Quantum Fisher Information}
\newacronym{qfim}{QFIM}{Quantum Fisher Information Matrix}
\newacronym{crb}{CRB}{Cram\'er-Rao bound}
\newacronym{qcrb}{QCRB}{Quantum Cram\'er-Rao bound}
\newacronym{povm}{POVM}{Positive Operator-Valued Measure}
\newacronym{qhbm}{QHBM}{Quantum Hamiltonian-Based Model}
\newacronym{qpngd}{QPNGD}{Quantum-Probabilistic Natural Gradient Descent}
\newacronym{ngd}{NGD}{Natural Gradient Descent}
\newacronym{qpmd}{QPMD}{Quantum-Probabilistic Mirror Descent}
\newacronym{ebm}{EBM}{Energy-Based Model}
\newacronym{qmhl}{QMHL}{Quantum Modular Hamiltonian Learning}
\newacronym{vqt}{VQT}{Variational Quantum Thermalization}
\newacronym{bkm}{BKM}{Bogoliubov-Kubo-Mori}
\newacronym{bh}{BH}{Bures-Helstrom}
\newacronym{qvartz}{QVARTZ}{Quantum Variational Recursive Time Evolution AnsatZe}
\newacronym{qgrass}{QGRASS}{Quantum Geometrically-Regularized Architecture Search}
\newacronym{meta-vqt}{Meta-VQT}{Meta-Variational Quantum Thermalization}
\newacronym{qspl}{QSPL}{Quantum-Stochastic Process Learning}
\newacronym{mcmc}{MCMC}{Markov-Chain Monte Carlo}
\newacronym{mc}{MC}{Morozova-\u{C}encov}
\newacronym{sld}{SLD}{Symmetric Logarithmic Derivative}
\newacronym{qnn}{QNN}{Quantum Neural Network}
\newacronym{cptp}{CPTP}{Completely-Positive Trace Preserving}
\newacronym{map}{MAP}{Maximum a Posteriori}
\newacronym{qphmm}{QPHMM}{Quantum-Probabilistic Hidden Markov Model}
\newacronym{tfim}{TFIM}{Transverse Field Ising Model}
\newacronym{sgd}{SGD}{Stochastic Gradient Descent}
\newacronym{qhea}{QHEA}{Quantum Hardware-Efficient Ansatz}
\newacronym{ode}{ODE}{Ordinary Differential Equation}
\newacronym{cp}{CP}{Completely Positive}
\newacronym{vqa}{VQA}{Variational Quantum Algorithm}
\newacronym{vqite}{VarQITE}{Variational Quantum Imaginary Time Evolution}
\newacronym{vff}{VFF}{Variational Fast Forwarding}
\newacronym{qef}{QEF}{Quantum Exponential Family}
\newacronym{eth}{ETH}{Eigenstate Thermalization Hypothesis}

\makeatletter
\def\namedlabel#1#2{\begingroup
    #2
    \def\@currentlabel{#2}
    \phantomsection\label{#1}\endgroup
}
\makeatother

\begin{document}

\title{\texorpdfstring{Provably efficient variational generative modeling of quantum many-body systems\\ via quantum-probabilistic information geometry \\}{Provably efficient variational generative modeling of quantum many-body systems via quantum-probabilistic information geometry}}

\affiliation{X, Mountain View, CA, USA}

\affiliation{Google Brain, Mountain View, CA, USA}

\affiliation{Department of Computer Science, UC Berkeley, Berkeley, USA}

\affiliation{Department of Mathematics, Stanford University, Stanford, USA}
\affiliation{Institute for Quantum Computing, University of Waterloo, ON, Canada}
\affiliation{Department of Applied Mathematics, University of Waterloo, ON, Canada}

\author{Faris M. Sbahi}
\email[contact: ]{farissbahi@google.com}
\affiliation{X, Mountain View, CA, USA}
\affiliation{Google Brain, Mountain View, CA, USA}

\author{Antonio J. Martinez}
\affiliation{X, Mountain View, CA, USA}
\affiliation{Institute for Quantum Computing, University of Waterloo, ON, Canada}

\author{Sahil Patel}
\thanks{These authors contributed equally.}
\affiliation{X, Mountain View, CA, USA}
\affiliation{Department of Computer Science, UC Berkeley, Berkeley, USA}

\author{Dmitri Saberi}
\thanks{These authors contributed equally.}
\affiliation{X, Mountain View, CA, USA}
\affiliation{Department of Mathematics, Stanford University, Stanford, USA}

\author{Jae Hyeon Yoo}
\affiliation{X, Mountain View, CA, USA}
\affiliation{Google Brain, Mountain View, CA, USA}

\author{Geoffrey Roeder}
\affiliation{X, Mountain View, CA, USA}
\affiliation{Department of Computer Science, Princeton University,
  Princeton, NJ, USA}

\author{Guillaume Verdon}

\email[contact: ]{gverdon@google.com}
\affiliation{X, Mountain View, CA, USA}

\affiliation{Institute for Quantum Computing, University of Waterloo, ON, Canada}
\affiliation{Department of Applied Mathematics, University of Waterloo, ON, Canada}

\date{\today}

\begin{abstract}
The dual tasks of quantum Hamiltonian learning and quantum Gibbs sampling are relevant to many important problems in physics and chemistry.  In the low temperature regime, algorithms for these tasks often suffer from intractabilities, for example from poor sample- or time-complexity.  With the aim of addressing such intractabilities, we introduce a generalization of quantum natural gradient descent to parameterized mixed states, as well as provide a robust first-order approximating algorithm, Quantum-Probabilistic Mirror Descent. We prove data sample efficiency for the dual tasks using tools from information geometry and quantum metrology, thus generalizing the seminal result of classical Fisher efficiency to a variational quantum algorithm for the first time. Our approaches extend previously sample-efficient techniques to allow for flexibility in model choice, including to spectrally-decomposed models like Quantum Hamiltonian-Based Models, which may circumvent intractable time complexities. Our first-order algorithm is derived using a novel quantum generalization of the classical mirror descent duality. Both results require a special choice of metric, namely, the Bogoliubov-Kubo-Mori metric.  To test our proposed algorithms numerically, we compare their performance to existing baselines on the task of quantum Gibbs sampling for the transverse field Ising model. Finally, we propose an initialization strategy leveraging geometric locality for the modelling of sequences of states such as those arising from quantum-stochastic processes. We demonstrate its effectiveness empirically for both real and imaginary time evolution while defining a broader class of potential applications.
\end{abstract}

\maketitle

\section{Introduction}

\begin{figure*}[ht]
  \centering
  \subfloat[][Using second-order information from a metric over quantum states can help navigate a non-convex loss landscape. However, with finite model samples, we will observe increased robustness in \eqref{eq:md-relation} compared to \eqref{eq:natural-gradient-update} due to the difficulty of estimating the metric tensor explicitly.]{\includegraphics[width=0.44175\textwidth]{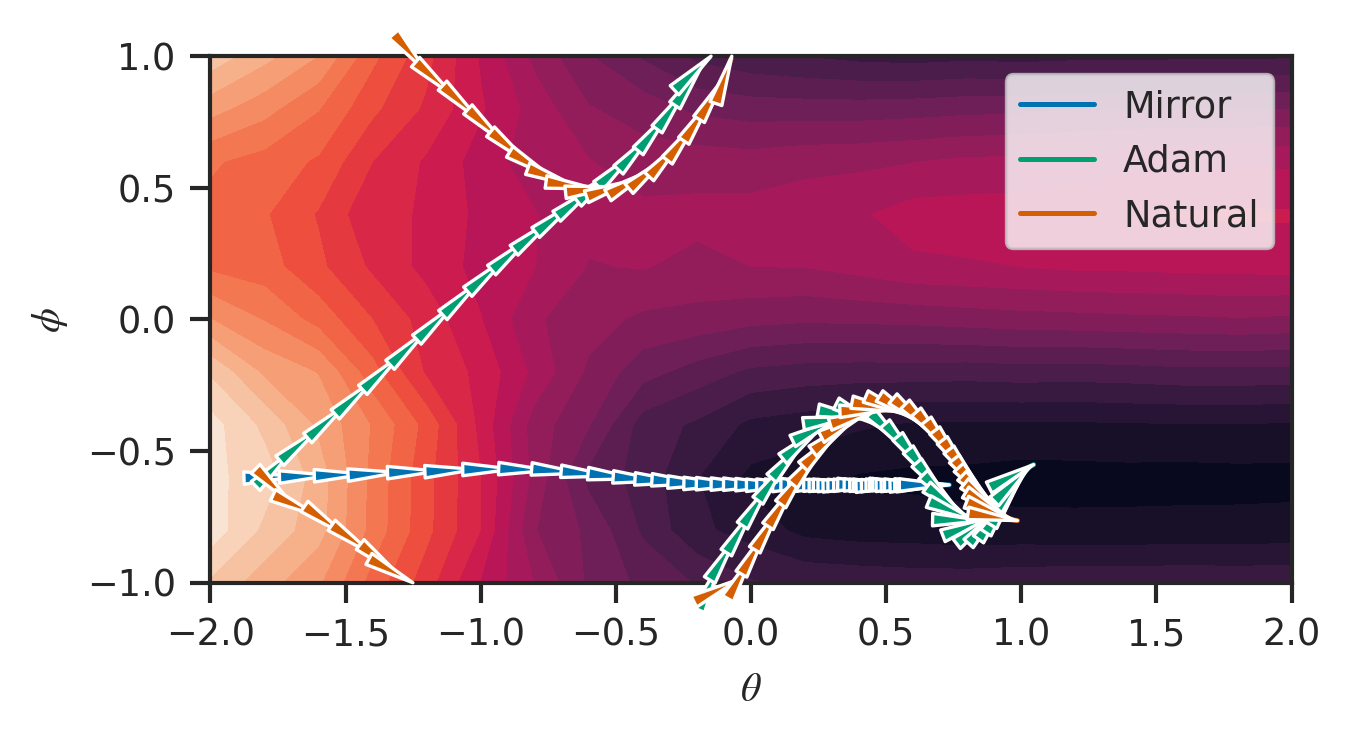}\label{fig:rgf}}
    \subfloat[][The Riemannian flow \eqref{eq:rgf-main} captures a notion of steepest descent over the density operator manifold, and so can diminish suboptimalities which arise due to choice of model parameterization. On the other hand, a clever choice of model parameterization may unlock time-complexity improvements. These two ideas together are key to our approach.]{\includegraphics[width=0.50825\textwidth]{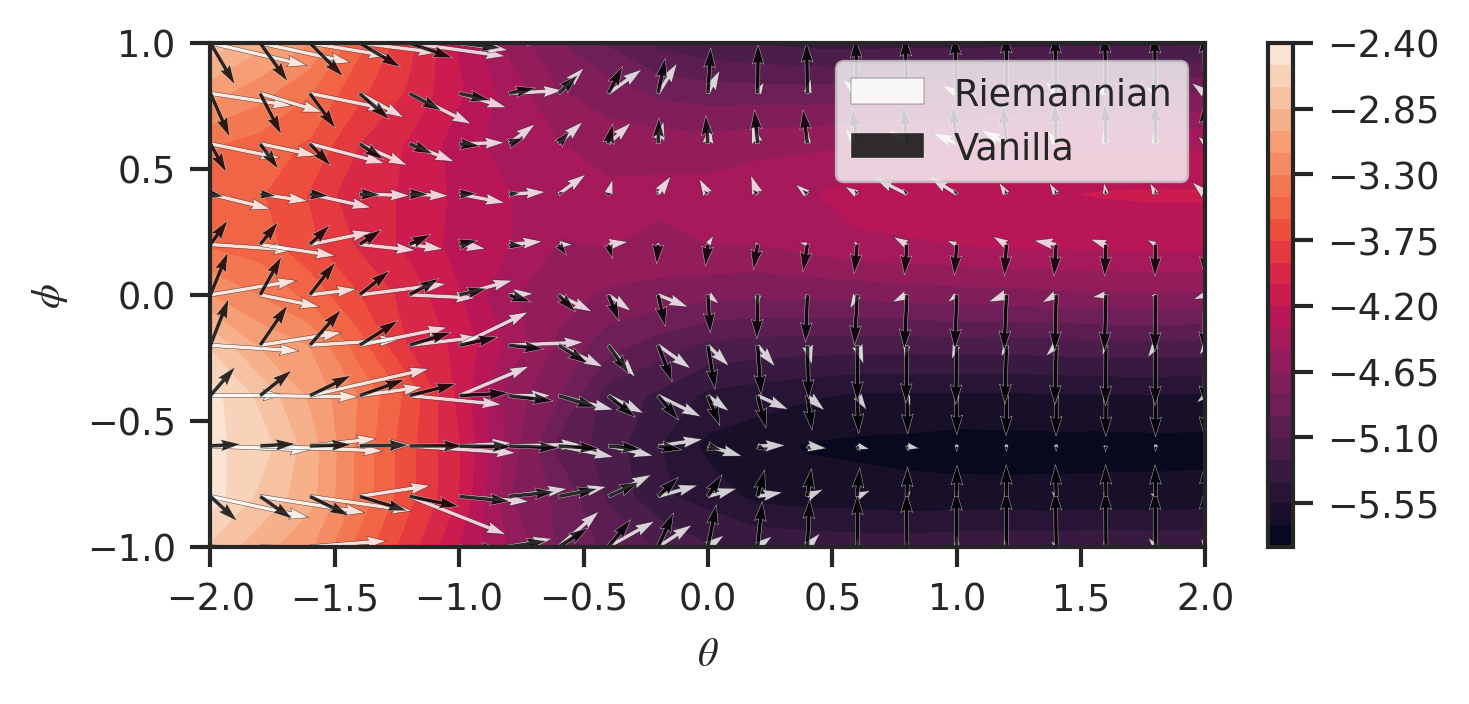}\label{fig:bounce}}
  \caption{Depiction of the information-geometric flow \eqref{eq:rgf-main} and the novel \Cref{alg:qngd,alg:qmd}, which are generally distinct discretizations of the flow. The loss is depicted as a heatmap, and $\bm \Omega = (\theta, \phi)$ gives the particular parameterization of quantum mixed states. This is obtained by freezing out all other parameters of the model used in \Cref{fig:VQT}. We note as well improved convergence behavior for \eqref{eq:md-relation} in a convex region about the optimum. This relates to our optimality result \Cref{thm:optimality-guarantee} as discussed in \Cref{sec:tuning-discrete}.}
  \label{fig:ngd-picture}
\end{figure*}

Quantum machine learning can be bifurcated as two principal research directions depending on the nature of the \textit{data} specification~\cite{anshu2022some,broughton2020tensorflow}. On one hand, the learning problem may be specified in terms of \textit{classical data} and so have associated classical algorithmic baselines. The prospect of \textit{quantum advantage} (when a quantum computer enables an exponential speedup over a classical baseline~\cite{wang2022quantum}) has sparked significant research efforts on quantum algorithms for classical data. For example, sparse matrix inversion is BQP-complete and so expected to admit such an advantage~\cite{harrow2009quantum}. However, a broader class of once anticipated advantages was shown to be an artifact of state preparation assumptions rather than following from the quantumness of the algorithms~\cite{tang2021quantum,tang2019quantum}. In particular, many speedups vanish if one has classical $l_2$-sampling access to the same data (which is at least as easy to obtain as quantum state preparation). Several additional theoretical~\cite{mcclean2018barren,wang2021noise} and practical~\cite{giovannetti2008quantum,aaronson2015read,broughton2020tensorflow} barriers have emerged along this direction, though key milestones in understanding have been reached~\cite{liu2021rigorous,arunachalam2017guest,arunachalam2022quantum}.

On the other hand, separate, exciting efforts are analyzing the problem of efficiently learning (or simulating) properties of unknown (or, respectively, known) quantum systems~\cite{aaronson2019shadow,huang2020predicting,huang2021quantum,huang2021information,huang2021provably,arute2019quantum} which serve as \textit{quantum data}. Hence, all relevant algorithms are quantum since, at the least, they require measuring (or preparing) a quantum system. For learning, the required data samples can scale as poorly as exponentially in the number of qubits~\cite{vogel1989determination}. Dealing with the quantum nature of the data in order to tame such scalings does not necessarily require entirely foreign algorithmic ideas compared to the classical; in many cases (e.g.,~\cite{brandao2019quantum,szegedy2004quantum,yung2012quantum,anshu2021sample,haah2021optimal,liu2019quantum}), so-called ``quantization" of classical techniques describes the fruitful methodology of (sometimes systematically) adjusting these techniques to integrate such structure. We will quantize several techniques and proofs in this way.

In this work, we propose a variational approach to a problem which has garnered significant attention, both theoretically~\cite{swingle2014reconstructing,bairey2019learning, qi2019determining,evans2019scalable,anshu2021sample,haah2021optimal,anshu2022some} and  experimentally~\cite{wang2017experimental,senko2014coherent}: \textit{learning} spatially local quantum many-body Hamiltonians.  In particular, such an algorithm should ideally be both time-efficient and sample-efficient, and apply to both low- and high-temperature Gibbs states.  The low-temperature learning problem arises in a variety of physics~\cite{kokail2021entanglement,dalmonte2022entanglement} and quantum technology settings, for example in relevance to distinguishing quantum phases~\cite{regnault2017entanglement,haag2012local} or holographic entanglement entropy~\cite{casini2011towards}. The classical analogue of this task, learning undirected graphical models or Markov random fields, is central to the machine learning and statistical inference communities~\cite{chow1968approximating,hinton1986learning,karger2001learning,abbeel2006learning,santhanam2012information,bresler2008reconstruction,bresler2015efficiently,vuffray2016interaction,klivans2017learning}.

While few formal results exist, recent trailblazing approaches have yielded provable data sample-efficiency~\cite{anshu2021sample,haah2021optimal}.  However, because their proofs assume a particular model\footnote{Anshu et al.~\cite{anshu2021sample} and Haah et al.~\cite{haah2021optimal} implicitly assume the \gls{qef}~\cite{yapage2008information}.  For more details see \Cref{sec:sample-efficient-many-body}.}, their methods require classically estimating gradients of the quantum log partition function.  This computation is known to suffer from the \textit{sign problem}~\cite{henelius2000sign,eisert2020Science}, and is generically time-inefficient in the low-temperature regime~\cite{harrow2020classical}.  Our proposed algorithm allows us to translate the estimation of a quantum log partition function gradient to a classical one by allowing measurements to be variational. Hence, as we will elaborate, broad classes of Hamiltonians (for example, satisfying the Eigenstate Thermalization Hypothesis~\cite{deutsch1991quantum,srednicki1994chaos}) may be amenable to our approach and not the aforementioned approaches. Our analysis proceeds by showing that our variational algorithm achieves an optimal parameter estimation limit, under certain assumptions, independent of the problem instance or model parameterization. No previous variational quantum algorithm (metric-aware or otherwise) is known to satisfy this criterion.

Furthermore, our result also applies to the historically important~\cite{feynman1982simulating, lloyd1996universal} and presently popular~\cite{arute2019quantum} reverse problem of \textit{simulation} of quantum states on quantum computers. In particular, we present a sample-efficient variational algorithm for quantum Gibbs sampling of states corresponding to a given Hamiltonian.  Quantum Gibbs sampling is important because it enables the study of non-zero-temperature physics on quantum computers~\cite{kassal2011simulating, motta2019qite, dallaire2016method, cohn2020minimal, sun2021qite, bauer2020quantum}.  It is also an important subroutine for a range of quantum algorithms seeking quantum advantage~\cite{brandao2017quantum, van2019improvements, brandao2019quantum, van2020quantum, brandao2022faster, montanaro2015quantum, harrow2020adaptive, lloyd2014quantum}. Our data-sample-efficiency result is the first such concrete statement for variational quantum Gibbs sampling, and carries the same model choice flexibility.

For both scenarios of learning and simulation, we observe numerical advantage over existing optimizers. In particular, such advantages are facilitated by a novel technique which translates the analogous second-order algorithm to a robust, first-order approximation. This is a quantization of a classical result \cite{raskutti2015information}. Finally, we will see empirically, and motivate in connection with our sample-efficiency result, that our approach is particularly conducive to learning or simulating \textit{sequences} of quantum states. This problem arises in a variety of contexts including molecular geometry optimization~\cite{nagaoka2003structure}, annealing~\cite{apolloni1989quantum,apolloni1990numerical,finnila1994quantum}, and time evolution~\cite{abrams1997simulation,kassal2008polynomial, mcardle2020quantum}.

\section{Overview of contributions}

\subsection{Efficient learning and sampling of many-body states}
\label{sec:on-optimality}

\begin{figure}
  \begin{algorithm}[H]
    \begin{algorithmic}[1]
      \For{$j=1,2,\ldots$}
      \State pick learning rate $\tfrac{1}{\lambda_j}$
      \State evaluate metric $\mathcal{I}(\bm\Omega_j)$
      \State compute pseudo-inverse $\mathcal{I}^{+}(\bm\Omega_j)$
      \State evaluate loss gradient $\nabla \mathcal{L}(\bm\Omega_j)$
      \State update model $\bm\Omega_{j+1} \gets \bm\Omega_j - \frac{1}{\lambda_j} \mathcal{I}^{+}(\bm\Omega_j) \nabla \mathcal{L}(\bm\Omega_j)$
      \EndFor
    \end{algorithmic}
    \caption{\gls{qpngd}}
    \label{alg:qngd}
  \end{algorithm}
\end{figure}

An important learning for quantum data problem is the \textit{quantum Hamiltonian learning} problem, which has recently received much attention in the quantum community~\cite{wiebe2014quantum,wiebe2014bhamiltonian,wang2017experimental,bairey2019learning, qi2019determining,evans2019scalable,bairey2020learning,anshu2021sample,haah2021optimal,anshu2022some}. The goal is to learn the Hamiltonian of a quantum many-body system from copies of its Gibbs state, which describes the equilibrium state of a system that is in contact with a heat bath.

More technically, consider a quantum system of $n$ qubits and a $\kappa$-local modular Hamiltonian $\hat K = \sum_{l = 1}^{p} \mu_l \hat E_l$ having coefficients $\mu_l, \lvert \mu_l \rvert \leq 1$ over a finite-dimensional lattice, and $\hat E_l$ are known non-identity Pauli operators. Furthermore, the interaction graph is assumed to be spatially local so that $p \equiv \lVert \bm{\mu}\rVert_0 = \mathcal{O}(n)$ in terms of the number of qubits.  Then, given copies of the Gibbs state $\hat \sigma_\beta$:
\begin{align}\label{eq:gibbs}
    \hat \sigma_\beta &= \frac{e^{- \beta \hat{K}}}{\mathcal{Z}_\beta}, \quad \mathcal{Z}_\beta := \tr[e^{- \beta \hat{K}}]
,\end{align}
where $\beta$ is inverse to temperature, the goal is to learn the coefficients $\mu_l$ to additive error $\eps$, or equivalently, to learn the vector $\bm \mu$ to error $\eps$ in $l_\infty$-norm.

Recently, Anshu, Arunachalam, Kuwahara, and Soleimanifar~\cite{anshu2021sample} studied the sample complexity of a simple approach which involves measuring in the $\{\hat E_l\}$ basis and then solving a classical optimization problem (the so-called maximum entropy problem). They found that this approach is learned to $l_\infty$ error $\eps$ with probability at least ${1 - \delta}$ using
\begin{align}
\label{eq:anshu-sample}
    \mathcal{O}\left(\operatorname{poly}(n) \log(\frac{n}{\delta}) \frac{e^{ \mathcal{O}(\beta^c)} }{\beta^{c'} \eps^2}\right)
\end{align}
data samples~\cite{haah2021optimal}, where $c, c' > 3$ are geometric constants\footnote{The existence of the factor exponential in $\beta$ can be attributed to the fact that finding ground states of local Hamiltonians is QMA-complete~\cite{kempe2006complexity}.}. Hence, this approach is sample-efficient in number of qubits and, in fact, best-known for general $\beta$. On the other hand, the algorithm implicitly assumes a particular model, the quantum exponential family\footnote{Also known as a quantum Boltzmann machine~\cite{yapage2008information, amin2018quantum}.  See \Cref{sec:sample-efficient-many-body,sec:online-loss-gradient}.}.  This assumption requires their algorithm to compute gradients of a quantum log partition function.  For low temperatures, this computation can become QMA-hard~\cite{wei2009interacting,aharonov2009power,harrow2020classical,alhambra2022quantum}, causing classical simulation to become intractable~\cite{temme2011quantum,yung2012quantum,tan2022sign}.

In our work, we design and analyze a variational approach to quantum Hamiltonian learning.  It is a quantum variety of online natural gradient descent with a special choice of metric over the quantum mixed state manifold.  Because it generalizes quantum natural gradient descent to quantum mixed states, we call it \glsfirst{qpngd}\footnote{The word ``probabilistic" is used to distinguish our algorithm from the pure state version described in existing literature}, displayed as \Cref{alg:qngd}.  It solves the same task using the polynomial scaling of Anshu et al.~\eqref{eq:anshu-sample}\footnote{For sufficiently small $\eps$, and assuming asymptotic convergence; see \Cref{sec:many-body}.}, while enabling flexibility in the choice of model (as we will elaborate).  One motivated choice is a diagonal ansatz which fits the spectral decomposition of the target state~\cite{verdon2019qhbm}; this choice turns the quantum log partition function into a classical one, so that roadblocks like the sign problem~\cite{eisert2020Science} encountered when attempting to sample from a quantum exponential family model can be circumvented\footnote{See e.g.~\cite{alhambra2022quantum} for a pedagogical review of the complexity of thermal states.}.  

As a motivating example, under the \gls{eth} assumption~\cite{deutsch1991quantum,srednicki1994chaos}, the quantum Metropolis sampling algorithm~\cite{temme2011quantum} has been shown to converge efficiently which is unexpected for classical methods~\cite{chen2021fast}. Quantum Metropolis sampling similarly diagonalizes the density operator in question so that the sampling becomes classical, but does so via the fault-tolerant quantum phase estimation algorithm. Hence, for our variational approach to be successful, a suitable fixed-depth diagonal ansatz must be posited, the difficulty of which is subject to recent, promising discussion~\cite{larose2019variational,cerezo2020variational}. Furthermore, as we will discuss in \Cref{sec:sample-efficient-many-body}, this flexibility may allow one to relax the prior information required to specify a sample-efficient model for certain problem instances.

To obtain the sample efficiency result, we consider the idea of \textit{Fisher efficiency}, and describe its quantum analogue \textit{quantum Fisher efficiency}.  These broadly applicable optimality properties would be a desirable feature of any variational update rule. Quantum and classical Fisher efficiency take the perspective of viewing descent rules as parameter estimation (metrological) strategies, and so we offer a bridge between the mature field of quantum metrology and the contemporary field of quantum variational optimization. As we will demonstrate, our particular metric-aware descent rule, under a specific choice of metric, is first-known to satisfy this criterion in quantum~\cite{stokes2020quantum,koczor2019quantum,van2020measurement} (\Cref{thm:optimality-guarantee}).

\subsection{Optimization robustness via convex duality}
\label{sec:tractability}

\begin{figure}
  \begin{algorithm}[H]
    \begin{algorithmic}[1]
      \For {$j=1,2,\ldots$}
      \State choose outer-loop learning rate $\tfrac{1}{\lambda_j}$
      \State evaluate loss gradient $\nabla_{\bm\Omega_{j}}\mathcal{L}(\bm \Omega_j)$
      \For {$k=1,2,\ldots,K$}
      \State choose inner-loop learning rate $\eta_k$
      \State \begin{varwidth}[t]{\linewidth} evaluate relative entropy gradient \par
        \hskip\algorithmicindent $\nabla_{\bm\Omega}D(\hat{\rho}_{\bm\Omega} \Vert \hat{\rho}_{\bm\Omega_j})\mid_{\bm\Omega = \bm\Omega_j^k}$
      \end{varwidth}
      \State \begin{varwidth}[t]{\linewidth} update $\bm\Omega_j^{k+1} \gets  \bm\Omega_j^k - \eta_k\big(\nabla_{\bm \Omega_j}\mathcal{L}(\bm \Omega_j)$ \par
        \hskip\algorithmicindent $+ \lambda_j \nabla_{\bm\Omega}D(\hat{\rho}_{\bm\Omega} \Vert \hat{\rho}_{\bm\Omega_j} )\mid_{\bm \Omega = \bm \Omega_j^k}\big)$
      \end{varwidth}
      \EndFor
      \State update model $\bm\Omega_{j+1} \gets \bm\Omega^{K+1}_j$
      \EndFor
    \end{algorithmic}
    \caption{\gls{qpmd}}
    \label{alg:qmd}
  \end{algorithm}
\end{figure}

There are important practical considerations required to achieve the efficient scaling and manageable constant factors in \Cref{eq:anshu-sample}. Perhaps most notably, \Cref{alg:qngd} requires estimating and inverting a second-order \textit{metric} quantity, a task which requires collecting quadratic-scaling samples\footnote{See \Cref{sec:qpmd} for the calculation.} from one's model at each optimization step.

In the classical literature, a result is known which translates our second-order optimization problem to a dual first-order one called \textit{mirror descent}~\cite{raskutti2015information}. As we will find, a special choice of metric over the quantum mixed state manifold -- the same choice which enables our efficiency result \eqref{eq:anshu-sample} --- allows us to derive a novel quantized version of this mirroring duality.  This duality is stated in \Cref{thm:md-ngd-duality}.  We leverage this result into a novel quantum optimization algorithm, \glsfirst{qpmd}, displayed as \Cref{alg:qmd}.  We find numerically (see \Cref{fig:VQT,fig:sequences}) that this algorithm is more robust than \gls{qpngd} during optimization.

To understand the utility of going to first order, it is useful to note the distinction between data and model sample complexities. For learning, this depends on whether a sample is drawn by re-preparing the ground truth (data) thermal state or the model thermal state\footnote{This is in tune with the classical distinction made implicitly by Amari~\cite{amari1998natural} in that one data sample is used at each optimizaton step for online natural gradient descent whereas, e.g., the information matrix (which depends only on the probabilistic model) is estimated without error.}. The efficient scaling \eqref{eq:anshu-sample} concerns the number of data thermal state re-preparations, also called data sample complexity, which may be the key resource of interest. On the other hand, there is an assumption that the metric quantity, which requires no data samples, is estimated sufficiently accurately. This may be costly in terms of the number of model state re-preparations required. Similarly, for simulation, the data sample complexity concerns only the number of measurements of the input Hamiltonian $\hat{K}$ on the prepared state.

Existing algorithms, both quantum~\cite{stokes2020quantum} and classical~\cite{martens2015optimizing,raskutti2015information}, which require estimating and inverting a metric quantity have recognized that taming the model sample scaling is critical for practical applications.  In quantum, block approximations~\cite{stokes2020quantum} have been considered as mechanisms to limit the computation of cross terms to pairs of parameters which are expected to be significantly correlated. Related types of inductive biases have been successful classically, for example assuming that the information matrix has a Kronecker product factorization~\cite{martens2015optimizing}.

The exact or approximate duality implies a potentially more robust update rule which exactly or approximately recovers the data samples scaling \eqref{eq:anshu-sample}. As we have noted, the improved model sample-efficiency is due to each iteration of the mirror descent update rule requiring solving a sub-problem which is first-order. This sub-problem requires computing gradients in the quantum relative entropy, a task to which the \gls{qhbm} ansatz\footnote{See \Cref{sec:review-qhbm} for a self-contained review.} of a Gibbs state is particularly amenable~\cite{verdon2019qhbm} and is non-trivial for general ansatze~\cite{amin2018quantum}, as we will discuss\footnote{See \Cref{sec:sample-efficient-many-body}.}.

\subsection{Efficient learning and Gibbs sampling of sequences of states}

In proving the quantum Fisher efficiency result, the neighborhood of ``fast convergence'' occurs when the third-order terms in the Taylor expansion of the loss go to zero sufficiently quickly (\Cref{sec:tuning-discrete}). Intuitively, this describes the regime for which the optimization problem is approximately convex quadratic.   In many important applications, the fineness of the path discretization is a control parameter and so we provide a criterion in \Cref{eq:metric_Var} by which one could choose a discretization where the optimality still holds (\Cref{sec:tuning-discrete}).

We can use this intuition to motivate one class of problems where these guarantees may then more consistently apply: the learning and Gibbs sampling of sequences of quantum states.  As we will discuss, this type of problem shows up in many contexts including molecular geometry optimization~\cite{nagaoka2003structure}, quantum annealing~\cite{apolloni1989quantum,apolloni1990numerical,finnila1994quantum}, and time evolution~\cite{abrams1997simulation,kassal2008polynomial, mcardle2020quantum}.  In these scenarios, after learning the first state in a sequence of sufficiently close\footnote{In the sense of quantum-statistical distance, quantified via the metric as in \Cref{sec:background-metric}} density operators, initializing one's learning at the previous optimum in the sequence may make optimization approximately convex quadratic.  We discuss sequence learning in detail in \Cref{eq:modeling-paths}, and observe that this straightforward initialization strategy leads to promising performance improvements.

\section{Background}
\label{sec:background}

\subsection{Loss functions for density operators}
\label{sec:background-loss}

Work in \glspl{vqa}~\cite{mcclean2016theory, preskill2018quantum, cerezo2021variational} has shown that it can be advantageous to allow the resources used for a quantum task to be tunable. For such algorithms, a fixed depth circuit is re-run as its parameters are tuned, rather than appending additional fixed circuits at each step of the algorithm.  Lending theoretical justification to this strategy is the result that physical states make up an exponentially small submanifold of Hilbert space~\cite{poulin2011quantum}, so that correspondingly, a quantum circuit ansatz may need only a small number of parameters to learn quantum states of interest.

To formalize these ideas, let $\mathcal{M}^{(N)}$ be the set of ${N \times N}$ density operators\footnote{Equivalently, density matrices or quantum mixed states.  We note that our notation in this section follows conventions in~\cite{bengtsson2017geometry}} .  Suppose we want to learn some unknown quantum state represented by density operator $\hat\sigma \in \mathcal{M}^{(N)}$.  We take the strategy of positing a \textit{parametric model}~\cite{murphy2012machine} $\hat\rho_{\Omega} \in \mathcal{M}^{(N)}$ with $P$ parameters $\bm\Omega \in \mathbb{R}^P$.  The parameters are to be tuned until $\hat\rho_{\bm\Omega} \approx \hat\sigma$.  Note that both quantum Hamiltonian learning and quantum Gibbs sampling can be phrased in terms of learning a parametric model for a target density operator~\cite{verdon2019qhbm}.

To measure how well our parameterized model $\hat\rho_{\bm\Omega}$ approximates $\hat\sigma$, we need to define a \textit{loss function}.  Generally speaking, a loss function is a map which uses information about the target data set to map a parameterized model to the real numbers, $\mathcal{L}: \mathbb{R}^P \rightarrow \mathbb{R}$.  In this paper we will let our loss functions be defined in terms of functions $\Phi$ acting on pairs of density operators,
\begin{equation}\label{eq:loss-function}
\mathcal{L}(\bm\Omega) = \Phi(\hat\rho_{\bm\Omega}, \hat\sigma).
\end{equation}
We require $\Phi$ to be a \textit{contrast functional}.  We say that $\Phi$ is a constrast functional\footnote{Our use of the term \textit{contrast functional} is consistent with~\cite{petz2002covariance, jencova2004generalized}.  Those are quantum generalizations of the related term ``contrast function" used in the classical literature~\cite{pfanzagl1973asymptotic, eguchi1983second}} if it is a non-negative smooth function~\cite{petz2002covariance} such that
\begin{align}
\label{eq:contrast-functional}
    \Phi(\hat\rho, \hat\sigma) = 0 \iff \hat\rho = \hat\sigma.
\end{align}
This property allows us to identify when we have perfectly learned the target state.  In the next section we describe how it allows us to relate metrics for quantum information geometry to loss functions on our parametric models.

Now we have all the components necessary to phrase state learning as a parameter optimization problem.  Given our parametric model $\hat\rho_{\bm\Omega}$, target data density operator $\hat\sigma$, and our choice of loss function $\mathcal{L}$, state learning is simply the task of finding $\bm\Omega^* = \operatorname{argmin}_{\bm\Omega} \mathcal{L}(\bm\Omega)$.  In other words, we find the minimum of the loss function.

In \glspl{vqa}, it can be advantageous to leverage gradient information to perform the minimization of the loss~\cite{harrow2021lowdepth}.  In the next section we discuss formal structures, \textit{Riemannian manifolds of density operators}, which will help us best leverage gradient information to optimize the loss.

\subsection{Riemannian manifolds of density operators}
\label{sec:background-metric}
The algorithms we will introduce in later sections depend on the mathematical concept of a \textit{Riemannian manifold}. In this subsection, we review some known results at the intersection of manifolds and density operators that we intend to leverage\footnote{See \Cref{sec:geometry} for further discussion.}.

Manifolds generalize calculus to sets of objects beyond vectors of real numbers~\cite{frankel2012geometry}.  Speaking loosely, a manifold is a continuous set of objects labelled by smooth coordinate functions which map one-to-one with Euclidean space. In our setting, the objects are given by the set of ${N \times N}$ density operators, $\mathcal{M}^{(N)}$. The parametric model which we have posited has a so-called \textit{hypothesis class} which spans some subset of $\mathcal{M}^{(N)}$. Informally, we can think about the parameters of this model as serving as the smooth coordinates. Formally, however, the one-to-one property is not anticipated to be met in general, particularly when certain inductive architectural biases are leveraged or the model specification includes a neural network\footnote{Nevertheless, it is possible to choose a model parameterization so as to describe a smooth embedding from classical (Euclidean) parameter space to $\mathcal{M}^{(N)}$  (\Cref{apdx:over-param}).}. This observation has noteworthy implications that we will discuss; for example, introducing a pseudo-inverse and breaking down exact invariance under re-parameterization for metric-aware descent (\Cref{alg:qngd}).

Given some new parameters $\bm\Omega'$ near $\bm\Omega$, we again need a quantitative way to distinguish the updated model $\hat\rho_{\bm\Omega^{'}}$ from $\hat\rho_{\bm\Omega}$. The mathematical tool for doing so on a manifold is the \textit{metric} $g_{\bm \Omega}(\cdot, \cdot)$ which defines an inner product taking arguments in the \textit{tangent space} $T_{\bm\Omega} \mathcal{M}^{(N)}$ about $\hat \rho_{\bm \Omega}$.  Adding a metric to our manifold of models turns it into a \textit{Riemannian manifold}.

\begin{figure}
\centering
\includegraphics[width=1.02\columnwidth]{potato_1_ultimate.png}
    \caption{An information-geometric perspective on quantum mixed state learning. A single metric-aware descent step \eqref{eq:natural-gradient-update} is depicted relative to model parameters $\bm \Omega_j$. The loss against target state $\hat \sigma_\beta$ is minimized subject to local updates constrained to a neighborhood of fixed statistical distinguishability $\Phi(\hat \rho_{\bm \Omega_j}, \hat \rho_{\bm \Omega_{j+1}}) = \eps^2$ so as to iterate towards an optimum $\bm{\Omega}^*$. The pre-image of this state manifold neighborhood forms an ellipsoid in the model parameter space whose principal radii are the eigenvectors of the inverse of the metric, $\mathcal{I}(\bm{\Omega}_j)$. By representing gradients of the loss $\nabla\mathcal{L}(\bm{\Omega_j})$ in terms of these eigenvectors, we can ensure consistent step sizes in terms of statistical distinguishability in the space of states. Note that a model, with any inductive bias, parameterizes only a subset of the overall mixed state manifold.
    }
    \label{fig:potato}
\end{figure}
    
A standard property to require of a metric over a statistical manifold is monotonicity. Recall that every physical process in quantum mechanics can be represented by a \gls{cptp} map, also called a \textit{quantum channel}~\cite{wilde2017quantum}.  Let $\hat H$ and $\hat H'$ be any two vectors in the tangent space at $\hat \rho_{\bm \Omega}$ and let $\mathcal{V}$ be any quantum channel. Then, a metric $g_{\bm \Omega}$ is \textit{monotone} if
\begin{align}
    g_{\bm \Omega}(\mathcal{V}(\hat H), \mathcal{V}(\hat H')) &\leq g_{\bm \Omega}(\hat H, \hat H')
\end{align}
is true for all $\bm \Omega$, $\hat H, \hat H'$, and $\mathcal{V}$~\cite{bengtsson2017geometry}.  Intuitively, a quantum channel may coarse-grain, or randomize, the state on which it acts; monotonicity says such randomization does not help in distinguishing states.

Furthermore, we will find it convenient to resolve the metric tensor to coordinates as a matrix. We will define the \textit{Information Matrix} $\mathcal{I}(\bm \Omega)$ as having matrix elements
\begin{align}
  \label{eq:monotone-info-matrix-main}
  [\mathcal{I}(\bm \Omega)]_{j, k} &:=  g_{\bm{{\Omega}}}(\partial_{\Omega_j} \hat \rho_{\bm \Omega}, \partial_{\Omega_k} \hat \rho_{\bm \Omega}).
\end{align}
This matrix is positive definite so long as $\{ \partial_{ \Omega_j} \hat \rho_{\bm \Omega}\}_j$ gives a basis for $T_{\bm\Omega} \mathcal{M}^{(N)}$, the tangent space of $\mathcal{M}^{(N)}$ at $\bm \Omega$ (\Cref{prop:pos-def-frame}). Then, we have the result that $\mathcal{I}$ is the Hessian of some contrast functional (\Cref{eq:contrast-functional}) $\Phi$~\cite{lesniewski1999monotone, jencova2004generalized},
\begin{align}
  \label{eq:hessian-of-some-potential}
  [\mathcal{I}({\bm \Omega})]_{j, k} &\equiv -\partial^2_{\Omega'_j \Omega_k} \Phi (\hat{\rho}_{\bm \Omega'}, \hat{\rho}_{\bm{\Omega}})\mid_{\bm \Omega'= \bm \Omega}.
\end{align}
Since our losses are defined in terms of contrast functionals (recall \Cref{eq:loss-function}), \Cref{eq:hessian-of-some-potential} tells us that taking the Hessian of our loss turns it into a valid monotone metric.

In the classical case, density operators are instead categorical probability distributions and $\Phi$ in \eqref{eq:constrained-opt} is uniquely the classical relative entropy\footnote{\u{C}encov's theorem~\cite{cencov2000statistical} says that, for categorical distributions, the Fisher-Rao metric is the unique metric (up to normalization) which satisfies the analogous monotonicity property. See \Cref{sec:geometry-monotone}.}. In this case, the information matrix is the classical Fisher information matrix. In the quantum case, the option set is broader and so we will find it valuable to consider the choice of metric carefully.

\Cref{fig:potato} illustrates the main features of our manifold perspective on quantum state learning.  It includes the full manifold of density operators, the submanifold spanned by our parametric model, the map between coordinate space $\bm\Omega$ and model space $\hat\rho_{\bm\Omega}$, and the information matrix $\mathcal{I}$.

\subsection{Natural gradient descent}
\label{sec:background-ngd}

In the previous section, we phrased the task of variational learning and Gibbs sampling as an optimization problem over a Riemannian manifold of parametric models.  Evidently, there are many different ways to perform such optimization. For example, we may seek an update rule which travels in a locally optimal fashion over the manifold in the sense of most rapidly reducing the loss. It is understood that \textit{natural gradient descent} offers such behavior, and we will re-express a well-known characterization~\cite{amari2016information} for completeness. Since such a rule depends only on the search space as opposed to the classically parameterizing space, we diminish~\cite{liang2019fisher} (and, in some cases, eliminate~\cite{amari1998natural}) dependencies on the choice of classical parameterization\footnote{See \Cref{sec:duality-qhbm}.}.

In particular, suppose we start at some parameter setpoint $\bm\Omega_j$ with associated loss function value $\mathcal{L}(\bm\Omega_j)$.  Optimizing our parametric model means changing the parameters by some update vector $\bm\delta_{j+1}$ to some new value $\bm\Omega_{j+1} = \bm\Omega_j + \bm\delta_{j+1}$ such that we expect $\mathcal{L}(\bm\Omega_{j+1}) < \mathcal{L}(\bm\Omega_j)$.  To find the direction of steepest descent, we first use the contrast functional $\Phi$ associated to our loss function $\mathcal{L}$ to fix a local neighborhood on our manifold anchored at $\hat\rho_{\bm\Omega_j}$.  Then, we optimize the direction of $\bm\delta_{j+1}$ restricted to that neighborhood:
\begin{align}
  \label{eq:constrained-opt}
  \bm{\delta}_{j + 1} & = \underset{\bm{\delta}:\ \eps^2 = \Phi(\hat{\rho}_{\bm{{\Omega}}_j }, \hat{\rho}_{\bm{{\Omega}}_j + \bm{\delta}}) }{\arg\min}\mathcal{L}(\bm{{\Omega}_j +\bm{\delta}})
\end{align}
where $\eps^2$ is a small constant.  Intuitively, as we can think of $\Phi$ (locally) as a distinguishability function over $\mathcal{M}^{(N)}$, this objective enforces that we update parameters so as to choose optimally from a ball of equally distinguishable operators relative to the current guess.

Writing the objective \eqref{eq:constrained-opt} as a Lagrangian with Lagrange multiplier $\lambda$ and expanding to first non-vanishing order can be shown to give (\Cref{sec:metric-aware-derivation})
\begin{align}
  \label{eq:natural-gradient-update}
  \tag{QPNGD}
  \bm{\Omega}_{j + 1} \leftarrow \bm{{\Omega}}_j -\tfrac{1}{\lambda}  (\mathcal{I} ({\bm{{\Omega}}_j}))^{+}\nabla_{\bm{{\Omega}}_j}\mathcal{L}(\bm{{\Omega}}_j)
,\end{align}
where $A^{+}$ denotes the Moore-Penrose pseudo-inverse of a matrix $A$. This is \Cref{alg:qngd}. Such update rules are known as \textit{natural gradient} or \textit{metric-aware} descent.  We show an example trajectory for this algorithm in \Cref{fig:ngd-picture}.

\Cref{eq:natural-gradient-update} aesthetically matches the so-called classical natural gradient update rule~\cite{amari2016information} and existing quantum generalizations ~\cite{stokes2020quantum, koczor2019quantum, van2020measurement}. However, in the quantum case, existing works have thus far assumed the choice of $\Phi$ corresponding to the so-called \gls{bh} metric. In~\cite{stokes2020quantum}, the authors explored this monotone metric in its relevance to pure state optimization where it becomes unique. For mixed states, existing literature has considered the metric induced by the \gls{qfi}~\cite{koczor2019quantum, van2020measurement}. Doing so may be motivated by a belief that the optimal parameters correspond to an approximately pure density operator, and so suitable low-rank approximations have been considered~\cite{koczor2019quantum,van2020measurement}. In our generic consideration, such near-purity assumptions are not necessary.

Hence, in the next section, we will motivate a different choice of $\Phi$, corresponding to the so-called \gls{bkm} metric, and find that it allows for so-called quantum Fisher efficient learning. 

\section{Efficient learning and sampling of many-body states}

In this section, we first describe the use of the \gls{bkm} metric in the QPNGD update rule \eqref{eq:natural-gradient-update}.  Then we discuss how this update rule, under natural assumptions, meets a desirable asymptotic optimality criteria which characterizes optimization convergence. This is analogous to a known classical result~\cite{amari1998natural}, but first-known for quantum. Finally, we interpret this optimality result in terms of many-body learning and simulation problems.

\subsection{Online quantum natural gradient for a special choice of metric}

Following \eqref{eq:natural-gradient-update}, for learning, consider an online metric-aware update rule with a particular choice of learning rate,
\begin{equation}
    \label{eq:qcrb-gradient-update}
    \bm{\Omega}_{j + 1} = \bm{{\Omega}}_j - \tfrac{1}{j}  \left(\mathcal{I}^{\BKM} ({\bm{{\Omega}}_j})\right)^{+} \tilde{h}(\bm{\Omega})
\end{equation}
where we define $\tilde{h}(\cdot)$ to be an online unbiased estimator obtained by the environment drawing a single pure state $\ket{\bm x}$ from the eigenstates of data density operator $\hat\rho_{\bm \Omega^*}$ (with probability of the corresponding eigenvalue) at each optimization step.  We choose $\tilde{h}(\cdot)$ such that
\begin{equation}
    \mathbb{E}_{\bm x}[\tilde{h}(\bm{\Omega})] = \nabla_{\bm \Omega} D(\hat \rho_{\bm \Omega^*} \Vert \hat \rho_{\bm{{\Omega}} })
,\end{equation}
with $D(\cdot \Vert \cdot)$ being the canonical quantum relative entropy~\cite{wilde2017quantum}, and
\begin{equation}
    [\mathcal{I}^{\BKM} ({\bm \Omega})]_{j, k} \equiv -\partial^2_{\Omega'_j \Omega_k} D
    (\hat{\rho}_{\bm \Omega'} \Vert \hat{\rho}_{\bm{\Omega}})\mid_{\bm \Omega'= \bm \Omega}
.\end{equation}
Thus, the metric potential, as in \eqref{eq:hessian-of-some-potential}, is given by the quantum relative entropy (in either direction since they are equivalent up to third-order c.f. \Cref{prop:forward-reverse-local-equiv}):
\begin{align}
   \label{eq:contrast-is-qre}
    \Phi^{\operatorname{BKM}}(\cdot, \cdot) \equiv D(\cdot \Vert \cdot) 
.\end{align}
Choosing the quantum relative entropy as our contrast functional leads to the \gls{bkm} metric~\cite{bengtsson2017geometry}.

Before we proceed, we discuss the experimental feasibility of the proposed algorithm. Roughly speaking, the implementation of \eqref{eq:qcrb-gradient-update} requires the ability to estimate first- and second-order derivatives in a relative entropy loss. For a general density operator ansatz, this is achievable but involves potentially intractable averaging over a quantum thermal distribution~\cite{amin2018quantum}. To this end, a particular variational ansatz class -- termed \glspl{qhbm}~\cite{verdon2019qhbm} -- admits unbiased estimators for these gradients (\Cref{sec:qhbm-applications}) and Hessians (\Cref{sec:estimating-bkm-tensor}) that circumvent such quantum averaging. This works because \glspl{qhbm} use the spectral representation of a density operator, and so the eigenvalue distribution is classical. Hence, the challenging problem of estimating gradients in the quantum log-partition function becomes a classical (and therefore sign problem free) problem and can be estimated offline from the quantum device. Particularly for low temperatures, this classical problem is still NP-hard in general~\cite{harrow2020classical,sly2010computational,sly2012computational}, but meaningful separation is expected for broad classes of problem instances~\cite{alhambra2022quantum} (see \Cref{sec:sample-efficient-many-body}). All in all, as a byproduct of diagonalization, \glspl{qhbm} decouple learning a mixed state into separable quantum and classical statistical learning problems. We refer the reader to Verdon et al.~\cite{verdon2019qhbm} for a comprehensive description of \glspl{qhbm}, though we provide an overview in \Cref{sec:review-qhbm} for completeness.

\subsection{Fisher efficiency}
\label{sec:nd-optimality-complexity}

In the classical case, attaining so-called Fisher efficiency roughly means that the asymptotic accuracy of an estimator, as measured by the error covariance matrix, attains the well-known classical \gls{crb} to first-order in the number of data samples utilized. We will follow in the steps of Amari~\cite{amari1998natural} and apply this idea to optimization.

We may think of an online optimization rule as a statistical estimator by saying that the latest parameters at step $j$ are the estimator given $\mathcal{O}\left(j\right)$ data samples. Fisher efficiency is met for classical online natural gradient descent and a particular choice of learning rate, assuming that the optimal parameters are eventually reached~\cite{amari1998natural}. This implies that, to first-order, such an update rule can achieve the best-case asymptotic measurement scaling which is usually associated to maximum likelihood estimation.

Correspondingly, the quantum analogue of Fisher efficiency, quantum Fisher efficiency, would attain the generalized \gls{qcrb} (\Cref{sec:geometry-qcrb}),
\begin{align}
    \label{eq:fisher-efficient}
    \tag{QFE}
    \Cov(\bm \Omega_j; \bm \Omega^*) &= \tfrac{1}{j} \left(\mathcal{I}(\bm \Omega^*)\right)^{-1} + \mathcal{O}\left(\tfrac{1}{j^2}\right),
\end{align}
to first-order in $j$, where $j$ is the number of optimization steps. Again, the term on the left-hand side can be thought of as a generalized error covariance relative to the optimal parameters.  

In what follows, when we call something ``quantum Fisher efficient", we mean that it attains the generalized \gls{qcrb} under the \gls{bkm} metric. The informal theorem statement which follows says that the learning rule \eqref{eq:qcrb-gradient-update} is optimal in the sense of \eqref{eq:fisher-efficient}, and similarly a corollary which says that the swapped loss -- the one used for variational Gibbs sampling -- meets the same optimality criteria. The formal statements and proofs are given in \Cref{res:qpngd-achieves-qcrb}.

\begin{theorem}
\label{thm:optimality-guarantee}
Suppose that $\mathcal{I}^{\BKM} ({\bm{{\Omega}}})$ is non-singular for all $\bm \Omega$. Furthermore, suppose that $\bm \Omega_j$ converges to the optimal parameters $\bm\Omega^*$ in expectation, i.e., $\mathbb{E}_{\bm x}[\bm\Omega_j] \to \bm\Omega^*$ as $j \to \infty$. In such a case, the learning rule \eqref{eq:qcrb-gradient-update} is quantum Fisher efficient i.e. satisfies \eqref{eq:fisher-efficient}.
\end{theorem}

\begin{corollary}
The equivalent statement holds when one swaps arguments of the loss, $D(\hat\rho_{\bm \Omega^*} \Vert \hat\rho_{\bm{{\Omega}}_j }) \rightarrow D(\hat\rho_{\bm{{\Omega}}_j } \Vert \hat\rho_{\bm \Omega^*})$.
\end{corollary}

Note that no particular coordinates or model structure have been chosen for this result. Since the online estimator achieves quantum Fisher efficiency, one can check that using more data samples at each optimization step can only improve convergence and therefore also achieves quantum Fisher efficiency. Again, the update rule \eqref{eq:qcrb-gradient-update} is simply the online \eqref{eq:natural-gradient-update} where the contrast functional which gives the \gls{bkm} metric \eqref{eq:contrast-is-qre} is taken to match the loss, $\mathcal{L}(\bm \Omega) \equiv D(\hat\rho_{\bm \Omega^*} \Vert \hat\rho_{\bm{{\Omega}}_j })$.  The corollary follows because $D(\hat\rho_{\bm{{\Omega}}_j } \Vert \hat\rho_{\bm \Omega^*})$ is symmetric to third-order and so induces the same local metric as $D(\hat\rho_{\bm \Omega^*} \Vert \hat\rho_{\bm{{\Omega}}_j })$\footnote{See \Cref{sec:quantum-relative-entropy} for details on the quantum relative entropy.}.

There are two intuitive reasons as to why we are able to do this with our particular descent rule and proof strategy. First, we take advantage of the fact that our variational loss will be given precisely by the contrast functional $\Phi$ and so the metric evaluated at the optimum is its curvature at the optimum\footnote{It is important to note that $\mathcal{I}(\bm\Omega)$ is \textit{not} the Hessian of the loss in general, since its definition \eqref{eq:hessian-of-some-potential} is evaluated at $\bm\Omega' = \bm\Omega$, rather than at distinct parameters $\bm\Omega^* \neq \bm\Omega$. Choosing the information matrix to be the Hessian of the loss would make \eqref{eq:natural-gradient-update} reduce to an update rule akin to Newton's method~\cite{amari2016information}.}. Hence, optimization steps respect the fundamental distinguishibility of density operators associated to the loss. The fact that the loss curvature and metric did not identify in this way was remarked during the construction of Stokes et al.~\cite{stokes2020quantum}\footnote{Although, a H{\"o}lder-like bound on their discrepancy was shown.}. The second idea follows the \gls{bkm} choice of metric and is technical; our proof strategy uses the fact that the derivative of quantum relative entropy in local coordinates can be seen as a quantum expectation of the metric's so-called logarithmic derivative\footnote{The interested reader may refer to the discussion surrounding and regarding \eqref{eq:cute-derivative}.}.

In contrast to classical natural gradient descent, no quantum Fisher efficiency result has been shown for prior constructions of quantum metric-aware descent rules~\cite{stokes2020quantum, koczor2019quantum, van2020measurement}. \Cref{thm:optimality-guarantee} provides such a guarantee. We leave open the possibility of showing quantum Fisher efficiency for other choices of $\Phi$\footnote{See the discussion in \Cref{sec:nd-optimality-complexity}.} through a differing proof strategy. Nevertheless, we will show a tractability advantage (\Cref{sec:geometric-regularizers}) which is specific to our choice in \Cref{eq:contrast-is-qre}.

In this sense, it is interesting to consider whether a similar optimality result can be found (through a differing proof strategy\footnote{As has been mentioned, and referring to \eqref{eq:cute-derivative} for details, our proof strategy uses the fact that the derivative of quantum relative entropy in local coordinates can be seen as a quantum expectation of the metric's so-called logarithmic derivative.}) for a distinct choice of contrast functional\footnote{This has been called a generalized quantum relative entropy~\cite{lesniewski1999monotone}.} \eqref{eq:hessian-of-some-potential} e.g. the one which induces the \gls{bh} metric. This would be compelling because, for example, the \gls{bh} instance of quantum Fisher efficiency \eqref{eq:fisher-efficient} gives the tightest asymptotic scaling guarantee, at least provably in the single-parameter case (\Cref{sec:geometry-qcrb}). To this end, in \Cref{apdx:prop:compatibility}, we show that there exists \textit{some} parameter estimation strategy which can attain the \gls{bh} scaling for \glspl{qhbm} since this is not in general guaranteed for an arbitrary parameterization. We leave open the possibility that, in particular, the \gls{bh} analogue of \eqref{eq:qcrb-gradient-update} achieves this scaling.

As in the classical case, the number of model parameters may in general exceed the dimension of the relevant mixed state manifold. In such a case, the information matrix is clearly guaranteed to be singular\footnote{See \Cref{apdx:lem:traceless-spanning} for a reminder.} and so \Cref{thm:optimality-guarantee} does not directly apply. In fact, convergence faster than \eqref{eq:fisher-efficient} becomes possible due to over-fitting~\cite{martens2014new}, implying a tradeoff with generalization. Nevertheless, classically and under realistic assumptions, convergence rate improvements akin to \Cref{thm:optimality-guarantee} that do not come at the expense of generalization have been shown for this case by demonstrating that such over-parameterized models behave like their local linear approximations\footnote{This is the well-known Neural Tangent Kernel~\cite{jacot2018neural} idea. Note that the neural network function behaves linearly, but not (in general) the loss.} (at the initial parameters) throughout optimization~\cite{zhang2019fast,cai2019gram,martens2014new}. We expect similar guarantees here.

Note as well that, as in the classical analogue, it is assumed that the model parameters eventually converge to the optimal ones. This will not hold in general for non-convex objectives, as are expected with \glspl{ebm} and \glspl{qnn}. However, in practice, a reasonable local optimum might be a sufficient proxy for the global optimum, in which case a property analogous to quantum Fisher efficiency may still (approximately) hold~\cite{martens2014new}.

\subsection{Applications}

\subsubsection{Sample-efficient learning of many-body states}
\label{sec:sample-efficient-many-body}

We now briefly discuss one concrete application of the prescribed update rule being quantum Fisher efficient. As described, for the quantum Hamiltonian learning problem\footnote{In the language of \glspl{qhbm}, this is the usual type of problem considered when optimizing the reverse relative entropy loss~\cite{verdon2019qhbm}.} (\Cref{sec:on-optimality}), it was shown in Anshu et al. that a non-trivial data sample complexity \eqref{eq:anshu-sample}, which is polynomial in the number of qudits, can be achieved for general quantum many-body Hamiltonians that are known to have a spatially local interaction graph~\cite{anshu2021sample}. This result analyzes an offline classical learning algorithm known as maximum entropy estimation paired with an assumption that, for qubits, one measures Pauli tensor operators with the appropriate locality.

A key component of the analysis was showing a strong convexity property in the log-partition function. Incidentally, the Hessian of the log-partition function is precisely the \gls{bkm} information matrix for a particular choice of coordinates, although this connection was not identified by them. Using this strong convexity, we work out in \Cref{sec:many-body} that, for a particular ansatz, quantum Fisher efficiency implies recovering the polynomial sample-efficiency \eqref{eq:anshu-sample}. Our result is worked out in \Cref{thm:sample-efficient}, and of course subject to the same concessions as the quantum Fisher efficiency result itself. Note that the assumption of eventual convergence generally holds for this scenario since it is strongly convex.

The ansatz plugged into quantum Fisher efficiency to achieve this result corresponds to performing the same local, Pauli tensor measurements. In particular, the ansatz is that of a quantum exponential family\footnote{See \Cref{apdx:mixture-exponential}.} with a strong inductive bias in that it is assumed that the relevant non-identity Pauli operators $\hat E_l$ are known (an assumption shared with Anshu et al.):
\begin{align}
\label{eq:inductive-bias-ansatz}
    \hat \rho_{\bm \Omega} = \tfrac{1}{\mathcal{Z}_{\bm \Omega}}e^{- \sum^p_{l=1} \Omega_l \hat E_l}
,\end{align}
using the notation of \Cref{sec:on-optimality}. The online gradients are worked out in \Cref{sec:online-loss-gradient}.

Existing variational approaches to learning Hamiltonians from Gibbs states have thus far not shown similar, concrete sample-efficiency guarantees~\cite{yapage2008information, kieferova2017tomography, amin2018quantum,verdon2019qhbm,zoufal2021variational,kokail2021quantum,larose2019variational,cerezo2020variational}.  Note that we use the same amount of enhancement as the fixed-measurement approaches\footnote{In other words, we do not perform joint measurement over several Gibbs state copies i.e. quantum enhancement~\cite{huang2021quantum}. In fact, for learning, the ansatz which we analyze measures the same observables (\Cref{sec:sample-efficient-many-body}).}.

To understand the impact of this result, we first observe that strong convexity in the loss function applied to the ansatz \eqref{eq:inductive-bias-ansatz} generally implies expected fast convergence even for vanilla gradient descent. However, we will see from \eqref{eq:rgf-main} that when we use the update rule \eqref{eq:qcrb-gradient-update} then the same guarantee holds approximately under any smooth, invertible re-parameterization. The ability to re-parameterize, i.e., choose a different model which parameterizes the same subset of $\mathcal{M}^{(N)}$ spanned by \eqref{eq:inductive-bias-ansatz}, unlocks new routes to tackling time-inefficiency. For example, we will be able to consider models with friendlier gradients that can be more tractably estimated classically. For example, for \glspl{qhbm}\footnote{Note that \glspl{qhbm} can be projected into proper coordinates via the method described in \Cref{apdx:over-param}.}, the fact that the model Gibbs state is always represented diagonally allows one to circumvent the log-partition function depending on quantum parameters; instead, the log-partition function depends only on classical parameters. Quantum algorithms which diagonalize density operators so as to translate the sampling problem to be classical are expected to be efficient in broad cases where directly classically sampling from the quantum thermal distribution is not~\cite{alhambra2022quantum} (for example, under the \gls{eth} assumption~\cite{chen2021fast}). Given that these quantum algorithms are non-variational, a natural follow-up is to characterize problem-specific time-complexity speedups that may be feasible by means of particular fixed-depth diagonal ansatze.

Even with precise knowledge of the specific relevant instance of \eqref{eq:inductive-bias-ansatz}, positing a distinct model which is an exact re-parameterization may be unrealistic in some cases. However, we may expect similar behavior given a sufficient approximation. In this way, when we do not have the strong prior of knowing the relevant $\{\hat E_l\}$, we may still be able to posit a reasonably approximate re-parameterization of the space spanned by \eqref{eq:inductive-bias-ansatz}. We note that for high temperature (above the phase transition point) a separate approach has been analyzed which delivers optimal sample- and time-complexity with sample complexity $S = O(\log n / (\beta \eps)^2)$ and time-complexity $\mathcal{O}(nS)$~\cite{haah2021optimal}.

\subsubsection{Sample-efficient modelling of Gibbs states}
A similar story can be told for the quantum Gibbs sampling literature.  In this case, we are given the Hamiltonian $\hat K$ as input and may seek to model its corresponding Gibbs state in as few iterations as possible (essentially variational inference~\cite{murphy2012machine}). Similar to the learning problem, at each iteration of variational optimization, we measure the closeness of the model state to the input Hamiltonian by re-preparing the latest model state at least once. For this reverse problem, we recover the same count of required measurements of the known Hamiltonian against the candidate Gibbs state \eqref{eq:anshu-sample}.  For more discussion on quantum Gibbs sampling, see \Cref{sec:numerics}. As with learning, a similar polynomial efficiency is expected for variational quantum Gibbs sampling combined with vanilla stochastic gradient descent, if we use the ansatz \eqref{eq:inductive-bias-ansatz} given the strong convexity~\cite{chowdhury2020variational}. However, again, our method allows flexibility in the choice of model.

\section{Quantum-Probabilistic Mirror Descent}
\label{sec:qpmd}

\subsection{Motivation}

In \Cref{sec:tractability}, we discussed the distinction between model and data samples. We noted in analyzing our online natural gradient update rule that there is an assumption that the metric quantity, which requires no data samples but quadratic-scaling model samples, is estimated sufficiently accurately. This matches the seminal analysis of Amari~\cite{amari1998natural}. In practice, while data state re-preparation may be more costly than model state re-preparation, both are likely to be considered finite resources. Hence, variance in estimation of the information matrix can lead to noisy and ineffective optimization behavior. To address this issue, we derive a robust \textit{mirror descent} algorithm~\cite{raskutti2015information} (\Cref{alg:qmd}) which approximates the optimization behavior of metric-aware updates (for a particular choice of metric) without computing any second-order terms directly. Establishing this duality is a first for quantum metric-aware algorithms. Our numerics will corroborate its performance impact (\Cref{fig:VQT,fig:sequences}).

To see the model sample complexity more explicitly, assume the \gls{qhbm} ansatz and that the unitary component is parameterized so as to be differentiable through parameter shifts. Recall that, in general, each parameter shift is measured as a separate observable and so the sample cost of $k$ parameter shifts will scale as $\mathcal{O}(k \eps^\alpha)$ for some small $\alpha$ given a desired (in-)accuracy $\eps$~\cite{knill2007optimal}. Then, estimating the information matrix \eqref{eq:hessian-of-some-potential}, which has $\Theta(d^2)$ elements, costs $\Theta(q^2)$ parameter shifts where $d = q + c$ and $q$ gives the number of quantum parameters. We work out the estimation of the \gls{bkm} metric tensor for \glspl{qhbm} in \Cref{sec:estimating-bkm-tensor}. On the other hand, we can perform each mirror descent update with $\Theta(kq)$ parameter-shifts where $k$ is the number of inner-loop steps in \Cref{alg:qmd}. We will see empirically (\Cref{fig:VQT}) that $k$ can be considerably favorable towards mirror descent \eqref{eq:natural-gradient-update}.

\subsection{Duality}
\label{sec:geometric-regularizers}

In \Cref{sec:nd-optimality-complexity}, we saw that our specification of metric-aware descent is quantum Fisher efficient when we choose the metric to be that of the \gls{bkm}. Now, we will see that this same choice of metric offers an interesting dual, equivalent first-order implementation of metric-aware descent.

\begin{theorem}[Dual relationship between mirror descent and metric-aware descent]
  \label{thm:md-ngd-duality}

There exists two choices of coordinates $\bm \varphi$ and $\bm \eta$ of $\mathcal{M}^{(N)}$ such that the \gls{bkm} metric-aware descent relation \eqref{eq:natural-gradient-update} in $\bm \varphi$ is equivalent to the mirror descent relation
\begin{align}
\label{eq:md-relation}
\tag{QPMD}
    \bm\eta_{j+1} &= \arg\min_{\bm\eta} \left[ \langle \bm \eta,  \nabla_{\bm \eta_j} \mathcal{L}( {\bm{{\eta}}_j } )\rangle + \lambda D  ( \hat\rho_{\bm \eta} \Vert \hat \rho_{\bm \eta_j} ) \right]
,\end{align}
where $\langle \bm x, \bm y \rangle := \bm x^T \bm y$ is the usual Euclidean inner product.

\end{theorem}

The result follows from the fact that the two special choices of parameterization can be considered dual to one another in the sense of being related by Legendre transform. It is known that the \gls{bkm} metric is the unique monotone metric for which an analogous duality exists\footnote{See \Cref{sec:md-duality}. Specifically, the \gls{bkm} metric is the unique monotone metric for which the mixture and exponential flat affine connections are mutually dual~\cite{hasegawa1997exponential,grasselli2001uniqueness}.} and so this allows us to recover a mirroring result (\Cref{thm:md-ngd-duality}). The formal statement and proof is given in \Cref{thm:dual-md=ngd}. 

By treating the minimization of \eqref{eq:md-relation} as a sub-problem (inner-loop) that is solved with gradient descent for each $j$, we obtain \Cref{alg:qmd}. In comparison with \Cref{alg:qngd}, no inversion is required and we have transformed a second-order method to be entirely first-order, thus requiring measurement of fewer parameter-shifted observables at each step.  Hence, assuming $k$ inner-loop gradient steps and $\mathcal{O}(d)$ gradient estimation model sample complexity for $d$ parameters, the model sample complexity becomes $\mathcal{O}(kd)$ for each update step.

We will notice empirically that $k$ scaling sublinearly in $d$ can suffice for achieving good convergences (\Cref{exp:ising-equilibrium}).  Further, given the equivalence from \Cref{thm:md-ngd-duality}, the analogous guarantees of \Cref{sec:nd-optimality-complexity} can apply to \Cref{alg:qmd} regarding optimality.  These results indicate that \eqref{eq:md-relation} can be leveraged as a sample-efficient approximation to \eqref{eq:natural-gradient-update} within the \gls{bkm} geometry.

As stated in the theorem, the equivalence holds for a special choice of coordinates which in general differ from e.g. the \gls{qhbm} parameterization. We may then ask how the two algorithms relate under coordinate re-parameterization. Indeed, for arbitrary smooth, injective re-parameterizations, \eqref{eq:md-relation} and \eqref{eq:natural-gradient-update} are two generally distinct discretizations of the same coordinate-invariant flow over $\mathcal{M}^{(N)}$:
\begin{align}
\label{eq:rgf-main}
\tag{QPRGF}
    \frac{d\bm \Omega(t)}{dt}  &= -\mathcal{I}^{-1}(\bm \Omega(t)) \nabla \mathcal{L}(\bm \Omega(t))
,\end{align}
paired with a choice of boundary condition, $\bm \Omega(0)$. Our derivation (\Cref{sec:duality-qhbm}) is specific to the \gls{bkm} choice, again.

For $\lambda \rightarrow \infty$, the discretizations approach the underlying flow. Interestingly, we remark in our derivation that \eqref{eq:md-relation} can be considered a “more accurate” discretization of \eqref{eq:rgf-main} in the sense of being more faithful to the geometry of the search space. The analogous remark has been made classically~\cite{gunasekar2021mirrorless}.

\section[Numerical results: Quantum Gibbs sampling]{\texorpdfstring{Numerical results:\\ Quantum Gibbs sampling}{Numerical results: Quantum Gibbs sampling}}
\label{sec:numerics}

\subsection{Motivation}
Calculating the properties of quantum systems in thermal equilibrium is an important task.  Properties of interest include simple expectation values~\cite{motta2019qite} and correlation functions~\cite{dallaire2016method, cohn2020minimal, sun2021qite}, which are relevant to chemistry~\cite{kassal2011simulating} and materials science~\cite{bauer2020quantum}.  Quantum Gibbs sampling is also an important subroutine in many quantum algorithms which seek quantum advantage on classical tasks such as semidefinite programs~\cite{brandao2017quantum, van2019improvements, brandao2019quantum, van2020quantum, brandao2022faster}, Monte Carlo integration~\cite{montanaro2015quantum}, Bayesian inference~\cite{harrow2020adaptive}, and principal component analysis~\cite{lloyd2014quantum}. When a quantum computer is used to sample and coherently post-process the thermal state of interest, the task is called \textit{quantum Gibbs sampling}.
 
Non-variational methods for quantum Gibbs sampling exist~\cite{terhal2000problem, temme2011quantum}, including algorithms with run times bounded in terms of temperature and accuracy.  However, many such time-bounded algorithms require quantum phase estimation~\cite{abrams1999quantum, poulin2009sampling, bilgin2010preparing, riera2012thermalization, yung2012quantum} or spectral gap amplification~\cite{somma2013spectral, chowdhury2016quantum}, so that their use must await fault tolerance.  Others impose restrictions, such as requiring the terms of the target Hamiltonian to commute~\cite{kastoryano2016quantum, ge2016rapid}, or assume the ability to directly apply quantum channels in hardware via engineered dissipation~\cite{brandao2019finite}.

Recently, \glspl{vqa} have begun to be developed for the task of quantum Gibbs sampling.  In this setting, the quantum computer is used to train a model for the relevant state.  Once a model is trained, samples can be drawn from the model and coherently post-processed.  Such variational algorithms fall into two categories, depending on whether their loss functions are built from time-dependent or equilibrium variational principles.

Given a Hamiltonian $\hat{K}$ and a state $\ket{\psi}_0$ describing a physical system, recall that the Schr{\"o}dinger equation specifies how $\ket{\psi}_0$ evolves into $\ket{\psi}_t$ over time~\cite{schrodinger1926undulatory}.  Approximate solutions to this equation can be found by postulating a parameterized trial wavefunction (equivalently, a parameterized quantum circuit), then optimizing it under McLachlan's variational principle~\cite{mclachlan1964variational}.  The same variational principle can be applied to the imaginary time version of the Schr{\"o}dinger equation~\cite{broeckhove1988equivalence} to yield variational ground states.  When quantum computers are used to optimize the McLachlan objective, we have \gls{vqite}~\cite{mcardle2019variational, zoufal2021variational}.  The solutions obtained with this method can be bootstrapped into quantum Gibbs samplers~\cite{white2009minimally, motta2019qite, sun2021qite} using methods adapted from the tensor network literature~\cite{verstraete2004matrix, zwolak2004mixed}.

To take the equilibrium perspective, recall that a system with Hamiltonian $\hat{K}$ in thermal equilibrium at inverse temperature $\beta$ is described by the Gibbs state \eqref{eq:gibbs}.  This state is the unique minimizer of the Helmholtz free energy~\cite{rau2017statistical}.  Thus to obtain an approximation to the Gibbs state, we can postulate a variational mixed state ansatz and optimize the parameters to minimize the free energy of that ansatz.  When quantum computers are used to optimize the free energy, we have \gls{vqt}~\cite{verdon2019qhbm}.  Some approaches descend approximations to the free energy~\cite{wu2019variational, wu2019variational, chowdhury2020variational, wang2021variational}, but if the right ansatz is used, then the free energy itself can be optimized via gradient-based optimizers~\cite{martyn2019product, verdon2019qhbm, liu2021solving}.  In the next section we study the numerical performance of both \gls{qpngd} and \gls{qpmd} for training models against a free energy loss.

\subsection{Equilibrium simulation of a Transverse-Field Ising Model}
\label{exp:ising-equilibrium}

\begin{figure*}[ht]
  \centering
     \subfloat[][]{\includegraphics[width=0.85\textwidth]{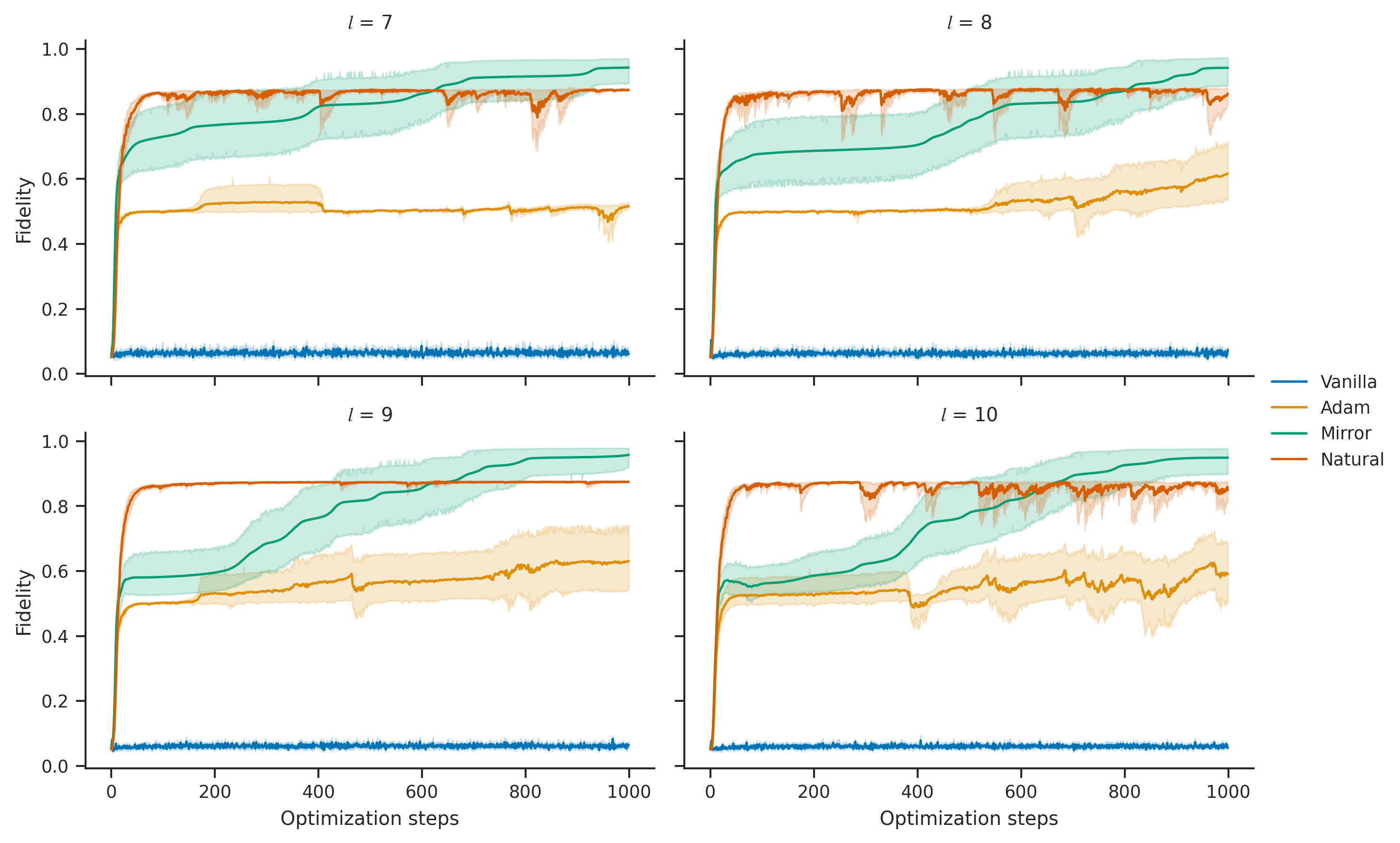}\label{fig:head-to-head-fids}}
  \caption{Simulation fidelities\footnote{See \Cref{fig:VQT-full} for the optimization loss curves.} (16 trials per algorithm) as a function of the number of optimization steps. We used 500 quantum data samples are used per optimization step for each algorithm. The \gls{qhbm} is parameterized with an $l$-layer \gls{qhea}~\cite{verdon2019qhbm} for the \gls{qnn} and a fully connected Boltzmann machine for the \gls{ebm}. We train against a 6-qubit \gls{tfim} \eqref{eq:tfim-h} at critical transverse field at inverse temperature $\beta = 2.0$ using the forward relative entropy loss. The number of required parameter shifts for a single iteration of Vanilla \gls{sgd}, \gls{qpngd}, and \gls{qpmd} is $2q$, $2q(q+1) + 2q$, and $2kq$, respectively. Here, the number of quantum parameters is $q = 17l$ and $k$ is the number of \gls{qpmd} inner loop steps. We use inverse learning rate $\lambda_j \equiv 1/0.1$ (see e.g. \Cref{alg:qngd,alg:qmd}) throughout. For \gls{qpmd}, $k \approx 20$ was required for consistent convergence, independent of $l$.}
\label{fig:VQT}
\end{figure*}

The \glsfirst{tfim} is a simplified model of many important physical systems~\cite{stinchcombe1973ising}.  The Hamiltonian of the model can be written as
\begin{align}
\label{eq:tfim-h}
\hat{H}_{\operatorname{TFIM}} = -J\sum_{\langle i, j \rangle}\hat{Z}_i\hat{Z}_j - \lambda\sum_i\hat{X}_i,
\end{align}
where $\hat{X}_j$ and $\hat{Z}_j$ are the X and Z single-qubit Pauli operators acting on the $j$th qubit, $J, \lambda$ are parameters chosen to model the system of interest, and $\langle i, j \rangle$ indexes pairs of qubits on some lattice.  The classical limit $\lambda=0$ was originally introduced to study ferromagnetic phase transitions~\cite{lenz1920beitrag, ising1924beitrag, ising2017fate}.  The quantized version was introduced by Heisenberg~\cite{heisenberg1928theorie}.  It has since been used to calculate properties of many physical systems displaying ferromagnetism, such as ferroelectric crystals~\cite{blinc1960isotopic, de1963collective} and rare earth magnets~\cite{wang1968collective}.  In these settings, $J$ depends on the distance between lattice sites in the crystal and $\lambda$ depends on the tunneling frequency of the particles between sites.

In this section we study the equilibrium properties of the \gls{tfim} using a simulation of quantum Gibbs sampling.  More specifically, given an inverse temperature $\beta$, the task is to simulate the associated thermal state 
\begin{align}
\label{eq:target-tfim}
    {\hat \sigma}_{\beta} = e^{-\beta \hat{H}_{\operatorname{TFIM}} } / {\mathcal{Z}_{\beta}} \quad \text{where} \quad \mathcal{Z}_{\beta} = \tr[ e^{-\beta \hat{H}_{\operatorname{TFIM}}}]
.\end{align}
We choose a \gls{qhbm} $\hat\rho_{\bm\Omega}$ as our ansatz and train it to represent $\hat{ \sigma}_\beta$ using the free energy as our loss function\footnote{This makes our strategy an instance of \gls{vqt}, see \Cref{par:vqt}}.  Equivalently, we minimize the forward quantum relative entropy $D(\hat\rho_{\bm\Omega} \Vert \hat{\sigma}_\beta)$.

For our investigation we choose the target Hamiltonian to be equation \eqref{eq:target-tfim} on a six qubit chain.  We choose ${J=\lambda=1}$, which is the quantum critical point of the system~\cite{elliott1970ising}; we choose a quantum critical point because these points often have rich entanglement structure and high quantum complexity~\cite{osborne2002entanglement}.  For our model we choose our unitary to be a \glsfirst{qhea} with $l$ layers and our energy function to be a fully connected Boltzmann machine.  We compare four different optimizers for minimizing the free energy loss: \glsfirst{sgd}, Adam, \glsfirst{qpngd} (\Cref{alg:qngd}), and \glsfirst{qpmd} (\Cref{alg:qmd}).  We also test four different ansatz depths.

The results are shown in \Cref{fig:VQT}.  We see that both of our proposed algorithms, \gls{qpngd} and \gls{qpmd}, outperform Adam (which does not converge usefully) while using a fixed resource of data samples per iteration.  While \gls{qpngd} ascends more quickly to a high fidelity, it jitters at a suboptimal fidelity.  In constrast, we see that \gls{qpmd} smoothly ascends to the optimum, despite using fewer model samples per iteration than \gls{qpngd}. As motivated theoretically in \Cref{sec:qpmd}, \gls{qpmd} can be considered a (model) sample-efficient approximation to \gls{qpngd}. In fact, Adam can be understood as an approximate version of classical \gls{ngd} leveraging an online estimator of the diagonal elements of the metric~\cite{kingma2014adam}. 

\section{Efficient generative modeling of quantum-stochastic processes}

\label{eq:modeling-paths}

\begin{figure}[ht]
  \centering
  \includegraphics[width=1\columnwidth]{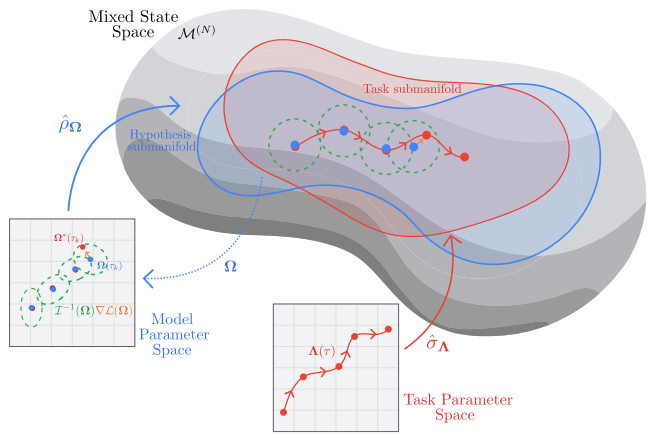}
  \caption{Intuitive depiction of variational generative modelling of target paths in Gibbs state space. Pictured above in blue is the model hypothesis class; $\{\hat{\rho}_{\bm{\Omega}}:\bm{\Omega}\in\mathbb{R}^P\}\subseteq\mathcal{M}^{(N)}$, and in red, the task space $\{\hat{\sigma}_{\bm{\Lambda}}:\bm{\Lambda}\in\mathbb{R}^R\}\subseteq\mathcal{M}^{(N)}$. Also pictured is a task parameter space path $\bm{\Lambda}(\tau)$ and its embedding $\hat{\sigma}_{\bm{\Lambda}(\tau)}$ into density operator space. By considering a sequence of points $\{\tau_k\}_k$ along this path, we obtain a sequence of target quantum states $\hat{\sigma}_{\bm{\Lambda}(\tau_k)}$. Via chained metric-aware optimizers, we can sequentially or recursively find the sequence of optimal parameters $\{\bm{\Omega}^*(\tau_k)\}_k$ for which our model approximates each corresponding state $\hat{\rho}_{\bm{\Omega}^*(\tau_k)}\approx \hat{\sigma}_{\bm{\Lambda}(\tau_k)}$.
  }
\label{fig:md-sequence}
\end{figure}

Quantum Fisher efficiency is an asymptotic result and so may bring untenable constant factors. From the proof, for a finite number of optimization steps, fast convergence in the error covariance occurs in the neighborhood where the loss is well-approximated by a convex quadratic function (\Cref{sec:tuning-discrete}). For this reason, initialization strategies which begin the optimization in proximity to the optimum would help achieve practical performance for many tasks of interest. As we motivate below, there are families of practical scenarios, which can be summarized as the modeling of sequences of mixed states that lie on a continuous path in some task space, which are conducive to a simple initialization strategy: initializing parameters at the optimum of the previous element in the task sequence. We call this \textit{chained initialization}. As we demonstrate empirically below, this strategy is advantageous (\Cref{sec:sequences-states}) for simulations of both real and imaginary time evolution, amongst other potential applications.

\subsection{Sequences of States}
\label{sec:sequences-states}

Often, one wishes to model the equilibrium distribution of a quantum system as certain parameters of the target Hamiltonian are continuously modified.  One such scenario is molecular geometry optimization, where parameters in the Hamiltonian represent inter-particle distances.  The goal is to minimize the free energy with respect to those parameters, to find what configuration the molecule takes on in thermal equilibrium~\cite{nagaoka2003structure}.  Another scenario is quantum annealing, where we wish to reach equilibrium as a parameter of the Hamiltonian is continuously tuned~\cite{apolloni1989quantum, apolloni1990numerical, finnila1994quantum}.  A valid schedule can be constructed using measurements at multiple parameter points, using for example the Bashful Adiabatic Algorithm~\cite{jarret2018quantum}.  From a learning perspective, if one has quantum measurement access to a physical system, then one may wish to learn a generative model of that system as some physical parameter is continuously varied.  One such scenario is learning the entanglement Hamiltonian of a quantum field during time evolution~\cite{wen2018entanglement}.

For such parametric families of tasks and sufficiently fine evolution, neighboring target states may be expected to be close to one another according to the metric over $\mathcal{M}^{(N)}$. On the other hand, their distance in parameter space may be arbitrarily far. Hence, optimizers such as Adam which are unaware of the metric may not meaningfully benefit from initializing model parameters at the previous optimum. However, through a basic chaining of \gls{qpmd} optimization loops, we can leverage their adjacency on  $\mathcal{M}^{(N)}$ while continuing not to refer explicitly to the Euclidean parameter space.

Concretely, suppose we have a parameterization  $\bm{\Lambda}\in \mathbb{R}^R$ of a task space.  The task parameters (which could be target Hamiltonian parameters, time, temperature, or others) specify the target state $\hat{\sigma}_{\bm{\Lambda}}$. Suppose we are interested in a \textit{path} $ \bm{\Lambda}(\tau)$ in this space of parameters where $\tau\in[0,T]$. We can consider a partitioning $\{\tau_{k}\}_{k=1}^M\subset [0,T]$ of the path which gives the sequence of target states $\{\hat{\sigma}_{\bm{\Lambda}({\tau_{k}})}\}_{k=1}^M$. We wish to learn a sequence of parameters so that $\hat{\rho}_{\bm{\Omega}^{*}(\tau_k)} \approx \hat{\sigma}_{\bm{\Lambda}({\tau_{k}})}$ for each $k$.

Assuming the aforementioned geometric locality between neighboring states motivates our initialization strategy.  Having trained model $\hat{\rho}_{\bm{\Omega}(\tau_k)}$ in the sequence, we simply propose initializing the parameters of the next model $\hat{\rho}_{\bm{\Omega}(\tau_{k+1})}$ at $\bm{ \Omega}^*(\tau_k)$ at the start of its optimization.  We call this \textit{chained initialization}, written explicitly in \Cref{alg:seq} (\Cref{sec:gen_seq}).  We can view this as a straightforward meta-learning for initialization \cite{nichol2018first,verdon2019learning} technique.

\subsubsection{Numerical results}

To test this initialization strategy numerically, we consider a simulation scenario for a sequence of states. In particular, we consider a sequence of thermal states defined by discretely varying the inverse temperature $\beta$ in \eqref{eq:target-tfim} from $0.5$ to $2.25$.  This means the sequence of target states undergoes a cooling process, also known as imaginary or Wick-rotated time evolution~\cite{goldberg1967integration}.  We additionally fix $J = \lambda = 1$.  As our model for the system at each temperature, we choose a \gls{qhbm} with seven \gls{qhea} layers.  This depth was chosen because it was found to be sufficient when learning the $\beta=2$ temperature in \Cref{fig:VQT}.  Similarly, the energy function is chosen to be a fully connected Boltzmann machine. 

\begin{figure*}[ht]
  \centering
  \subfloat[][Simulating 6-qubit states corresponding to varying $\beta$. 500 optimization steps are used for the first state in the sequence, and 100 thereafter.]{\includegraphics[width=0.9\textwidth]{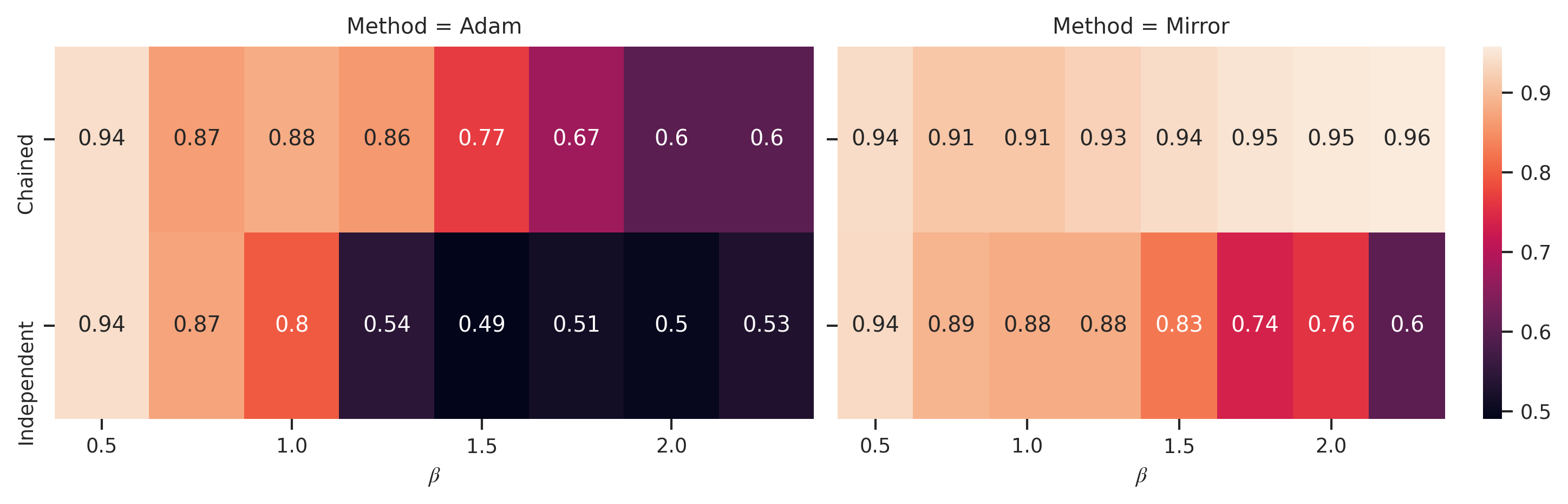}\label{fig:metavqt-heat}}
  \\
  \subfloat[][Learning 4-qubit states generated by a sequence of time-evolution maps. A known model state is given as input, and 100 optimization steps are utilized for each recursive application of the map. A fixed $\beta = 2.0$ is chosen.]{\includegraphics[width=0.9\textwidth]{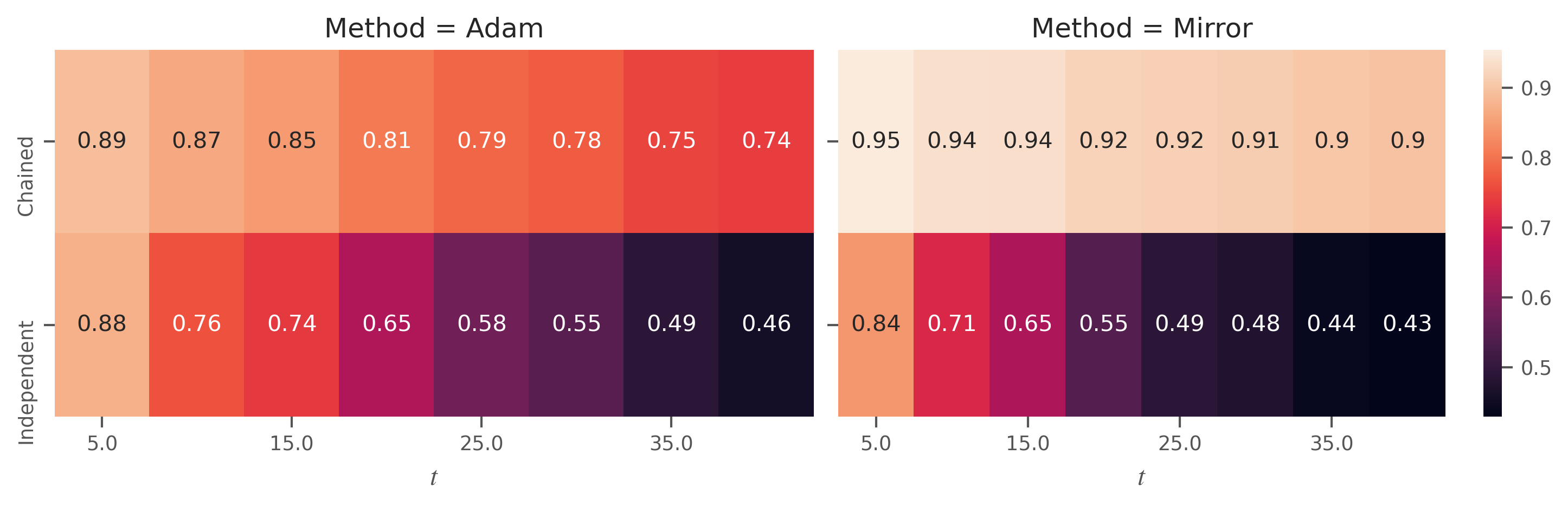}\label{fig:qvartz-heat}}
  \caption{Fidelities for sequences experiments (16 trials per cell). Fidelity is the average over the last 10 steps. Full curves with confidence intervals shown in \Cref{fig:metavqt-curves,fig:qvartz-curves}. 500 quantum data samples are used per optimization step for each algorithm. The \gls{qhbm} is parameterized with an $7$-layer \gls{qhea}~\cite{verdon2019qhbm} for the \gls{qnn} and a fully connected Boltzmann machine for the \gls{ebm}.}
  \label{fig:sequences}
\end{figure*}

In \Cref{fig:metavqt-heat}, we compare two different initialization strategies across two different optimization algorithms, applied to the cooling process described above.  The ``independent" row in each subfigure corresponds to randomly initializing the parameters of the model at each temperature step, while ``chained" corresponds to initializing the parameters at the previous optimum.  For both tested optimization algorithms, Adam and \gls{qpmd}, chained initialization helps find better optima.  We see that \gls{qpmd} outperforms Adam under either initialization strategy, corroborating the benefits seen in \Cref{fig:VQT}.  Figures displaying full training curves and confidence intervals at each temperature step can be found in the appendix (\Cref{fig:metavqt-curves}). Again, chaining is particularly useful here when combined with \gls{qpmd} because it is aware of locality between neighboring states in the sense of statistical distance. Indeed, neighboring states may be arbitrarily far in parameter space and so initialization at the previous optimum is not sufficient for other optimizers (such as Adam) which are unaware of the fundamental statistical distance.

\subsection{Sequences of Maps}
\label{sec:sequences-maps}

In the previous section, we talked about simulating and learning sequences of Hamiltonians.  In this section, we describe a different scenario: simulating the evolution of an initial state under a known mapping.  Such mappings are one realization of \textit{quantum-stochastic processes}\footnote{See \Cref{app:qsto_dyn} for background material on open quantum system time evolution.}~\cite{milz2021quantum}.

Calculating the time evolution of quantum systems is an important problem in physics~\cite{abrams1997simulation} and chemistry~\cite{kassal2008polynomial, mcardle2020quantum}, and it is known that time evolution of both open and closed quantum systems could be simulated efficiently using quantum computers~\cite{lloyd1996universal, kliesch2011dissipative}.  Given a Hamiltonian $\hat{H}$, the formulas of Trotter~\cite{trotter1959product} and Suzuki~\cite{suzuki1976generalized} inform how to approximately time evolve states under $\hat{H}$ on a quantum computer~\cite{nielsen2010quantum}.  However, the resulting quantum circuits grow linearly with the simulation time, making them infeasible to implement on near term hardware.

Here, we describe an approach whose circuit depth does not grow with time.  Suppose we are given access to copies of an initial quantum state $\hat{\sigma}_0$ and a known quantum channel $\mathcal{V}_t$ that applies an open quantum system evolution for time $t$.  Our goal is then to simulate the evolution of the initial quantum state under the action of $\mathcal{V}_t$ up to some final simulation time $T$.

We could naively use quantum channel Trotter decomposition to simulate $\mathcal{V}_T$~\cite{kliesch2011dissipative}, but this would take circuit depth scaling with $T$.  Instead, suppose we discretize $\mathcal{V}_t$ into a sequence of short time evolutions $\mathcal{V}_{t_{k+1}, t_{k}}$ for $\{t_k\}_{k=1}^{M}$.  We now define a recursive algorithm in terms of these shorter channels.  Assume we can generate copies of the learned state $\hat\rho_{\bm \Omega^*(t_{k})}$ at step ${k + 1}$ (for $k=1$, this is simply the initial state $\hat{\sigma}_0$).  Then we can apply the $k$th channel $\mathcal{V}_{t_{k+1}, t_{k}}$ to samples from $\hat\rho_{\bm \Omega^*(t_{k})}$, and train the next model $\hat\rho_{\bm \Omega(t_{k+1})}$ against it using the reverse relative entropy loss.  The resulting optimized state then serves as approximation of the true evolved state,
\begin{align}
    \label{eq:thing}
    \hat{\sigma}(t_{k+1}) \approx \mathcal{V}_{t_{k+1}, t_{k}}(\hat \rho_{\bm \Omega^*(t_{k})}) \approx \hat\rho_{\bm \Omega^*(t_{k+1})}.
\end{align}
We may intuitively view this approach as \textit{checkpointing}\footnote{We generalize to sequences of \gls{cptp} maps and provide more background in \Cref{sec:qvartz}. Pseudocode for the generalized form is given in \Cref{alg:recurs}.} of the quantum dynamics of a system relative to an initial state, saving the the information in the \textit{classical} parameters of a model. As a result, our quantum circuit depth requirements can remain constant with respect to $T$, assuming a fixed upper bound to the quantum complexity~\cite{haferkamp2022linear} of states along the trajectory. We refer to this approach as \gls{qvartz}.

Other variational approaches exist for simulating time evolution.  One promising method, called \gls{vff}, uses quantum compilation to learn a diagonalized ansatz for the time evolution operator~\cite{cirstoiu2020variational}.  Like checkpointing, the circuit depth of \gls{vff} is fixed independently of $t$.  However, often one is only interested in applying time dynamics to a limited set of initial states.  We hypothesize that learning the dynamics for all possible input states is generally more difficult than learning the dynamics of a few specific input states; in this case checkpointing may have an advantage over \gls{vff}.  In another approach, a differential equation for the time evolution of the parameters of the model circuit is developed~\cite{li2017efficient};  then, a quantum computer is used to calculate terms in the differential equation, while a classical computer performs the integration.  The checkpointing approach we described above has previously been proposed for pure states~\cite{lin2021real, benedetti2021hardware}.

\subsubsection{Numerical results}

We test our approach by applying unitary time evolution to a mixed initial state.  We take the initial state to be the Gibbs state corresponding to the \gls{tfim} Hamiltonian in equation \eqref{eq:tfim-h}.  We choose the same parameters as in the equilibrium numerics of \Cref{fig:VQT}: $\beta = 2$, ${J = \lambda = 1}$, but now supported on a chain of four qubits.  To enable querying the initial state and evolving it through time, we start each trial of the experiment by training a \gls{qhbm} $\hat{\rho}_{\bm\Omega_0}$ via \gls{vqt} against the Hamiltonian.

For our time evolution we choose a uniform partition $\{t_k\}_{k=0}^{M}$ with ${M = 8}$, such that $t_0 = 0$ and $t_M = 40$ (we work in nondimensional units).  Next, recall the \gls{tfim} Hamiltonian in \Cref{eq:tfim-h}.  We set up two independent Gaussian processes, $p(J, t)$ and $p(\lambda,t)$, with exponential quadratic kernels\footnote{From \href{https://www.tensorflow.org/probability/api_docs/python/tfp/math/psd_kernels/ExponentiatedQuadratic}{the TFP documentation}, we have $k(x, y) = A^2 \text{exp}(-||x - y||^2 / (2 s^2))$.  We used $A=s=1$ for both the $J$ and $\lambda$ processes.}.  We sample each process once for each time interval $(t_k, t_{k+1})$, yielding sequences $\{J_k\}_{k=1}^{M-1}$ and $\{\lambda_k\}_{k=1}^{M-1}$ for the parameters, and $\Delta t_k = t_{k+1}-t_k$ for the time interval lengths.  Letting $\hat{H}_{\textsc{tfim}}^k$ be the \gls{tfim} Hamiltonian with $J = J_k$ and $\lambda = \lambda_k$,  our chosen time evolution operator during the interval $(t_k, t_{k+1})$ was $\smash{\mathcal{V}_{t_{k+1}, t_{k}} = \text{exp}(-i\Delta t_k\hat{H}_{\textsc{tfim}}^k)}$.  We approximate this evolution using Trotterization.  In summary, the sequence of channels mimics time evolution under a low frequency noise applied to the equilibrium Hamiltonian.

In \Cref{fig:qvartz-heat} we compare random initialization against chained initialization under optimization with either Adam or \gls{qpmd}.  Chained initialization helps find better optima under both optimizers.  We also see that \gls{qpmd} outperforms Adam under either initialization strategy, similar to the results in \Cref{fig:metavqt-heat}.  Figures displaying full training curves and confidence intervals at each temperature step can be found in the appendix (\Cref{fig:qvartz-curves}).

\subsection{Tuning Path Discretization Fineness}
\label{sec:tuning-discrete}

In both the state and map sequence scenarios, there is still the question of choosing appropriate step sizes in the task parameter (e.g. $\beta$ in the case of imaginary time evolution).

We recognize that \Cref{thm:optimality-guarantee} is asymptotic. Hence, for a finite and small number of optimization steps, convergence in the error covariance may not be meaningfully efficient. However, from the proof of the theorem, the neighborhood of the optimal parameters for which the asymptotic quantum Fisher efficiency \eqref{eq:fisher-efficient} holds is a region for which the loss converges to a convex quadratic function sufficiently quickly so that, as a necessary condition\footnote{See \Cref{sec:non-asymptotic}.},
\begin{align}
\label{eq:metric_Var_og}
\lVert \mathcal{I}(\bm \Omega_j) - \mathcal{I}(\bm \Omega^*) \rVert_F =\mathcal{O}\left(\frac{1}{j}\right)
,\end{align}
for all $j > J$ where $J$ is a finite constant, and where $\lVert \cdot \rVert_F$ is the Frobenius norm. Hence, since the chained initialization introduced in \Cref{sec:sequences-states} involves initialization at the previous optimum, $\bm{\Omega}_{0}(\tau_{k+1}) := \bm{ \Omega}^*(\tau_k)$, it is natural to prefer a path discretization which satisfies 
\begin{align}
\label{eq:metric_Var}
\lVert \mathcal{I}(\bm{\Omega}^*(\tau_{k+1})) - \mathcal{I}(\bm{ \Omega}^*(\tau_{k})) \rVert_F \leq L
,\end{align}
for some sufficiently small $L$ that evidently controls the constant factor in \eqref{eq:metric_Var_og}, and so serves as a useful empirical indicator.

To select a task space discretization which meets this initialization criteria it is then necessary to be able to translate from task space to model space coordinates. To this end, we work out the relevant expressions specific to real-time evolution, imaginary time-evolution, and generalizations of the latter to any generic path in parametric mixed state space in \Cref{sec:cts-sequence-limit}.

\section{Discussion \& Outlook}

In this paper, we have discussed a general variational algorithm, \gls{qpngd}, for learning and simulating quantum Gibbs states. This algorithm is a generalization of metric-aware descent to quantum. However, we showed that a specific choice of metric (the \gls{bkm} i.e. Hessian of self-relative entropy) that had not been discussed in prior work on optimization leads to several interesting properties; extending any of these results to additional choices of metric is of interest. 

For one, we showed that, for general ansatze, this choice leads to an asymptotic optimality termed quantum Fisher efficiency. This is a novel result for quantum variational algorithms and is analogous to a celebrated classical result~\cite{amari1998natural}.  It describes convergence as a function of data samples in terms of optimal parameter estimation. We suggested extending the applicability of this result, for example to non-convex landscapes and over-parameterized models, as an interesting follow-up with classical precedent~\cite{martens2014new}.

As an application of our quantum Fisher efficiency result, we studied the theoretically and experimentally popular quantum Hamiltonian learning problem~\cite{swingle2014reconstructing,bairey2019learning, qi2019determining,evans2019scalable,anshu2021sample,haah2021optimal,anshu2022some,wang2017experimental,senko2014coherent} and observed that we recover an any-temperature sample-efficient scaling akin to the best-known one of Anshu et al.~\cite{anshu2021sample}. We discussed an analogous scaling result for the ubiquitous Gibbs sampling (simulation) task~\cite{feynman1982simulating, lloyd1996universal,kassal2011simulating, lloyd2014quantum, montanaro2015quantum, dallaire2016method, brandao2017quantum, brandao2019quantum, arute2019quantum, van2019improvements, motta2019qite, cohn2020minimal, van2020quantum, bauer2020quantum, harrow2020adaptive, sun2021qite, brandao2022faster}, the first-known guarantee of its kind to be stated.  Improving our complexity analysis is a desirable follow-up. Finding further applications of quantum Fisher efficiency is also of interest.

Furthermore, we distinguished data samples from model samples and noted that our update rule (\gls{qpngd}) may be costly in terms of number of model thermal state preparations. Hence, again for the same choice of metric, we established a dual first-order algorithm (\gls{qpmd}) which approximates the same underlying coordinate-invariant flow. This extended two popular results in the classical literature~\cite{raskutti2015information,gunasekar2021mirrorless} to the realm of quantum machine learning. 

We leveraged the fact that this metric indicates such a coordinate-invariant flow to suggest carrying over the aforementioned sample-efficiency result to an ansatz amenable to computational efficiency improvements and that acts as a re-parameterization of the same model subset of $\mathcal{M}^{(N)}$. We discussed in depth the \gls{qhbm} ansatz which, through diagonalization, can translate the relevant thermal sampling problem to be classical instead of quantum, thus opening to expected exponential time complexity advantages for broad problem classes (for example, under the \gls{eth} assumption). Hence, this prescribes a general approach which makes headway into the wide-open question~\cite{anshu2022some} of practical algorithms for quantum Hamiltonian learning that are both time- and sample-efficient. Theoretical investigation into specific problem instances and variational diagonal ansatze which admit such advantage are desired.

Having designed and analyzed our main algorithm and its tractable approximation, we then identified, in connection with asymptotic quantum Fisher efficiency, a broad class of scenarios where non-asymptotic convergence may be fast. This class of scenarios is either learning or simulating \textit{sequences} of quantum Gibbs states, where the states in the sequence are in information-geometric proximity.  Such proximity allows metric-aware optimizers like \gls{qpmd} to take advantage of the simple strategy of initializing at the previous optimum.  Sequence scenarios can arise by varying the inverse temperature $\beta$ or the Hamiltonian parameters of a target Gibbs state, or by tracking quantum states through open or closed time evolutions.

For inference on sequences of maps, checkpointing (\gls{qvartz}) converts a problem whose circuit depth scales linearly with time into a problem whose circuit depth saturates at the quantum complexity of the evolved state itself.  This motivates further study of the learnability and representability of quantum states generated by low-depth quantum circuits~\cite{farhi2014quantum,peruzzo2014variational,bravyi2018quantum,cerezo2021variational, haferkamp2022linear,brandao2021models}. We believe our approach to time evolution, coupled with quantum neural architecture search methods~\cite{zhang2020differentiable,du2020quantum,bilkis2021semi,li2020quantum}, may enable the estimation of the quantum complexity of states during time evolution, a topic of central interest in the study of quantum gravity~\cite{brown2018second,
brown2017quantum,
stanford2014complexity,
brown2015complexity,
susskind2020three,
brown2016complexity,
susskind2018black} and holographic condensed matter theory~\cite{jian2021complexity,das2017three}.

In our numerics, we observed a particularly strong characteristic to \gls{qpmd}, and strong performance advantage for sequences, tying together all of the theoretical results discussed above. Further studies into more complex, and especially the higher-difficulty non-stoquastic~\cite{bravyi2014monte} many-body systems would be desirable. To tackle such harder classes of Hamiltonians, numerical studies which invoke ansatze (which may be specific instances of \glspl{qhbm}) that are non-trivially time-efficient in their optimization for specific problem instances are of interest.

Note that simulations presented in this paper were all assuming noiseless quantum computations. For further extensions of this work, we plan to address this limitation by exploring ways to extrapolate and interpolate state paths via representations of quantum dynamical maps as flows in model parameter space, potentially yielding novel variants of fast-forwarding and error mitigation~\cite{van2022probabilistic} strategies which are native to our quantum-probabilistic representations. Demonstrating the effectiveness of such a quantum-probabilistic error mitigation scheme would be a stepping stone towards a real-world experimental demonstration of quantum advantage for time evolution via \acrshort{qvartz} on a noisy, near-term device~\cite{preskill2018quantum,preskill2021quantum}, an objective for future work.

For our core proofs, consistent with the literature taken as baseline, we assumed strong inductive biases in our model which matched quite tightly the physical scenario at hand (though discussed that relaxation is feasible with our approach). In realistic scenarios, one often knows symmetries or general properties of the system, but not the exact space spanned by the Hamiltonian parameterization. As such, explorations of ansatze with physical-context-aware inductive biases are of interest to ensure time efficiency for scenarios with less sharp priors over the Hamiltonian structure. To this end, recent advances towards a theory of quantum geometric deep learning~\cite{verdon2019quantum,larocca2022group,meyer2022exploiting} are interesting directions in which to extent our current framework.

\begin{acknowledgments}
We thank James Martens, Marcin Jarzyna, Josh Dillon, Will Grathwohl, Thomas Wang, Trevor McCourt, and Patrick Coles for helpful discussions.  We thank Joshua Greenberg for artistic illustrations.   All numerical simulations in this paper were performed using our \href{https://github.com/google/qhbm-library}{open source QHBM library}\footnote{Accessible at \url{https://github.com/google/qhbm-library}.}, built on a combination of TensorFlow Quantum~\cite{broughton2020tensorflow} and TensorFlow Probability~\cite{dillon2017tensorflow}.  X, formerly known as Google[x], is part of the Alphabet family of companies, which includes
Google, Verily, Waymo, and others (\url{www.x.company}).
\end{acknowledgments}

\bibliographystyle{unsrt}
\bibliography{references}


\cleardoublepage
\onecolumngrid
\appendix

\section{Additional experiments and figures}

\begin{figure*}[!ht]
  \centering
  \subfloat[][]{\includegraphics[width=0.85\textwidth]{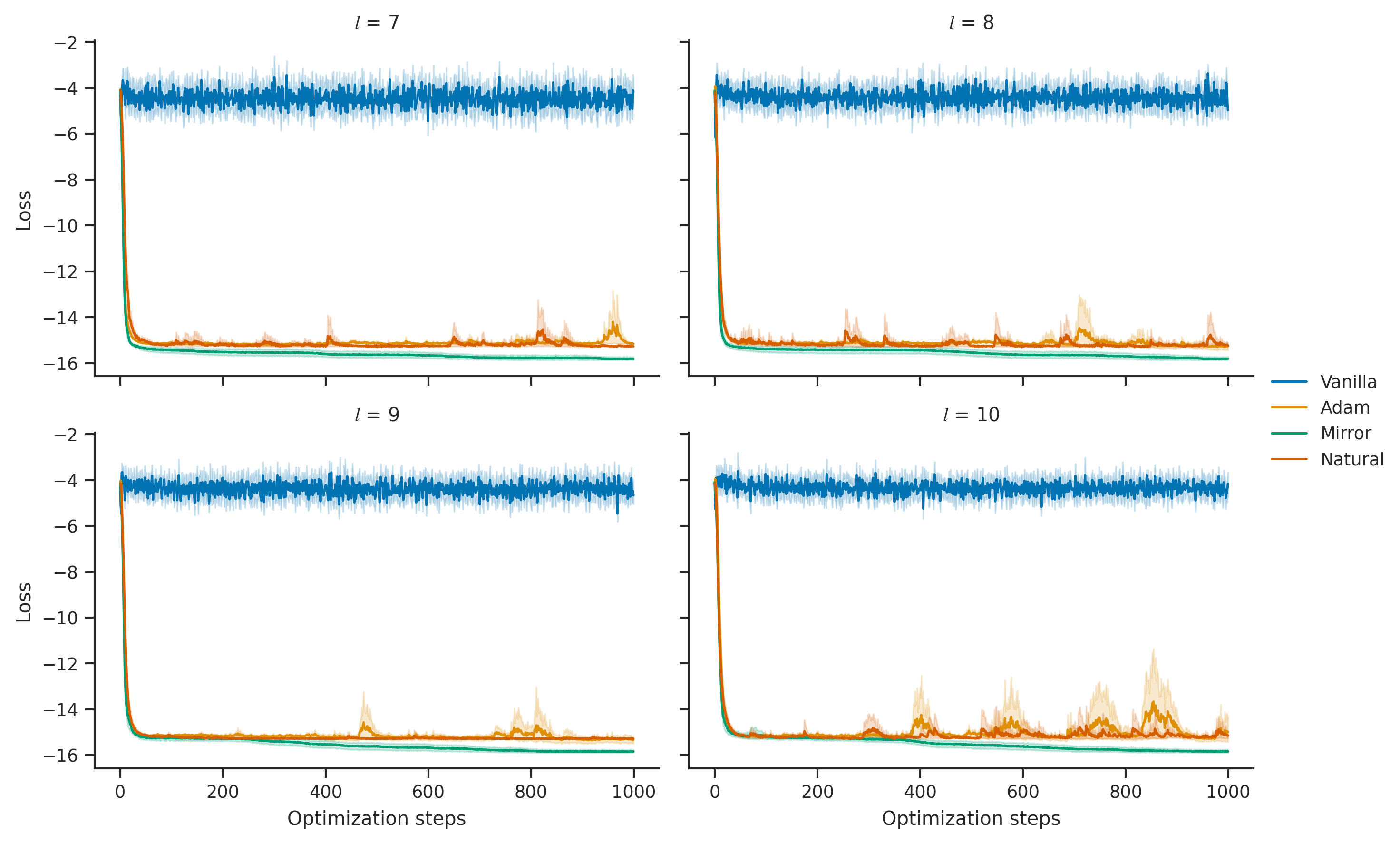}\label{fig:head-to-head-full}}
  \caption{Optimization losses corresponding to \Cref{fig:VQT}.}
\label{fig:VQT-full}
\end{figure*}

\begin{figure*}[!ht]
  \centering
  \subfloat[][]{\includegraphics[width=0.85\textwidth]{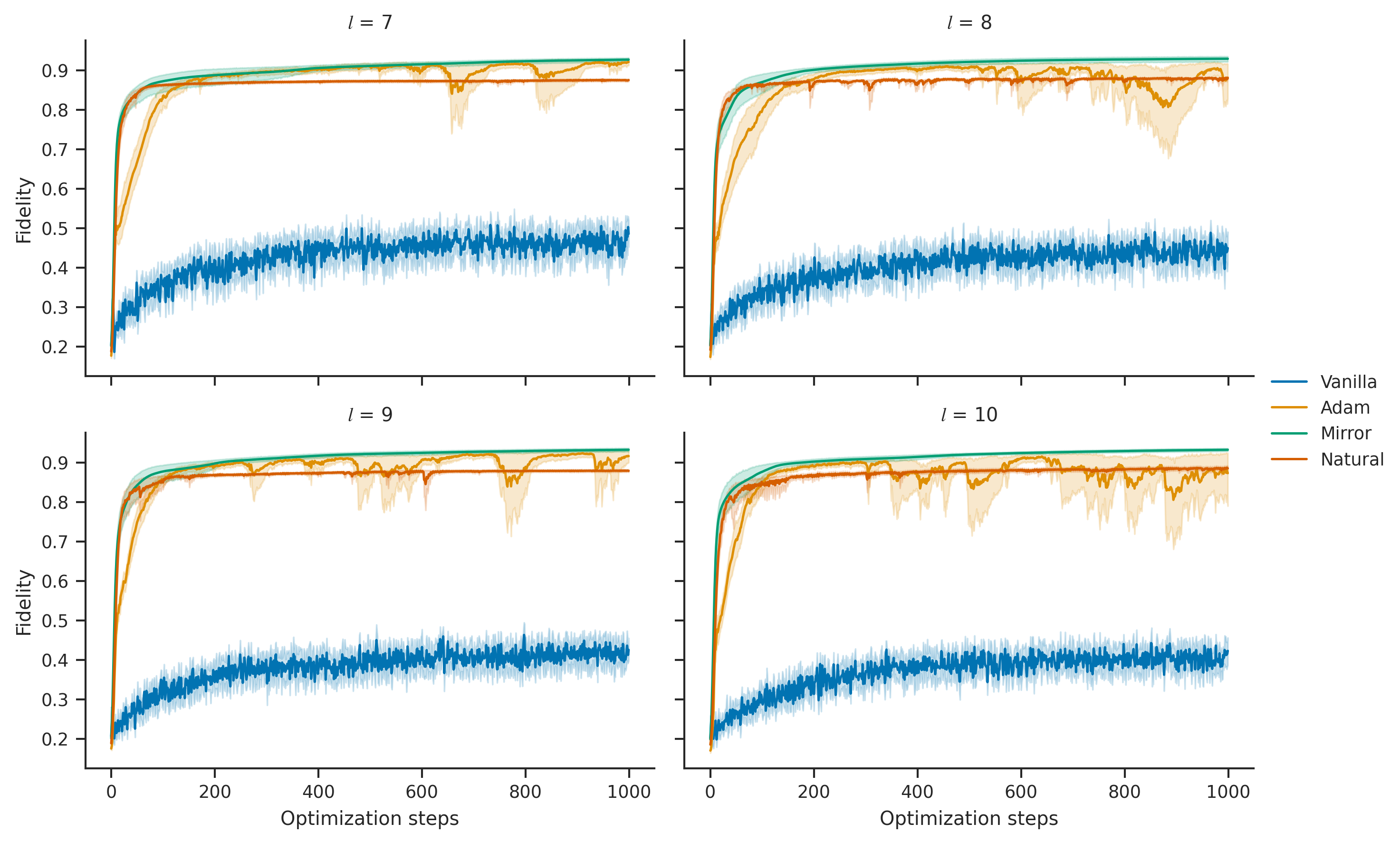}\label{fig:head-to-head-fids-beta-1}}
  \\
  \subfloat[][]{\includegraphics[width=0.85\textwidth]{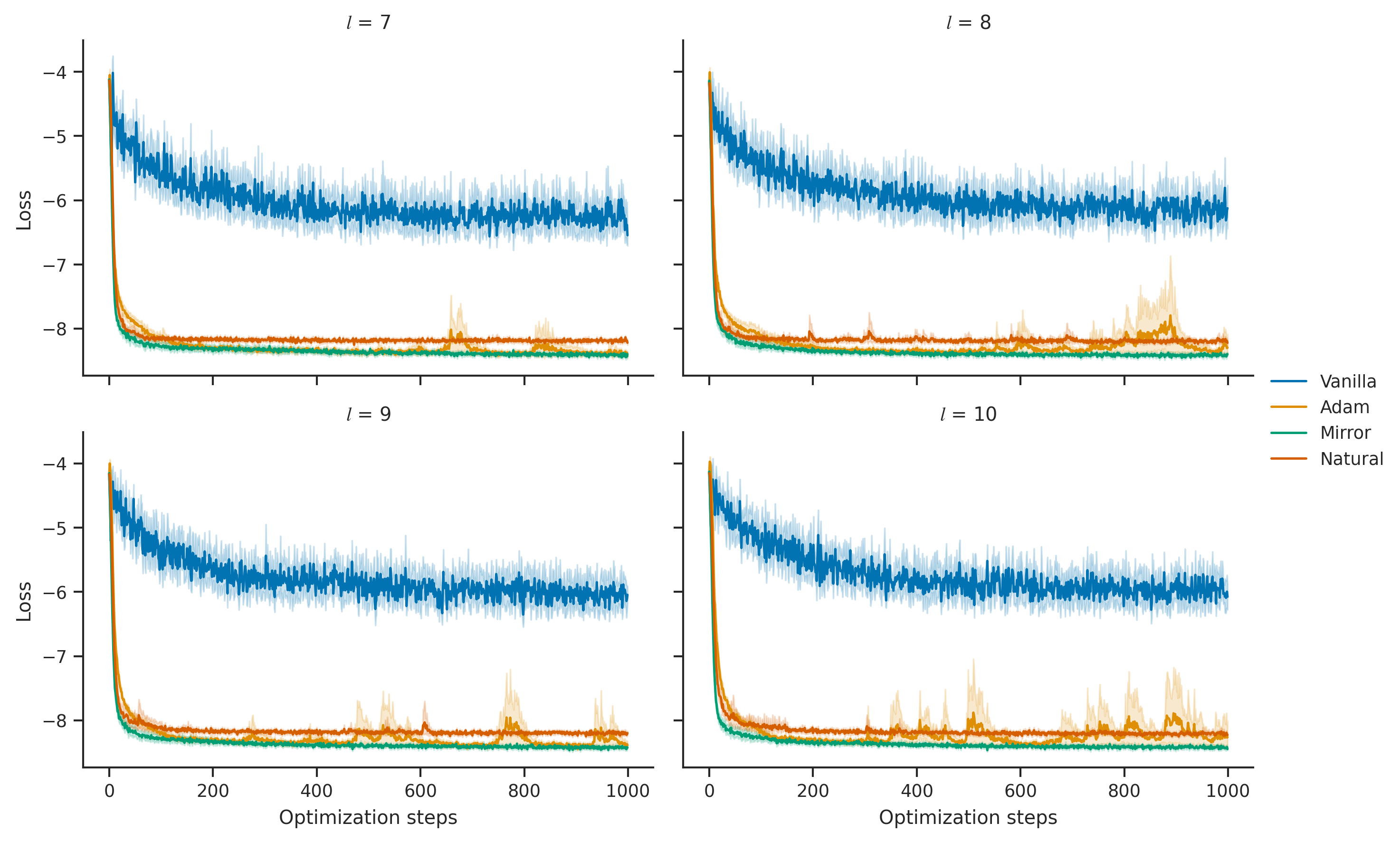}\label{fig:head-to-head-loss-beta-1}}
  \caption{Variation of \Cref{fig:VQT} with $\beta = 1.0$.}
\label{fig:VQT-beta-1}
\end{figure*}

\begin{figure*}[!ht]
  \centering
  \subfloat[][]{\includegraphics[width=0.5\textwidth]{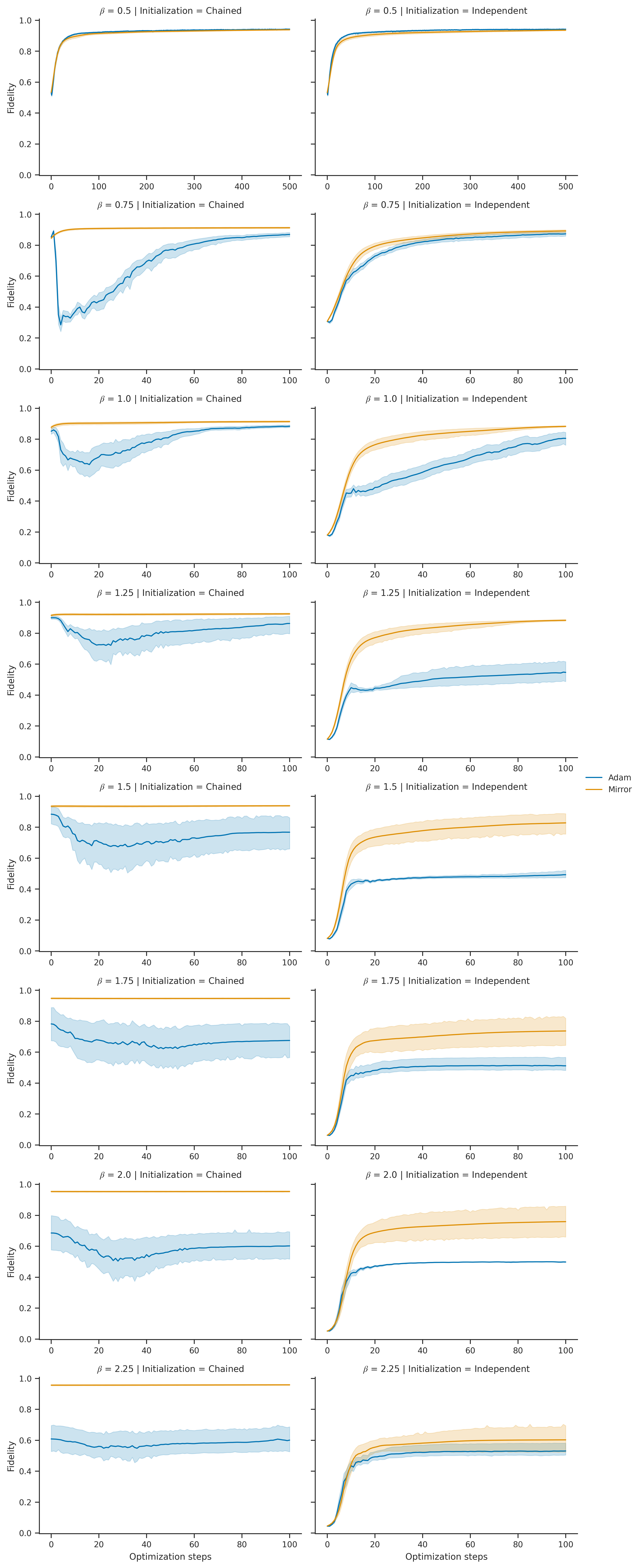}\label{fig:metaqvqt-fids}}
  \subfloat[][]{\includegraphics[width=0.5\textwidth]{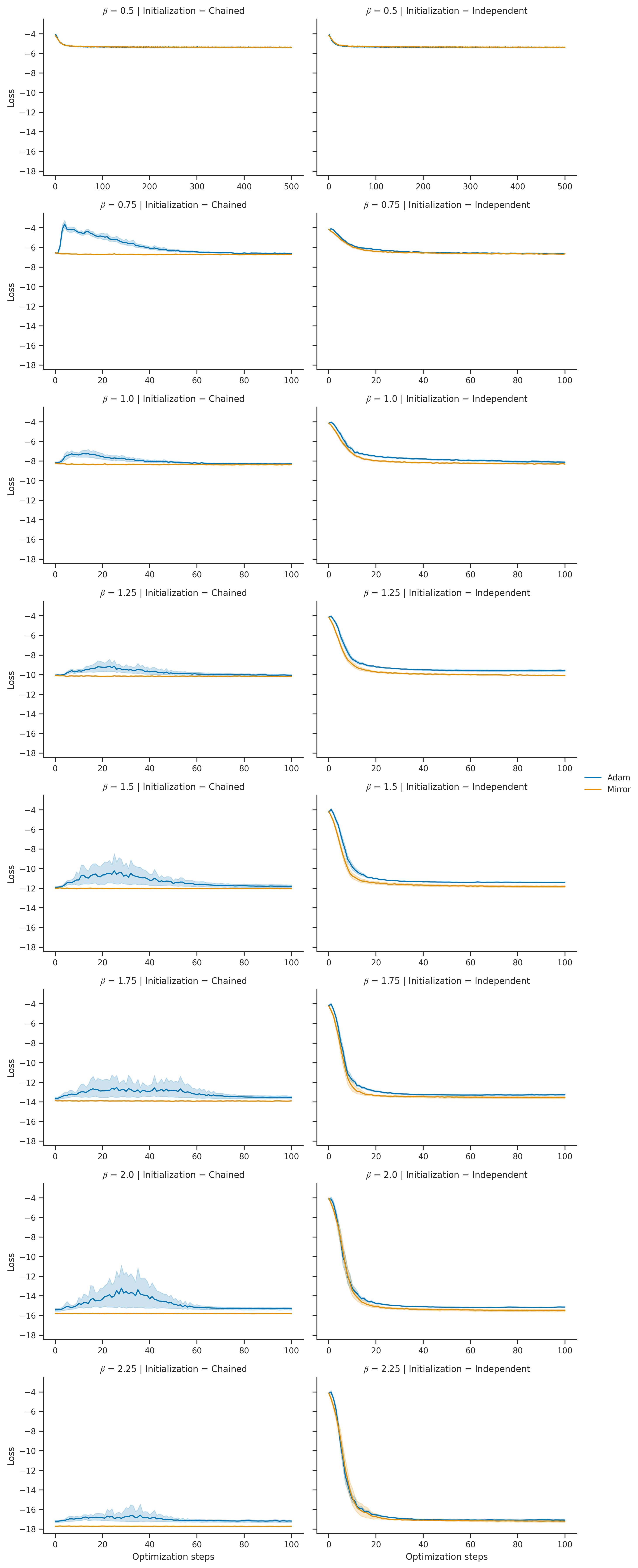}\label{fig:metaqvqt-losses}}
  \caption{Fidelity and optimization loss curves corresponding to \Cref{fig:metavqt-heat}.}
\label{fig:metavqt-curves}
\end{figure*}

\begin{figure*}[!ht]
  \centering
  \subfloat[][]{\includegraphics[width=0.5\textwidth]{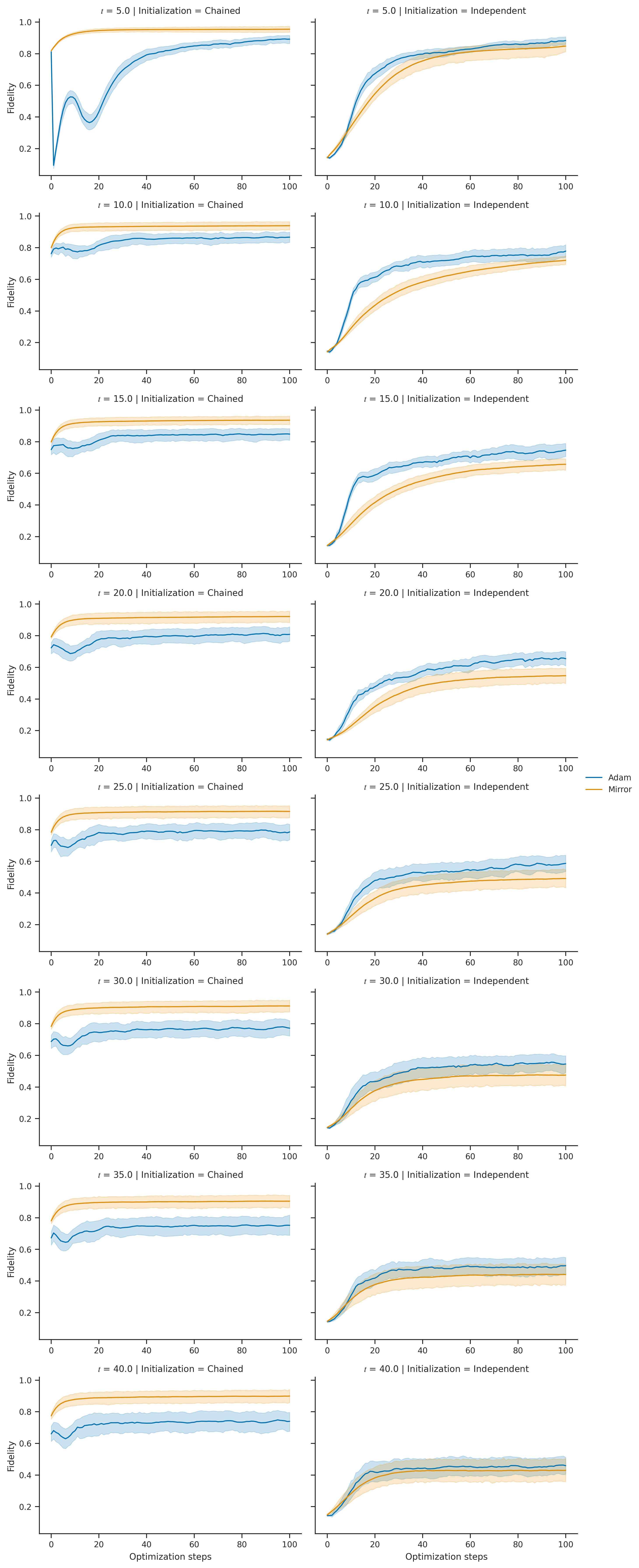}\label{fig:qvartz-fids}}
  \subfloat[][]{\includegraphics[width=0.5\textwidth]{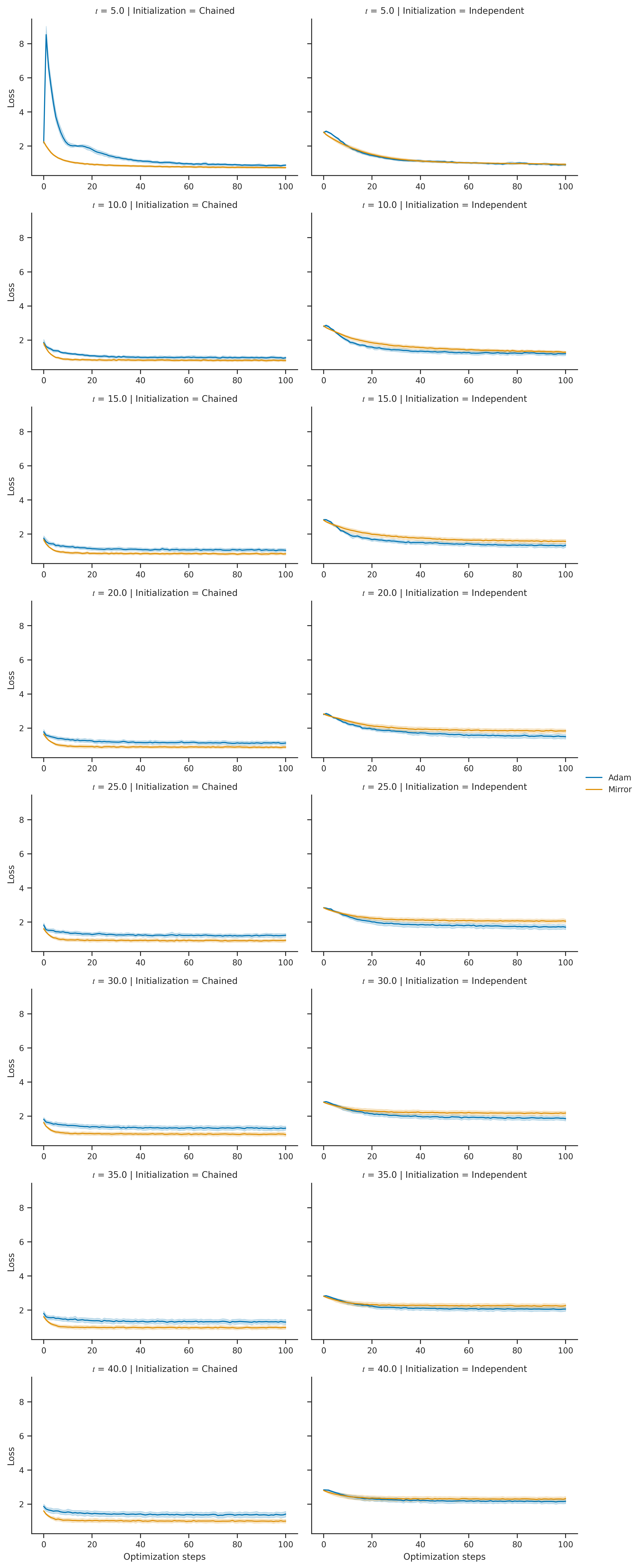}\label{fig:qvartz-losses}}
  \caption{Fidelity and optimization loss curves corresponding to \Cref{fig:qvartz-heat}.}
\label{fig:qvartz-curves}
\end{figure*}

\clearpage

\section{Geometry of the space of quantum mixed states}
\label{sec:geometry}

In this section, we discuss aspects of information geometry needed for the design and analysis of \Cref{alg:qngd,alg:qmd} in \Cref{sec:metric-aware,sec:md-duality}. We begin with a background assumption.

\begin{assumption}
  \label{assum:full-rank}
Our working Hilbert space is of dimension $N = 2^n$. The corresponding density operator space, $\mathcal{M}^{(N)}$, consists of non-singular\footnote{In this work, we will identity the Riemannian metric structure over $\mathcal{M}^{(N)}$ with certain information matrices (as in \cref{eq:monotone-info-matrix}). However, it has been shown that this identification is possible so long as the rank of $\hat \rho_{\bm \Omega}$ below does not depend on its classical parameterization \cite{vsafranek2017discontinuities,naudts2018quantum}, since such information matrices otherwise become discontinuous. We will see that this positivity restriction asks that $\bm\theta$ parameterize classical categorical distributions with full support over all bitstrings. } $N \times N$ density operators and $\dim \mathcal{M}^{(N)} = N^2 - 1$. We write a variational density operator as\footnote{We note that this is simply the spectral decomposition, valid for any density operator: $\ket{\bm\phi(x)}$ are eigenstates of $\hat \rho_{\bm \Omega}$ with corresponding eigenvalues $p_{\bm \theta}(x)$, and $x$ is just a labelling of those eigenstates.  The separation of ${\bm \Omega}$ into distinct parameter sets $\bm\theta$ and $\bm\phi$ foreshadows our favored choice of ansatz; see \Cref{sec:review-qhbm} for more information.}
\begin{align}
\hat \rho_{\bm \Omega} = \sum_{x=1}^N p_{\bm \theta}(x) \ket{\bm\phi(x)}\bra{\bm \phi(x)}
,\end{align}
where $\hat \rho_{\bm \Omega} \in \mathcal{M}^{(N)}$.
\end{assumption}

Now, in \Cref{sec:geometry-monotone} we discuss a theorem characterizing the set of Riemannian \textit{monotone} metrics over $\mathcal{M}^{(N)}$.  In \Cref{sec:geometry-examples} we introduce two important examples of such metrics, which we will use frequently throughout this paper.  We introduce several useful tools and terms in \Cref{sec:geometry-tools}.  Then, in \Cref{sec:geometry-pullback}, we will pull our example metrics back to a parameter space geometry over variational models of quantum mixed states. In the main text (\Cref{sec:background-ngd}), we discussed how associating geometric structures to parameterized models can be beneficial for gradient-based optimization.

\subsection{Characterizing monotone Riemannian metrics}
\label{sec:geometry-monotone}
We argued for the utility of monotone metrics in \Cref{sec:background}.  We restate the property formally below.  From \cite{bengtsson2017geometry} we have the following definition:

\begin{definition}[Monotone Riemmanian metric over $\mathcal{M}^{(N)}$]
  \label{def:monotone}
  A distance $d_{\operatorname{mon}}(\cdot, \cdot)$ over $\mathcal{M}^{(N)}$ is monotone if it does not grow under the action of a \gls{cptp} map (i.e., a quantum channel) $\mathcal{V}$,
  \begin{align}
    d_{\operatorname{mon}}(\mathcal{V}(\hat\rho), \mathcal{V}(\hat\sigma)) \leq d_{\operatorname{mon}}(\hat\rho, \hat\sigma)
  \end{align}
  and if a monotone distance is geodesic then the corresponding metric is called monotone.
\end{definition}

Monotonicity is a natural property to consider because it intuitively encodes the idea that, in general, stochastic maps are randomizing (i.e. coarse-graining should result in information loss) and so should draw distributions nearer to one another. Given the desirability of this property, in what follows we restrict our consideration to monotone metrics.

Which metrics are monotone?  The classical (commutative) analogoue of a \gls{cptp} map is a stochastic matrix \cite{bengtsson2017geometry}, and \u{C}encov's theorem \cite{cencov2000statistical} says that, for categorical distributions, the Fisher-Rao metric is the unique metric (up to normalization) which satisfies classical monotonicity.  In contrast, for the manifold of density operators $\mathcal{M}^{(N)}$, there are infinitely many monotone metrics \cite{petz1996monotone}.  In this quantum setting, the Morozova-\u{C}encov-Petz theorem \cite{bengtsson2017geometry,morozova1989markov,petz1996monotone} acts as the quantum generalization of \u{C}encov's theorem.  It characterizes the set of monotone metrics in terms of a mapping between (self-inversive and unital) operator monotone functions, $f$, and monotone Riemannian metrics (see \Cref{sec:geometry-pullback} for examples). As a reminder, a \textit{metric} $g_{\bm \Omega}(\cdot, \cdot)$ defines an inner product taking arguments in the \textit{tangent space} $T_{\bm\Omega} \mathcal{M}^{(N)}$ about $\hat \rho_{\bm \Omega}$. In particular, the tangent vectors on $\mathcal{M}^{(N)}$ are traceless, Hermitian matrices (reminded in \Cref{apdx:lem:traceless-spanning}).
\begin{theorem}
\label{thm:mcp}
(Morozova-\u{C}encov-Petz) Let $\hat{\rho}_{\bm \Omega} \in \mathcal{M}^{(N)}$. Then, for any monotone metric $g_{\bm \Omega} : T_{\bm \Omega}\mathcal{M}^{(N)} \times T_{\bm \Omega}\mathcal{M}^{(N)} \to \mathbb{C}$, there exists a constant $C$ and function $c : \mathbb{R}^2 \to \mathbb{R}$ such that for any traceless Hermitian $H, H' \in T_{\Omega} \mathcal{M}^{(N)}$,
\begin{align}
\label{eq:monotone-metric-diagonal-density}
g_{\bm \Omega}(\hat H, \hat H') &=  \frac{1}{4}\left[C \sum_{x} \frac{\bra{\bm\phi(x)} \hat H\ket{\bm\phi(x)}\bra{\bm\phi(x)} \hat H'\ket{\bm\phi(x)}}{p_{\bm \theta}(x)} + 2\sum_{x<y} c(p_{\bm \theta}(x), p_{\bm \theta}(y))\bra{\bm\phi(x)} \hat H\ket{\bm\phi(y)}\bra{\bm\phi(y)} \hat H'\ket{\bm\phi(x)}\right].
\end{align}
Here, $c$ is symmetric  $\ensuremath{(i.e., }c(x,y) = c(y,x)\ensuremath{)}$ and obeys $c(sx,sy) = s^{-1}c(x,y)$. Furthermore, the function $f(t) := \frac{1}{c(t,1)}$ is operator monotone.  
\end{theorem}
This theorem shows that there are many non-trivially different monotone metrics on the space of density operators.  Note that, if we were to restrict consideration to the set of \textit{pure} quantum states, we recover a natural uniqueness in the choice of metric.  Thus existing works which consider metric-aware descent for pure state optimization have circumvented the question of choosing from among the various quantum monotone metrics \cite{stokes2020quantum}.

We can also make the reverse implication of deriving a monotone metric $g_{\bm \Omega}$ given a function $c$ satisfying the criteria above.  For that, we first make a definition:
\begin{definition}
A \gls{mc} function is an operator monotone map $f:\mathbb{R}_+\to\mathbb{R}_+$ satisfying $f(1/t) = f(t)/t$ and $f(1) = 1.$
\end{definition}
It is not hard to show that \Cref{thm:mcp} yields a bijective correspondence between monotone metrics and \gls{mc} functions; given a monotone metric $g$, the function $f(t) := 1/c(t,1)$ is \gls{mc}. Conversely, if $f(t)$ is \gls{mc}, one defines $c(x,y) = 1/(yf(x/y))$, which by definition of \gls{mc} functions, satisfies the properties of $c(x,y)$ in \Cref{thm:mcp}. Hence, inserting our $c(x,y)$ into \eqref{eq:monotone-metric-diagonal-density} (and fixing $C=1$) yields a monotone metric.

\subsection{Examples of monotone Riemannian metrics}
\label{sec:geometry-examples}
We pause the theoretical exposition to give two key examples of monotone metrics, defining them in terms of their corresponding \gls{mc} functions.

\subsubsection{\texorpdfstring{\acrfull{qfi} and the \acrfull{bh} metric}{QFI and the BH metric}}
\label{sec:bh-metric}

In the case of the \gls{bh} metric, the \gls{mc} function is given by
\begin{align}
\label{eq:bh-f}
    f(t) &= \frac{t + 1}{2}
,\end{align}
and so $c(x, y) = \tfrac{2}{x + y}$ which is the reciprocal of the arithmetic mean. Amongst all operator monotone functions on $[0, +\infty)$ which are self-inversive and unital, there exists a minimal and maximal function, i.e., \gls{mc} functions that are \textit{everywhere} smaller (resp. larger) than \textit{all} \gls{mc} functions \cite{kubo1980means}. The maximal function is in fact \eqref{eq:bh-f}. It is the case that $\mathscr{L}^{\operatorname{BH}}_{\bm \Omega}(\hat H) = 2 \int_{0}^\infty e^{-s \hat \rho_{\bm \Omega} } \hat H e^{-s \hat \rho_{\bm \Omega} } ds$ and so $g^{\operatorname{BH}}_{\bm \Omega}(\hat H, \hat H')    =  2 \int_{0}^\infty \tr[e^{-s \hat\rho_{\bm \Omega} } \hat H e^{-s \hat \rho_{\bm \Omega} } \hat H'] ds$.

\subsubsection{\texorpdfstring{Quantum Relative Entropy and the \acrfull*{bkm} metric}{Quantum Relative Entropy and the BKM metric}}
\label{sec:qxe-bkm}

In the case of the \gls{bkm} metric,
\begin{align}
\label{eq:bkm-f}
    f(t) &= \lim_{t' \rightarrow t}\frac{t' - 1}{\log t'}
,\end{align}
and so $c(x, y) = \tfrac{\log x - \log y}{x - y }, x \neq y$ and $c(x,x) = \tfrac{1}{x}$ which together is the reciprocal of the logarithmic mean. It is the case that in mixture coordinates (\Cref{def:mixture-coords}) $\mathscr{L}^{\operatorname{BKM}}_{\bm \Omega}(\hat{H}) = \int_0^\infty (\hat{\rho}_{\bm \Omega}+s\mathbbm{1})^{-1} \hat{H} (\hat{\rho}_{\bm \Omega}+s\mathbbm{1})^{-1} ds$ and so $\mathscr{L}^{\operatorname{BKM}}_{\bm \Omega}(\hat{H})(\partial_{\Omega_j} \hat \rho_{\bm \Omega}) = \partial_{\Omega_j} \log \hat\rho_{\bm \Omega}$. Although $f$ is neither minimal nor maximal, we will use the facts that the implied metric is the Hessian of the canonical quantum relative entropy, $g^{\operatorname{BKM}}_{\bm \Omega}(\hat H, \hat H') = -\partial^2_{\alpha\beta} D(\hat\rho_{\bm \Omega} + \alpha \hat H \| \hat \rho_{\bm \Omega} + \beta \hat H')\mid_{\alpha = \beta = 0}$ (\cref{eq:bkm-hessian-canonical}), and a duality property (\Cref{sec:app:duality}) to motivate its usefulness. In exponential coordinates, the raising and lowering operators are swapped, a hint toward the duality property.

\subsection{Tools for working with monotone metrics}
\label{sec:geometry-tools}

We will apply several mathematical tools involving monotone metrics in \Cref{sec:metric-aware,sec:md-duality}.  We introduce some of these tools below.

\subsubsection{Positivity of the metric}
First, based on \Cref{thm:mcp}, we have the fact that all monotone metrics over $\mathcal{M}^{(N)}$ in our consideration are positive definite. For pedagogical purposes, we check this explicitly.

\begin{proposition}[Riemannian metric signature]
\label{cor:positive-definite}
All monotone metrics over $\mathcal{M}^{(N)}$ are positive definite under \Cref{assum:full-rank}.
\begin{proof}
From \eqref{eq:monotone-metric-diagonal-density}, we see that
\begin{align}
\label{eq:metric-sig}
    g(\hat H, \hat H) &= \sum_x \frac{\lvert \bra{\bm \phi(x)} \hat H \ket{\bm \phi(x)}\rvert^2}{p_{\bm \theta}(x)} + 2\sum_{x < y} c(p_{\bm \theta}(x), p_{\bm \theta}(y)) \lvert \bra{\bm \phi(x)}\hat H \ket{\bm \phi(y)}\rvert^2
.\end{align}
Because $\hat\rho_{\bm \Omega}$ is strictly positive (\Cref{assum:full-rank}), $\lvert \bra{\bm \phi(x)} \hat{H} \ket{\bm \phi(y)}\rvert^2 >0, \forall, x, y$. Furthermore, $c(p_{\bm \theta}(x), p_{\bm \theta}(y)) > 0$. Hence, $g(\hat H, \hat H) = 0$ only if $\hat H = 0$. Therefore, $g(\cdot, \cdot)$ is a positive definite Riemannian metric.
\end{proof}
\end{proposition}
We can then relate the positive definiteness of the metric to the positive definiteness of the metric resolved to a matrix in an arbitrary basis:
\begin{corollary}
\label{prop:pos-def-frame}
Under \Cref{assum:full-rank}, any monotone metric over $\mathcal{M}^{(N)}$ is positive definite when resolved to a matrix in terms of an arbitrary tangent space basis.
\begin{proof}
Choosing a basis of traceless, Hermitian matrices, or frame, $h = \{ \hat\sigma_i \}_{i=1}^{N^2 - 1}$ we can write the Riemannian metric tensor as a matrix,
\begin{align}
[\hat G_h]_{j,k} &= g(\hat\sigma_j, \hat\sigma_k) 
\end{align}
so that arbitrary traceless, Hermitian matrices $v[h] = \sum_j v^j \hat\sigma_j, w[h] = \sum_j w^j \hat\sigma_j$ have value
\begin{align}
g(v, w) &= w[h]^\dag \hat G_h v[h]
,\end{align}
and so $\hat G_h$ is positive definite for all $h$ whenever $g(v, w)$ is a positive definite Riemannian metric in that $w[h]^\dag G[h] w[h] > 0, \forall w \neq 0$.
\end{proof}
\end{corollary}

\subsubsection{Information matrix}
It will be convenient to resolve the metric tensor to coordinates as a matrix. We will define the \textit{Information Matrix} $\mathcal{I}^f(\bm \Omega)$ as having matrix elements
\begin{align}
  \label{eq:monotone-info-matrix}
  [\mathcal{I}^f(\bm \Omega)]_{j, k} &:=  g^f_{\bm{{\Omega}}}(\partial_{\Omega_j} \hat \rho_{\bm \Omega}, \partial_{\Omega_k} \hat \rho_{\bm \Omega}),
\end{align}
i.e. the pullback of the metric tensor to parameter space\footnote{Other definitions in the literature may include an inconsequential factor of 4, $[\mathcal{I}^f(\bm \Omega)]_{j, k} :=  4g_{\bm{{\Omega}}}(\partial_{\Omega_j} \hat \rho_{\bm \Omega}, \partial_{\Omega_k} \hat \rho_{\bm \Omega})$.}. Here, $f$ denotes the choice of \gls{mc} function which we saw to uniquely specify the metric at $\bm \Omega$. We work the information matrix out in a particular ansatz class for the two metrics in consideration in \Cref{sec:estimating-bkm-tensor,sec:estimating-bh-tensor}.

We note that $\partial_{\Omega_j} \hat \rho_{\bm \Omega}$ is a traceless Hermitian matrix and so describes a tangent vector on $\mathcal{M}^{(N)}$ at $\hat\rho_{\bm \Omega}$ (\Cref{apdx:lem:traceless-spanning}). This matrix is positive definite so long as $\{ \partial_{ \Omega_j} \hat \rho_{\bm \Omega}\}_j$ gives a basis for the tangent space of $\mathcal{M}^{(N)}$ at $\bm \Omega$ (\Cref{prop:pos-def-frame}). It is possible to choose a model parameterization so as to either describe a smooth embedding or global diffeomorphism from $\mathcal{M}^{(N)}$ to classical parameter space (\Cref{apdx:over-param}). However, this is not anticipated in general, particularly when strong inductive architectural biases are leveraged or $\theta$ consists of neural network parameters.

As we will see, in the $\hat\rho_{\bm \Omega}$-basis,

\begin{align}
\label{eq:mc-info-matrix}
  [\mathcal{I}^f(\bm \Omega)]_{j, k} &= \sum_{l, m}  c(p_{\bm\theta}(l) , p_{\bm\theta}(m))
 \bra{l} \partial_{\Omega_j} \hat \rho_{\bm \Omega} \ket{m}\bra{m} \partial_{\Omega_k} \hat \rho_{\bm \Omega} \ket{l},  \\
  \intertext{where,}
  \label{eq:c-from-f}
  c(x, y) &= \frac{1}{yf(x/y)}.
\end{align}

\subsubsection{Raising and lowering the metric}

We now describe another useful representation of the metric inner product. The choice of $f$ fixes a linear operator $\mathscr{L}^{f}_{\bm \Omega}$ which acts as the metric with lowered indices. Hence, it allows us to write the metric inner product as the Hilbert-Schmidt inner product
\begin{equation}
\label{eq:metric-lowering}
  g_{\bm \Omega}^{f}(\hat H, \hat H') = \tr[ \hat H \mathscr{L}^{f}_{\bm \Omega}(\hat H') ]
\end{equation}
for arbitrary tangent vectors $\hat{H}, \hat{H}'$.  We can also define raising operators for the class of monotone metrics on $\mathcal{M}^{(N)}$:

\begin{definition}
The linear operator $\mathscr{R}^{f}_{\bm \Omega}$ acts as the metric with raised indices and is defined so that, in the $\hat \rho_{\bm \Omega}$-basis \cite{petz2011introduction},
\begin{align}
\label{eq:raising-operator-mcp}
    \bra{j} \mathscr{R}^{f}_{\bm \Omega}(\hat{H}) \ket{k} &=
    \begin{cases}
    \frac{1}{c(p_{\bm \theta}(j) , p_{\bm \theta}(k))}\bra{j} \hat{H} \ket{k}  , & c(p_{\bm \theta}(j) , p_{\bm \theta}(k)) \neq 0 \\
    0,& \text{otherwise}
    \end{cases}
\end{align}
and satisfies $\mathscr{R}^{f}_{\bm \Omega} := (\mathscr{L}^{f}_{\bm \Omega})^{-1}$,
\begin{align}
  \bra{j} \mathscr{L}^{f}_{\bm \Omega}(\hat{H}) \ket{k} &= c(p_{\bm \theta}(j) , p_{\bm \theta}(k))\bra{j} \hat{H} \ket{k} 
\end{align}
Hence, for example, 
\begin{align}
\label{eq:info-matrix-calculation}
   \tr[\hat H \mathscr{L}^{f}_{\bm \Omega}(\hat H')] &= \sum_{j} \bra{j} \left(\sum_{k, l} \bra{k}\hat H\ket{l}\ket{k}\bra{l}\right) \left(\sum_{k, l} c(p_{\bm \theta}(k) , p_{\bm \theta}(l))\bra{k} \hat{H}' \ket{l}\ket{k}\bra{l}\right)  \ket{j} \\
   &= \sum_{j, k} c(p_{\bm \theta}(j) , p_{\bm \theta}(k)) \bra{j}\hat{H}\ket{k}\bra{k}\hat{H}' \ket{j}
,\end{align}
which matches the result of \Cref{thm:mcp} with $c(x,y) = \tfrac{1}{yf(x/y)} \implies c(x,x) \equiv \frac{1}{x}$ and $C \equiv 1$.

\end{definition}

\subsection{Monotone metrics over \texorpdfstring{$\mathcal{M}^{(N)}$}{M(N)}}
\label{sec:geometry-pullback}
Here we apply the metric raising and lowering operators to our two example metrics, \gls{bh} and \gls{bkm}.  This lets us represent those metrics in model coordinates as generalized information matrices.

\begin{example}[Bures-Helstrom]
Choose $f(t) = \frac{t + 1}{2}$ as in \eqref{eq:bh-f} and so $c(x, y) = \frac{2}{x+y}$. Then, by \eqref{eq:raising-operator-mcp}, the linear operator $\mathscr{R}^{\operatorname{BH}}_{\bm \Omega}(\cdot)$ writes
\begin{align}
    \mathscr{R}^{\operatorname{BH}}_{\bm \Omega}(\hat{H}) & = \frac{1}{2} \{ \hat \rho_{\bm \Omega}, \hat{H} \}
.\end{align}    
Hence, in the $\hat \rho_{\bm \Omega}$-basis, the lowering operator $\mathscr{L}^{\operatorname{BH}}_{\bm \Omega} = (\mathscr{R}^{\operatorname{BH}}_{\bm \Omega})^{-1}$ is given by
\begin{align}
    \mathscr{L}^{\operatorname{BH}}_{\bm \Omega}(\hat{H}) & = 2 \sum_{x, y} \frac{\bra{\bm\phi(x)} \hat{H} \ket{\bm\phi(y)}}{p_{\bm\theta}(x) + p_{\bm\theta}(y)} \ket{\bm\phi(x)}\bra{\bm\phi(y)},
\end{align}
where $\hat \rho_{\bm \Omega} = \sum_x p_{\bm \theta}(x) \ket{\bm\phi(x)}\bra{\bm\phi(x)}$. Therefore, the \gls{qfim} $\mathcal{I}^{\operatorname{BH}}(\bm \Omega)$ is given by
\begin{align}
\label{eq:bh-info-mcp}
    [\mathcal{I}^{\operatorname{BH}}(\bm \Omega)]_{j, k} & = \begin{aligned}[t]\sum_{x} &\frac{ \bra{\bm\phi(x)} \partial_{\bm \Omega_j} \hat \rho_{\bm \Omega}\ket{\bm\phi(x)}\bra{\bm\phi(x)} \partial_{\bm \Omega_k} \hat \rho_{\bm \Omega}\ket{\bm\phi(x)}}{p_{\bm \theta}(x)} \\
    &+ 4\sum_{x< y} \frac{1}{p_{\bm\theta}(x) + p_{\bm\theta}(y)} \bra{\bm \phi(x)} (\partial_{\bm \Omega_j} \rho )\ket{\bm \phi(y)} \bra{\bm \phi(y)} (\partial_{\bm \Omega_k} \hat{\rho} ) \ket{\bm \phi(x)}
    ,\end{aligned}
\end{align}
which we will resolve block-wise for diagonalized models in \Cref{sec:estimating-bh-tensor}.
\end{example}

\begin{example}[Bogoliubov-Kubo-Mori]
Choose 
\begin{align*}
f(t) = \lim_{t' \rightarrow t}\frac{t' - 1}{\log t'}   
\end{align*}
as in \eqref{eq:bkm-f}, meaning that 
\begin{align*}
c(x, y) = \begin{cases}\frac{\log x - \log y}{x - y} & x \neq y \\ \frac{1}{x} & x=y \end{cases} 
\end{align*}
Hence, we say that $c(x, y) := \frac{1}{\mu_{\log}(x, y)}$ where $\mu_{\log}(x, y)$ is the logarithmic mean. In this case, $\mathscr{R}^{\operatorname{BKM}}_{\bm \Omega}(\cdot)$ writes
\begin{align}
    \mathscr{R}^{\operatorname{BKM}}_{\bm \Omega}(\hat H) & = \int_0^1 \hat \rho_{\bm \Omega}^s \hat{H}  \hat \rho_{\bm \Omega}^{1-s} ds.
\end{align}
Hence, in the $\hat \rho_{\bm \Omega}$-basis,
\begin{align}
    \mathscr{L}^{\operatorname{BKM}}_{\bm \Omega}(\hat H) &= \sum_x \frac{\bra{\bm \phi(x)}\hat H\ket{\bm \phi(x)}}{p_{\bm\theta}(x)}\ket{\bm\phi(x)}\bra{\bm\phi(x)} + 2 \sum_{x < y} \frac{1}{\mu_{\log}(p_{\bm \theta}(x), p_{\bm \theta}(y))}  \bra{\bm\phi(x)} \hat H \ket{\bm\phi(y)} \ket{\bm \phi(x)}\bra{\bm\phi(y)} \\
    \label{eq:bkm-lowering}
    & = \int_0^\infty (\hat\rho_{\bm \Omega}+s\mathbbm{1})^{-1} \hat H (\hat\rho_{\bm \Omega}+s\mathbbm{1})^{-1} ds.
\end{align}
Therefore, the \gls{bkm} information matrix $\mathcal{I}^{\operatorname{BKM}}(\bm \Omega)$ is given by
\begin{align}
    \label{eq:info-bkm}
    [\mathcal{I}^{\operatorname{BKM}}(\bm \Omega)]_{j, k} & = \begin{aligned}[t]\sum_{x} &\frac{ \bra{\bm\phi(x)} \partial_{\bm \Omega_j} \hat \rho_{\bm \Omega}\ket{\bm\phi(x)}\bra{\bm\phi(x)} \partial_{\bm \Omega_k} \hat \rho_{\bm \Omega}\ket{\bm\phi(x)}}{p_{\bm \theta}(x)} \\
    &+ 2\sum_{x < y} \frac{1}{\mu_{\log}(p_{\bm \theta}(x), p_{\bm \theta}(y))}  \bra{\bm \phi(x)} \partial_{\bm \Omega_j} \hat \rho \ket{\bm \phi(y)} \bra{\bm \phi(y)} \partial_{\bm \Omega_k} \hat \rho \ket{\bm \phi(x)}
    ,\end{aligned}
\end{align}
which we will resolve block-wise for diagonalized models in \Cref{sec:estimating-bkm-tensor}.
\end{example}

\section{State discrimination bounds}
\label{sec:geometry-qcrb}

In this section, we discuss the relationship between information geometry and metrology, leading to the concept of generalized \gls{qfi} and the generalized \gls{qcrb}. Significant literature \cite{liu2019quantum} is dedicated to the fundamental metrological question of parameter estimation, especially in relation to the well-studied \gls{qcrb} as it usually relates to the \gls{bh} geometry. In the applications of quantum Hamiltonian learning and variational Gibbs sampling, we are essentially interested in estimating the parameters of an unknown quantum mixed state. We will discuss how searching for such parameters through a metric-aware gradient-based optimization can saturate fundamental asymptotic limits of parameter estimation (\Cref{res:qpngd-achieves-qcrb}) whereas this is not generally the case for other gradient-based strategies. In fact, other quantum natural gradient strategies have fallen short of such proven guarantees because of, intuitively, the discrepancy between the utilized curvature in state space and curvature of the objective \cite{stokes2020quantum}.

The construction of a \gls{qcrb} which generalizes beyond the \gls{bh} geometry essentially involves a straightforward application of Cauchy-Schwarz for Hermitian operators. We provide the construction in \Cref{apdx:thm:general-qcrb}. In particular, take $\hat{\bm A} = (\hat A_1, \cdots, \hat A_d)$ as a collection of quantum observables. Then, in the multi-parameter case, the generalized \gls{qcrb} relates a covariance matrix $\Cov^{f}(\hat {\bm A}; \bm \Omega)$ at $ \hat \rho_{\bm \Omega}$ to the \gls{qfim} $\mathcal{I}^f(\bm \Omega)$ as,
\begin{align}
    \label{eq:qcrb-apdx}
    \Cov^{f}(\hat {\bm A} ; \bm \Omega) \succeq \frac{1}{m} J^\dag (\mathcal{I}^f(\bm \Omega))^{+} J,
\end{align}
where the generalized covariance is
\begin{align}
\label{eq:generalized-covariance}
    [\Cov^{f}(\hat {\bm A}; \bm \Omega)]_{j, k} & = \tr[ \hat A_j \mathscr{R}^{f}_{\bm \Omega}(\hat A_k)] - (\Tr[\rho_{\bm \Omega}\hat A_j])(\Tr[\rho_{\bm \Omega}\hat A_k])
,\end{align}
and the bias matrix is
\begin{align}
\label{eq:bias-matrix}
 [J]_{j, k} &:=  \frac{\partial}{\partial \Omega_j} \tr[\hat{\rho}_{\bm \Omega} \hat A_k]\mid_{ \bm \Omega = \bm \Omega^*}
,\end{align}
where $m$ is an arbitrary number of collected measurement samples of $\hat \rho_{\bm \Omega^*}$, and $\hat A_j^{+}$ denotes the Moore-Penrose pseudo-inverse of a matrix $\hat A_j$. The linear operator $\mathscr{R}^{f}_{\bm \Omega}$ is defined to satisfy $\mathscr{R}^{f}_{\bm \Omega} := (\mathscr{L}^{f}_{\bm \Omega})^{-1}$. 
  
Consider if we assume that $\hat {\bm A}$ is now a collection of quantum observables that are constructed as locally unbiased estimators to the optimal parameters $\bm \Omega^*$. Note that $\hat A_j$ is a locally unbiased estimator of $\Omega^*_j$ at $\bm \Omega^*$ if
\begin{align}
\label{eq:local-unbiasedness}
    \frac{\partial}{\partial \Omega_j} \tr[\hat{\rho}_{\bm \Omega} \hat A_k]\mid_{ \bm \Omega = \bm \Omega^*} &= \delta_{j,k}
,\end{align}
which holds if $\tr[\hat \rho_{\bm \Omega^*} \hat A_j] = \Omega^*_j$, the stronger condition of global unbiasedness. Clearly, in such a case, we have that $J = \mathbbm{1}$. However, Corollary 7.7.10 of \cite{horn2012matrix} taken with \eqref{eq:qcrb-apdx} implies that this can only be the case if the information matrix is non-singular. Therefore, locally unbiased estimators require that the parameterization gives a locally non-singular information matrix. So, in this case, we recover a simpler inequality,
\begin{align}
\label{eq:qcrb-unbiased}
   \Cov^{f}(\hat{\bm A } ; \bm \Omega) \succeq \frac{1}{m} (\mathcal{I}^f(\bm \Omega))^{-1}. 
\end{align}

Now, we can use the theory of \gls{mc} functions to rewrite the covariance and information matrices. In the $\hat\rho_{\bm \Omega}$-basis \cite{petz2011introduction},
\begin{align}
  [\Cov^f(\hat{\bm A }; \bm \Omega)]_{j, k} &= \sum_{l, m}  \frac{1}{c(p_\theta(l) , p_\theta(m))}
  \bra{l} \hat A_j \ket{m}\bra{m} \hat A_k \ket{l} - \sum_{l} \bra{l} \hat A_j \ket{l} \sum_m \bra{m} \hat A_k \ket{m},
\end{align}
with $c(x, y) = \frac{1}{yf(x/y)}$ again as in \eqref{eq:c-from-f}.

Recall from \Cref{sec:bh-metric} that the \gls{mc} function $f$ is maximal for the \gls{bh} metric. Hence, for a single-parameter $\bm \Omega = (\Omega_0)$, the \gls{bh} instance of \Cref{apdx:thm:general-qcrb} gives the tightest guarantees with respect to the then scalar $\Cov^{\operatorname{BH}}(A; \bm \Omega)$ given that $\min_f \mathcal{I}^f(\bm \Omega) = \mathcal{I}^{\operatorname{BH}}(\bm \Omega)$ and $\max_f \Cov^{f}(\bm \Omega ; A) =  \Cov^{\operatorname{BH}}(\bm \Omega; A)$. This is consistent with the fact that $\tfrac{2}{p_\theta(j) + p_\theta(k)} \leq \tfrac{\log p_\theta(j) - \log p_\theta(k) }{p_\theta(j) - p_\theta(k)}$ which verifies the claim for the two metrics we have detailed.

Finally, attaining the \gls{bh} scaling (which is tightest for single-parameter), $\Cov^{\operatorname{BH}}(A ; \bm \Omega) \succeq \frac{1}{m} (\mathcal{I}^{\operatorname{BH}}(\bm \Omega))^{-1}$, is not always possible in the multi-parameter case. In proving the generalized bound, one implicitly assumes that each of the $\hat A^*_j$, the optimal parameter-estimating observables in $\Omega_j$, are informationally compatible. In other words, a necessary and sufficient condition to saturate the bound is that $\tr[\hat\rho_{\bm \Omega} [\hat A^*_j, \hat A^*_k]] = 0, \forall j,k$ so that parameters can be estimated optimally independently of one another. It turns out that this condition holds for certain ansatzes as we show in \Cref{apdx:prop:compatibility}.

Following \eqref{eq:generalized-covariance}, we can write the generalized covariance between two observables as the inner product,
\begin{align}
    \langle \hat H, \hat H'\rangle^{f}_{\bm\Omega} & = \tr [\mathscr{R}^{f}_{\bm \Omega}(\hat H) \hat H']
.\end{align}    
In such a case, it is useful to define an object termed the logarithmic derivative given by
\begin{align}
\label{eq:logarithmic-derivative}
    \hat{L}^{f}_{\bm \Omega, j} & = \mathscr{L}^{f}_{\bm\Omega}(\partial_{\bm \Omega_j} \hat \rho_{\bm \Omega})
.\end{align}
which then induces the relation
\begin{align}
\label{eq:cov-information}
    [\mathcal{I}^f(\bm \Omega)]_{j, k}
      & = \langle \hat{L}^{f}_{\bm \Omega, j}, \hat{L}^{f}_{\bm \Omega, k}\rangle^{f}_{\bm \Omega}  = g^{f}_{\bm \Omega}(\partial_{\bm \Omega_j} \hat \rho_{\bm \Omega}, \partial_{\bm \Omega_k} \hat \rho_{\bm \Omega})
.\end{align}
We will use these relations to write a generalized \gls{qcrb}, below.

\begin{theorem}[Generalized \gls{qcrb}]
\label{apdx:thm:general-qcrb}

The generalized \gls{qcrb} \eqref{eq:qcrb-apdx} holds.

\begin{proof}
We generalize the proof from \cite{petz2011introduction} to allow for a singular information matrix, as is anticipated for over-parameterized models like classical neural networks. Showing the inequality for a single sample $m=1$ is sufficient to verify the scaling.

For simplicity, we drop the metric specification, $f$. Without loss of generality, we assume that $\bm \Omega = 0$ and we will employ the block-matrix method. Hence, we may assume that we $\dim \bm \Omega =2$ and write the estimators of \eqref{eq:qcrb-apdx} as $\bm A = (A_1, A_2)$ and the logarithmic derivatives at $\bm \Omega = 0$ as in \eqref{eq:logarithmic-derivative} as $L_{1}, L_2$. We write the inner product of \eqref{eq:cov-information} at $\bm \Omega = 0$ plainly as $\langle \cdot, \cdot \rangle$.

Hence, consider the positive semi-definite matrix
\begin{align}
B := \begin{bmatrix} \langle \hat{A}_1, \hat{A}_1 \rangle & \langle \hat{A}_1, \hat{A}_2 \rangle & \langle \hat{A}_1, \hat{L}_1 \rangle & \langle \hat{A}_1, \hat{L}_2 \rangle \\
\langle \hat{A}_2, \hat{A}_1 \rangle & \langle \hat{A}_2, \hat{A}_2 \rangle & \langle \hat{A}_2, \hat{L}_1 \rangle &\langle \hat{A}_2, \hat{L}_2 \rangle  \\
\langle \hat{L}_1, \hat{A}_1 \rangle & \langle \hat{L}_1, \hat{A}_2 \rangle & \langle \hat{L}_1, \hat{L}_1 \rangle & \langle \hat{L}_1, \hat{L}_2 \rangle \\
\langle \hat{L}_2, \hat{A}_1 \rangle & \langle \hat{L}_2, \hat{A}_2 \rangle & \langle \hat{L}_2, \hat{L}_1 \rangle & \langle \hat{L}_2, L_2 \rangle \end{bmatrix} \succeq \hat 0
.\end{align}
Using \eqref{eq:bias-matrix}, we have that
\begin{align}
B = \begin{bmatrix} \Cov(\hat{\bm A}) & J \\ 
J^\dag & \mathcal{I} \end{bmatrix}
,\end{align}
where $\Cov(\hat{\bm A})$ is the covariance matrix at $\bm \Omega = 0$ as in \eqref{eq:generalized-covariance} and $\mathcal{I}$ is the information matrix at $\bm \Omega = 0$ as in \eqref{eq:cov-information}.

The positive semi-definiteness of $B$ implies the theorem using the fundamental identity for the Schur complement\footnote{See e.g. \cite{kreindler1972conditions} or Theorem 7.7.9 of \cite{horn2012matrix}.}.
\end{proof}
\end{theorem}

We now state a necessary and sufficient condition for saturation of the \gls{qcrb} under the \gls{bh} choice of metric.

\begin{lemma}[\cite{liu2019quantum}]
A necessary and sufficient condition for the saturation of the \gls{bh} multi-parameter \gls{qcrb} is
\begin{align}
\tr(\hat{\rho}[\hat{L}^{\operatorname{BH}}_{\bm \Omega, j},\hat{L}^{\operatorname{BH}}_{\bm \Omega, k}]) = 0 \hspace{0.3cm} \forall j,k.
\end{align}
\end{lemma}

We'll refer to the above as a compatibility condition for the logarithmic derivatives. Armed with this result, we can show that \glspl{qhbm} (\Cref{sec:review-qhbm}) are parameterized in such a way that the tigtest \gls{qcrb} can always be saturated.

\begin{proposition}
\label{apdx:prop:compatibility}
Let $\bm{\Omega} \in \mathbb{R}^d$ be coordinates for a \gls{qhbm} $\hat{\rho}_{\bm{{\Omega}}}$. Then, the compatibility condition $$\tr[\hat{\rho}_{\bm{\Omega}}[\hat{L}_j, \hat{L}_k]] = 0, \hspace{0.2cm} j,k \in \{1, \dots, d\}$$ 
holds. As a consequence, optimal parameter estimation for \glspl{qhbm}, with \glspl{qnn} amenable to parameter shifts, asymptotically saturates the Quantum multi-parameter Cramer-Rao bound.

\begin{proof}
We break the proof into a few different cases. Let $d = d_1 + d_2$, so that $\bm{\Omega} = (\bm\theta, \bm{{\phi}})$, with $\bm\theta\in\mathbb{R}^{d_1}$ and $\bm{{\phi}} \in \mathbb{R}^{d_2}$. Then, for all $j,k \in \{1, \dots, d_1\}$, we have
\begin{align}
\hat{L}_{\theta_i} = \hat{\rho}_{\bm{\Omega}}^{-1} \partial_{\theta_i}\hat{\rho}_{\bm{\Omega}}.
\end{align}
Henceforth, we write $\hat{\rho} := \hat{\rho}_{\bm{\Omega}}$ for notational convenience. A simple computation gives
\begin{align}
&\tr[\hat{\rho} [\hat{L}_{\theta_j}, \hat{L}_{\theta_k}]] \\
&= \tr[\hat{\rho}\left(\hat{\rho}^{-1}(\partial_{\theta_j}\hat{\rho})\hat{\rho}^{-1}\partial_{\theta_k}\hat{\rho} - \hat{\rho}^{-1}(\partial_{\theta_k}\hat{\rho})\hat{\rho}^{-1}\partial_{\theta_j}\hat{\rho}\right)] \\
&= \tr[(\partial_{\theta_j}\hat{\rho})\hat{\rho}^{-1}\partial_{\theta_k}\hat{\rho} - (\partial_{\theta_k}\hat{\rho})\hat{\rho}^{-1}\partial_{\theta_j}\hat{\rho}] \\
&= \tr[\hat{\rho}^{-1}\partial_{\theta_k}\hat{\rho}(\partial_{\theta_j}\hat{\rho}) - (\partial_{\theta_k}\hat{\rho})\hat{\rho}^{-1}\partial_{\theta_j}\hat{\rho}] \\
&= \tr[[\hat{\rho}^{-1}, \partial_{\theta_k} \hat{\rho}]\partial_{\theta_j}\hat{\rho}] = 0,
\end{align}
where in the last line we used $[\hat{\rho}^{-1}, \partial_{\theta_k} \hat{\rho}]=0$, which holds because the \gls{qhbm} is diagonalized.

Now, we consider the condition for the $\bm{\phi}$ parameters. We will see in \eqref{SLD-unitary-params} the form of the SLD,
\begin{align}
\hat{L}_{\phi_i} = \sum_{x,y} \frac{2}{p_\theta(x) + p_\theta(y)} (\bra{x}\hat{\rho}_{\theta (\phi + \Delta_i)}\ket{y} - \bra{x}\hat{\rho}_{\theta (\phi - \Delta_i)}\ket{y})\ket{x}\bra{y}.
\end{align}
For simplicity, denote $\hat{\rho}^+_i := \hat{\rho}_{\theta(\phi+\frac{\pi}{4}\hat{e}_i)}$ and $\hat{\rho}^-_i := \hat{\rho}_{\theta(\phi-\frac{\pi}{4}\hat{e}_i)}$. Then, we have:
\begin{align}
[\hat{L}_{\phi_j}, \hat{L}_{\phi_k}] =\!\!\!\!\sum_{x,y,x',y'}\!\! \frac{4}{(p_{\theta}(x)+p_{\theta}(y))(p_{\theta}(x') + p_{\theta}(y'))}((\hat{\rho}^+_j)_{xy} - (\hat{\rho}^-_j)_{xy})((\hat{\rho}_k^+)_{x'y'} - (\hat{\rho}_k^-)_{x'y'}) \left(\delta_{x',y}|x\rangle \langle y'| - \delta_{x,y'} |x'\rangle\langle y|\right).
\end{align}
Splitting the above sum into two via the Kronecker deltas, then relabeling indices, yields the simple conclusion $
[\hat{L}_{\phi_j}, \hat{L}_{\phi_k}] = 0.
$ Our final task is to prove compatibility between the $\theta$ and $\phi$ parameters, i.e., that $$\tr(\hat{\rho}[\hat{L}_{\theta_j}, \hat{L}_{\phi_k}]) = 0.$$
We expand this assertion:
\begin{align}
 \tr[\hat{\rho}[\hat{L}_{\theta_j}, \hat{L}_{\phi_k}]] &= \sum_{x,y} \frac{2}{p_{\bm\theta}(x) + p_{\bm\theta}(y)}((\hat{\rho}^+_k)_{xy} - (\hat{\rho}^-_k)_{xy}) \tr[\hat{\rho}[\hat{\rho}^{-1}\partial_{\theta_j}\hat{\hat{\rho}}, |x\rangle\langle y|]]\\
 &= \sum_{x,y} \frac{2}{p_{\bm\theta}(x) + p_{\bm\theta}(y)}((\hat{\rho}^+_k)_{xy} - (\hat{\rho}^-_k)_{xy}) \left(\langle y|\partial_{\theta_j}\hat{\rho}|x\rangle - \langle y|\hat{\rho}^{-1}(\partial_{\theta_j}\hat{\rho}) \hat{\rho}|x\rangle\right) \\
 &= \sum_{x,y} \frac{2}{p_{\bm\theta}(x) + p_{\bm\theta}(y)}((\hat{\rho}^+_k)_{xy} - (\hat{\rho}^-_k)_{xy}) \left(\langle y|\partial_{\theta_j}\hat{\rho}|x\rangle - \langle y|\partial_{\theta_j}\hat{\rho}|x\rangle\right) = 0.
\end{align}
Importantly, we used $[\partial_{\theta_j}\hat{\rho}, \hat{\rho}] = 0.$ Thus, the desired compatibility condition holds and we are done.
\end{proof}
\end{proposition}

\section{Metric-aware gradient descent}
\label{sec:metric-aware}

\subsection{Derivation}
\label{sec:metric-aware-derivation}

In this subsection, we derive a general form for quantum metric-aware gradient descent where the monotone metric over parameter space geometry is left as a degree of freedom (see \Cref{sec:geometry}). That is, we will derive that the metric-aware parameter update is the choice of parameter update which yields the steepest descent in $\mathcal{M}^{(N)}$ towards the minimum of a loss function while maintaining a predetermined step size in $\mathcal{M}^{(N)}$. 

The chosen monotone metric provides a notion of distance and length in parameter space which is a better representation of how changes in parameter space affect changes in state space. As such, evaluating this metric can be used to augment standard gradient descent strategies to obtain \textit{metric-aware gradient descent}. Such a method can be considered a second-order method (as it uses second-order information about $\mathcal{M}^{(N)}$). Although it can be more computationally costly per iteration, in many cases the descent procedure can converge on a significantly smaller number of iterations to the optimal parameters.

We can phrase our notion of steepest descent mathematically as the choice of $\bm {\Omega}_{j + 1}$ ($:= \bm {\Omega}_j + \bm \delta_{j+1}$) at gradient step ${(j+1)}$ following,
\begin{align}
  \bm{\delta}_{j + 1} & = \underset{\bm{\delta}:\ \varepsilon^2 = \Phi(\hat{\rho}_{\bm{{\Omega}}_j }, \hat{\rho}_{\bm{{\Omega}}_j + \bm{\delta}}) }{\arg\min}\mathcal{L}(\hat{\rho}_{\bm{{\Omega}}_j +\bm{\delta}})
,\end{align}
where $\mathcal{L}$ is our loss functional and $\varepsilon^2$ is a constant. Note that object of special interest is $\Phi$ which describes a metric-related potential that is symmetric and bilinear up to third-order. In particular, we use the fact that every monotone metric is the Hessian of some potential \cite{lesniewski1999monotone},
\begin{align}
  [\mathcal{I}({\bm \Omega})]_{j, k} &\equiv -\partial^2_{\Omega'_j \Omega_k} \Phi (\hat{\rho}_{\bm \Omega'}, \hat{\rho}_{\bm{\Omega}})\mid_{\bm \Omega'= \bm \Omega},
\end{align}
where $\mathcal{I}_{\bm \Omega}$ is the information matrix of the metric, as in \Cref{eq:monotone-info-matrix}. Furthermore, the potential corresponds to a locally perturbed member of the family of so-called monotone relative entropies \cite{lesniewski1999monotone}. It is also the case that a monotone relative entropy obeys that $\Phi(\hat{\rho}, \hat{\sigma})$ is at a minimum of zero when $\hat{\rho} = \hat{\sigma}$ implying that the first two orders of an expansion about $\hat\rho=\hat\sigma$ vanish. An explicit computation then shows that (\Cref{cor:mixed-hessian-rear-hessian}):
\begin{align}
\label{eq:expand-potential}
\Phi(\hat{\rho}_{\bm{{\Omega}}_j }, \hat{\rho}_{\bm{{\Omega}}_j + \bm{\delta}})  &= \frac{1}{2} \left \langle \bm{\delta}, \mathcal{I}({\bm{{\Omega}}_j}) \bm{\delta}\right\rangle + \mathcal{O}(\lVert \bm\delta \rVert^3)
\end{align}
Notice \eqref{eq:constrained-opt} is a constrained minimization problem, we can thus consider the relaxed Lagrangian,
\begin{align}
  \label{eq:ngd_fund}
  \bm{\delta}_{j + 1} = \underset{\bm{\delta}}{\arg\min}\left[ \mathcal{L}(\hat{\rho}_{\bm{{\Omega}}_j + \bm{\delta}}) + \lambda \left( \Phi(\hat{\rho}_{\bm{{\Omega}}_j}, \hat{\rho}_{\bm{{\Omega}}_j + \bm{\delta}}) - c^2 \right)\right].
\end{align}
We may expand both terms to their first non-vanishing order in $\bm{\delta}$ and remove constants which do not depend on $\bm \delta$. The loss function will be expanded to first-order and the potential term to second-order,
\begin{align}
\label{eq:discrete-diff-eq}
  \bm{\delta}_{j + 1}  &= \underset{\bm{\delta}}{\arg\min}\left[\left\langle \nabla_{\bm{{\Omega}}_j}\mathcal{L}(\hat{\rho}_{\bm{{\Omega}}_j}), \bm{\delta}\right\rangle + \frac{\lambda}{2} \left \langle \bm{\delta}, \mathcal{I}({\bm{{\Omega}}_j}) \bm{\delta}\right\rangle \right]
  + \mathcal{O}(\lVert \bm\delta \rVert^3)
  .\\
  \intertext{
Equivalently,}
  \bm{\Omega}_{j + 1} &\approx \underset{\bm{{\Omega}}}{\arg\min}\left[  \bm{{\Omega}}_j + \left\langle \nabla_{\bm{{\Omega}}_j}\mathcal{L}(\hat{\rho}_{\bm{{\Omega}}_j}), \bm{\Omega} - \bm{{\Omega}}_j \right\rangle 
              + \frac{\lambda}{2} \left\langle \bm{\Omega} - \bm{\Omega}_j, \mathcal{I}({\bm{{\Omega}}_j}) (\bm{\Omega} - \bm{\Omega}_j) \right\rangle \right] \\
             \label{eq:ngd-inner-product}
             &= \underset{\bm{{\Omega}}}{\arg\min}\left[ \left\langle \nabla_{\bm{{\Omega}}_j}\mathcal{L}(\hat{\rho}_{\bm{{\Omega}}_j}), \bm{\Omega} \right\rangle 
             + \frac{\lambda}{2} \left\langle \bm{\Omega} - \bm{\Omega}_j, \mathcal{I}({\bm{{\Omega}}_j}) (\bm{\Omega} - \bm{\Omega}_j) \right\rangle \right]
\end{align}
Differentiating \eqref{eq:ngd-inner-product} yields the metric-aware gradient update,
\begin{align}
  \bm{\Omega}_{j + 1} \leftarrow \bm{{\Omega}}_j -\tfrac{1}{\lambda}  \mathcal{I}^{+} ({\bm{{\Omega}}_j})\nabla_{\bm{{\Omega}}_j}\mathcal{L}(\hat{\rho}_{\bm{{\Omega}}_j}),
\end{align}
where $\eta := \tfrac{1}{\lambda}$ will be our effective learning rate. 

The objective \eqref{eq:constrained-opt} seems to indicate that the update rule is invariant under re-parameterization given its exclusive dependence on density operator arguments (assuming that $\varepsilon$ is sufficiently small, so that the relevant approximations are sufficiently precise). This is indeed the case when the model parameterization describes a global diffeomorphism from $\mathcal{M}^{(N)}$ (or a submanifold thereof) to classical parameter space\footnote{See \Cref{sec:duality-qhbm} for a relevant discussion.}. However, the introduction of the pseudo-inverse breaks this invariance in general because, for example in the case where the model is over-parameterized, the pseudo-inverse re-introduces a coordinate dependence by outputting the solution $\bm x$ to $\mathcal{I}(\bm \Omega_j)\bm x = \nabla_{\bm \Omega_j} \mathcal{L}(\hat \rho_{\bm \Omega_j})$ with minimum 2-norm. A common strategy to circumvent the pseudo-inverse is to regularize the information matrix $\mathcal{I}(\bm \Omega_j) \rightarrow \mathcal{I}(\bm \Omega_j) + \varepsilon\mathbbm{1}$ choosing a sufficiently large $\varepsilon$. This has the interpretation of adding an independent Gaussian prior over parameter space as we will see from \eqref{eq:laplace} and therefore likewise introduces a coordinate dependence.

\subsection{Optimality guarantees}
We now move to a setting of proving asymptotic optimality of the metric-aware gradient descent algorithm with the \gls{bkm} metric. Hence, for simplicity, we will drop the $\operatorname{BKM}$ qualifier throughout:
\begin{align}
    \mathcal{I}(\cdot) &\equiv \mathcal{I}^{\operatorname{BKM}}(\cdot) \\
    \Phi(\cdot, \cdot) &\equiv \Phi^{\operatorname{BKM}}(\cdot, \cdot) \\
    \Cov(\cdot ; \cdot) &\equiv \Cov^{\operatorname{BKM}}(\cdot ; \cdot)
\end{align}
In particular, we will show that the \gls{bkm} metric-aware update rule achieves the corresponding generalized \gls{qcrb} in the limit of many steps. To this end, we provide the context for this result by first giving our assumptions.

\begin{assumption}
\label{assum:convergence-for-optimality}

Assume that $\bm\Omega \mapsto \hat{\rho}_{\bm\Omega}$ is a smooth immersion $\mathbb{R}^d \to \mathcal{M}^{(N)}$, hence defining smooth local coordinates on a submanifold of $\mathcal{M}^{(N)}$.
Assume that the \gls{bkm} metric-aware update rule
\begin{align}
\label{apdx:eq:qcrb-gradient-update}
    \bm{\Omega}_{j + 1} &= \bm{{\Omega}}_j - \eta_j  \left(\mathcal{I} ({\bm{{\Omega}}_j})\right)^{-1}  \tilde{h}(\bm{\Omega_j})\\
    \intertext{where,}
    \label{eq:online-gradient-estimator}
    \mathbb{E}_{\bm x}[\tilde{h}(\bm{\Omega})] &= \nabla_{\bm \Omega} \Phi(\bm \Omega^* , {\bm{{\Omega}} })
,\end{align}
converges to the optimal parameters $\bm\Omega^*$ in expectation, i.e., $\mathbb{E}_{\bm x}[\bm\Omega_j] \to \bm\Omega^*$ as $j \to \infty$. Here, $\tilde{h}(\cdot)$ is an online unbiased estimator of the quantum relative entropy loss gradient obtained by the environment drawing a single pure state $\ket{\bm x}$ from the eigenstates of $\hat\rho_{\bm \Omega^*}$ (with probability of the corresponding eigenvalue) at each optimization step\footnote{Any measurement on a target density operator can be viewed as a stochastic operation on its eigenstates.}.
\end{assumption}

The following proposition quantizes the formulation of \cite{murata1998statistical} (see, in particular, eq. 48).

\begin{proposition}
\label{prop:convergence}

Suppose the learning rate, for the update rule \eqref{apdx:eq:qcrb-gradient-update} taken in expectation,
\begin{align}
\label{apdx:eq:qcrb-gradient-update-expected}
    \bm{\Omega}_{j + 1} &= \bm{{\Omega}}_j - \eta_j  \left(\mathcal{I} ({\bm{{\Omega}}_j})\right)^{-1}  \mathbb{E}_{\bm x}[\tilde{h}(\bm{\Omega_j})]
\end{align}
is chosen as $ \eta_j = \frac{1}{j + 1}$ with initial parameters $\bm \Omega_1$. Then, under the conditions of \Cref{assum:convergence-for-optimality}, we have 
\begin{align}
\label{eq:convergence-linear}
    \lVert \bm \Omega_{j} - \bm \Omega^* \rVert_2 = \frac{1}{j} \lVert \bm \Omega_{1} - \bm \Omega^* \rVert_2
,\end{align}
asymptotically.
\begin{proof}

We may Taylor expand the loss function as
\begin{align}
\Phi( \bm \Omega^*, \bm \Omega_j ) &= \Phi( \bm \Omega^*, \bm \Omega^*) + \langle \nabla_{\bm \Omega}\Phi( \bm \Omega^*, \bm \Omega )\mid_{\bm \Omega = \bm \Omega^*} , \bm \Omega_j - \bm \Omega^* \rangle + \frac{1}{2} \left \langle \bm\Omega_j - \bm \Omega^*, \mathcal{I}({\bm\Omega^*}) (\bm \Omega_j - \bm \Omega^*)\right\rangle + \mathcal{O}(\eps_j^3) \\
\label{eq:convex-bowl-expansion}
&= \frac{1}{2} \left \langle \bm\Omega_j - \bm \Omega^*, \mathcal{I}({\bm\Omega^*}) (\bm \Omega_j - \bm \Omega^*)\right\rangle + \mathcal{O}(\eps_j^3)
,\end{align}
where we used that a generalized relative entropy and its gradient vanish at the optimum, and defined $\eps_j := \lVert \bm \Omega_j - \bm \Omega^* \rVert_2$.
As a consequence,
\begin{align}
\label{eq:cute-expansion}
\nabla_{\bm\Omega_j} \Phi( \bm \Omega^*, \bm \Omega_j ) &= \langle \mathcal{I}({\bm\Omega^*}) , \bm \Omega_j - \bm \Omega^* \rangle + \mathcal{O}(\eps_j^2).
\end{align}
Similarly to \eqref{eq:convex-bowl-expansion}, now expanding the first argument,
\begin{align}
\label{eq:convex-bowl-other}
   \Phi( \bm \Omega^*, \bm \Omega_j ) &= \frac{1}{2} \left \langle \bm\Omega_j - \bm \Omega^*, \mathcal{I}({\bm\Omega_j}) (\bm \Omega_j - \bm \Omega^*)\right\rangle + \mathcal{O}(\eps_j^3).
\end{align}
So, from \cref{eq:convex-bowl-expansion,eq:convex-bowl-other},
\begin{align}
\label{eq:info-zeroth-1}
    \mathcal{I}(\bm \Omega_j) &=  \mathcal{I}(\bm \Omega^*) + \mathcal{O}\left(\eps_j\right).
\end{align}
\Cref{apdx:eq:qcrb-gradient-update-expected} can be written
\begin{align}
    \bm \Omega_{j+1} &=  \bm \Omega_j - \eta_j \mathcal{I}^{-1}(\bm \Omega_j)\nabla_{\bm \Omega_j} \Phi(\bm \Omega^* , {\bm{{\Omega}}_j }). \\
     \intertext{From \eqref{eq:cute-expansion},}
     &= \bm \Omega_j - \eta_j \mathcal{I}^{-1}(\bm \Omega_j) \mathcal{I}(\bm \Omega^*)(\bm \Omega_j - \bm \Omega^*) + \mathcal{O}(\eta_j \eps_j^2). \\
     \intertext{From \eqref{eq:info-zeroth-1},}
    \label{eq:metric_approx_assum}
    &= \bm \Omega_j - \eta_j (\bm \Omega_j - \bm \Omega^*) + \mathcal{O}(\eta_j \eps_j^2) \\
     &=   \eta_j \bm \Omega^* + (1- \eta_j) \bm \Omega_j + \mathcal{O}(\eta_j \eps_j^2).
     \end{align}
     Within some neighborhood of the optimum, the approximations \cref{eq:convex-bowl-expansion,eq:convex-bowl-other} are good enough so that $\eps_j = \mathcal{O}(\eta_j)$\footnote{This neighborhood may be arbitrarily small. See \Cref{rmk:1-j}.}.
     We have chosen $\eta_j = \tfrac{1}{j+1}$. Therefore, neglecting the third order $\mathcal{O}(\tfrac{1}{j^3})$ terms,
     \begin{align}
   \bm \Omega_j &= \bm \Omega^* + \prod_{k=2}^j \left(1 - \frac{1}{k}\right) (\bm \Omega_1 - \bm \Omega^*) \\
   &= \bm \Omega^* + \frac{1}{j} (\bm \Omega_1 - \bm \Omega^*)
 ,\end{align}
 for $j > 1$.
\end{proof}
\end{proposition}

Note that, for simplicity in notation, it is assumed that the learning algorithm begins at $j = 2$ with initial parameters $\bm \Omega_1$.

\begin{remark}
\label{rmk:1-j}
The convergence \eqref{eq:convergence-linear} is $\mathcal{O}\left(\tfrac{1}{j}\right)$ asymptotically which holds for any $\eta_j = \mathcal{O}(\frac{1}{j})$. We have used the fact that there exists some neighborhood of the optimum where $\eps_j := \lVert \bm \Omega_j - \bm \Omega^* \rVert_2 = \mathcal{O}(\eta_j)$. This neighborhood can be arbitrarily small, so long as it is reached in a finite number of steps, since we still use that convergence occurs as $j \rightarrow \infty$. This assumption is shared with Amari in his classical work \cite{amari1998natural}, and so we follow this precedent. Providing a more practical non-asymptotic convergence rate generally requires specifying more information about the parameterized state $\hat{\rho}_{\bm\Omega}$ or the target $\hat \rho_{\bm \Omega^*}$. Agnostic to this type of specification, we leverage \Cref{prop:convergence} to prove one of our main results, \Cref{res:qpngd-achieves-qcrb}. However, we will revisit this point in \Cref{sec:non-asymptotic}.
\end{remark}

The proof and theorem statement below should be contrasted with the classical analogue \cite{amari2016information} which is proven with less machinery.

\begin{theorem}
\label{res:qpngd-achieves-qcrb}

Suppose \Cref{assum:convergence-for-optimality} holds. Then, the update rule \eqref{apdx:eq:qcrb-gradient-update} with learning rate $\eta_j = \frac{1}{j}$ induces a dynamical equation on any arbitrary initial choice of a collection of quantum observables, acting as estimators, such that the generalized \gls{qcrb} \eqref{eq:qcrb-unbiased} is attained asymptotically. The equivalent statement holds when one swaps arguments of the loss, $\Phi(\bm \Omega^* , {\bm{{\Omega}}_j },) \rightarrow \Phi( {\bm{{\Omega}}_j }, \bm \Omega^* ) $.
  
  \begin{proof}
We will denote the $k$th element of $\bm \Omega_j$ as $\Omega_j^k$. We will attempt to map the metric-aware update rule \eqref{apdx:eq:qcrb-gradient-update} in parameter space to a latent dynamical equation (induced by quantum expectation) in quantum observables. This will allow us to tie back to the parameter estimation language and bounds of \Cref{sec:geometry-qcrb}. So, let $\bm \Omega^*$ give the true value of $\bm\Omega$, and assume that $\bm \Omega^* = 0$ without loss of generality. Following \eqref{eq:qcrb-apdx}, initialize a collection of quantum observables $\hat{\bm{A}}_0 = \{\hat{{A}}_0^k \}_{k=1}^d$ that satisfy $\tr[\hat\rho_{\bm \Omega^*} \hat{{A}}_j^k] = \Omega_j^k$, so that $\hat{\bm{A}}_j$ is an unbiased estimator of $\bm \Omega_j$. In such a case, we can rewrite \eqref{apdx:eq:qcrb-gradient-update} in expectation (as in \eqref{apdx:eq:qcrb-gradient-update-expected}) element-wise as
  \begin{align}
      \tr[\hat\rho_{\bm \Omega^*} \hat{{A}}_{j+1}^k] & = \tr[\hat\rho_{\bm \Omega^*} {\hat{A}}_{j}^k] - \tfrac{1}{j}  \sum_{l} [\mathcal{I}^{-1}({\bm{{\Omega}}_j})]_{k,l}  \partial_{\bm{{\Omega}}_j^l} \Phi( \bm \Omega^*, {\bm{{\Omega}}_j } ),
  \end{align}
  noting that this corresponds to the minimization of the reverse relative entropy loss. We have seen that $-\partial^2_{\bm \Omega_l', \bm \Omega_m} \Phi( \bm \Omega', \bm \Omega)\mid_{\bm \Omega' = \bm \Omega} = [\mathcal{I}(\bm \Omega_j)]_{l,m} = \tr[ (\partial_{\bm \Omega_l} \hat \rho_{\bm \Omega}) \mathscr{L}^{f}_{\bm \Omega}(\partial_{\bm \Omega_m} \hat \rho_{\bm \Omega}) ]$. For the \gls{bkm}, we can check from \eqref{eq:def-relative-entropy} that the first derivative satisfies,
  \begin{align}
  \label{eq:cute-derivative}
    -\partial_{\bm \Omega_l} \Phi( \bm \Omega', \bm \Omega ) &=  \tr[ \hat \rho_{\bm \Omega'} \mathscr{L}_{\bm \Omega}(\partial_{\bm \Omega_l} \hat \rho_{\bm \Omega}) ]
  \end{align}
  Now, given \eqref{eq:cute-derivative},
  \begin{align}
  \label{eq:projected-dynamics}
     \tr[\hat\rho_{\bm \Omega^*} \hat{{A}}_{j+1}^k] &=  \tr[\hat\rho_{\bm \Omega^*} \hat{{A}}_{j}^k]
      + \tfrac{1}{j}  \sum_{l} [\mathcal{I}^{-1}({\bm{{\Omega}}_j})]_{k,l}   \tr[ \hat \rho_{\bm \Omega^*} \mathscr{L}_{\bm \Omega_j}(\partial_{\bm{{\Omega}}_j^l} \hat \rho_{\bm \Omega_j}) ]
  .\end{align}
  With this in hand, we see that the metric-aware descent update \eqref{eq:projected-dynamics} is the quantum expectation\footnote{As in \Cref{assum:convergence-for-optimality}, the algorithmically relevant expectation $ [\mathcal{I}^{-1}({\bm{{\Omega}}_j})]_{k,l}   \tr[ \hat \rho_{\bm \Omega^*} \mathscr{L}_{\bm \Omega_j}(\partial_{\bm{{\Omega}}_j^l} \hat \rho_{\bm \Omega_j}) ]$ can be estimated unbiasedly, for each $j$, using a single pure state sample from the eigenstates of the data $\hat\rho_{\bm \Omega^*}$ (sampled with probability of the corresponding eigenvalue). Of course, the tangent vector and information matrix require additional samples to estimate, but these depend only on the model. This is a quantum generalization of \textit{online} learning.

      } of latent dynamical equation
  \begin{align}
  \label{eq:latent-dynamical}
  \hat{{A}}_{j+1}^k &=   \hat{{A}}_{j}^k
  + \tfrac{1}{j}  \sum_{l} [\mathcal{I}^{-1}({\bm{{\Omega}}_j})]_{k,l}   \mathscr{L}_{\bm \Omega_j}(\partial_{\bm{{\Omega}}_j^l} \hat \rho_{\bm \Omega_j})
  .\end{align}
  Hence, in expectation, \eqref{eq:latent-dynamical} describes how our collection of quantum observables, acting as estimators, updates at each descent step. So, we seek to bound the associated error covariance matrix corresponding to the metric given element-wise by,
  \begin{align}
 V_j &:= [\Cov(\hat{\bm{A}}_j  ; \bm \Omega^*)]_{k, l} \equiv   \tr[\mathscr{R}_{\bm \Omega^*}\left(\hat{{A}}_j^k\right) \hat{{A}}_j^l]
  \end{align}
  Using \Cref{prop:convergence} again, $\mathscr{R}_{\bm \Omega^*}(\cdot)  = \mathscr{R}_{\bm \Omega_j}(\cdot)  + \mathcal{O}\left(\tfrac{1}{j}\right)$. Furthermore, using both sides of \eqref{eq:latent-dynamical} and that $\mathscr{L}_{\bm \Omega} = (\mathscr{R}_{\bm \Omega})^{-1}$ and $ \tr[A\mathscr{R}_{\bm \Omega}(B)] = \tr[B^\dag \mathscr{R}_{\bm \Omega}(A^\dag)]$ \cite{petz2011introduction},
  \begin{align}
      [V_{j+1}]_{k, l} &= \begin{aligned}[t]&[V_j]_{k, l}
      + \frac{2}{j} \sum_{l'} [\mathcal{I}^{-1}({\bm{{\Omega}}_j})]_{l,l'} \tr[\mathscr{R}_{\bm \Omega^*}(\hat{{A}}_{j}^k) \mathscr{L}_{\bm \Omega_j}\left(\partial_{\bm \Omega_j^{l'}} \hat \rho_{\bm \Omega_j}\right)]  \\
      &+ \frac{1}{j^2} \sum_{k', l'} [\mathcal{I}^{-1}({\bm{{\Omega}}_j})]_{k,k'}
      \tr[(\partial_{\bm \Omega_j^{k'}} \hat \rho_{\bm \Omega_j}) \mathscr{L}_{\bm \Omega_j}(\partial_{\bm \Omega_j^{l'}} \hat \rho_{\bm \Omega_j})]
     [\mathcal{I}^{-1}({\bm{{\Omega}}_j})]_{l,l'} \\
     &+ \mathcal{O}\left(\tfrac{1}{j^3}\right)
     ,\end{aligned} \\
     \intertext{
     So, using \cref{eq:cute-expansion,eq:cute-derivative},}
    \tr[\mathscr{R}_{\bm \Omega^*}(\hat{{A}}_{j}^k) \mathscr{L}_{\bm \Omega_j}\left(\partial_{\bm \Omega_j^{l'}} \hat \rho_{\bm \Omega_j}\right)] &= -\sum_{m} [\mathcal{I}(\bm \Omega^*) ]_{l', m} \tr[\mathscr{R}_{\bm \Omega^*}(\hat{{A}}_{j}^k) \hat{{A}}_{j}^m] + \mathcal{O}\left(\tfrac{1}{j^2}\right) \\
    &= -[\mathcal{I}(\bm \Omega^*) V_j]_{l', k} + \mathcal{O}\left(\tfrac{1}{j^2}\right) \\
                V_{j+1} & = V_j - \tfrac{2 }{j}   ({\mathcal{I}}^{-1}_{\bm \Omega_j} ){\mathcal{I}}_{\bm \Omega^* }V_j + \tfrac{1}{j^2} ({\mathcal{I}}_{\bm \Omega_j} )^{-1}+ \mathcal{O}\left(\tfrac{1}{j^3}\right). \\
      \intertext{Finally, applying \Cref{prop:convergence} once last time gives that}
      \label{eq:info-zeroth}
      ({\mathcal{I}}_{\bm \Omega_j} )^{-1} &= \left({\mathcal{I}}_{\bm \Omega^*} \right)^{-1} + \mathcal{O}\left(\tfrac{1}{j}\right) \\
      \intertext{and so}
        V_{j+1}   & = V_j - \tfrac{2 }{j}V_j + \tfrac{1}{j^2} ({\mathcal{I}}_{\bm \Omega^*} )^{-1}+ \mathcal{O}\left(\tfrac{1}{j^3}\right)
      .\end{align}
    Therefore, $V_j = \frac{1}{j} ({\mathcal{I}}_{\bm \Omega^*} )^{-1} + \mathcal{O} \left(\frac{1}{j^2} \right)$ asymptotically, assuming that $\bm\Omega_j$ converges to $\bm \Omega^*$\footnote{This scaling is reminiscent of the Bernstein-von Mises theorem \cite{le2012asymptotic}.}. Note that the same asymptotic result holds if we flip the argument of our loss function $\Phi(\bm \Omega^* , {\bm{{\Omega}}_j },) \rightarrow \Phi( {\bm{{\Omega}}_j }, \bm \Omega^* ) $ because quantum relative entropy is symmetric up to third-order (\Cref{prop:forward-reverse-local-equiv}).
  \end{proof}
\end{theorem}

The meaning of an online gradient estimator for simulation is clarified in \Cref{sec:online-loss-gradient}.

\subsection{Practical considerations of optimality guarantees}

\subsubsection{Non-asymptotic optimization}
\label{sec:non-asymptotic}

The result of \Cref{res:qpngd-achieves-qcrb} is asymptotic (see \Cref{rmk:1-j}). Hence, we may question the practical convergence properties (e.g. in mean and variance) when restricting to a small number of steps. We may do so from the lens of the expansions used in its derivation. From the proof of \Cref{prop:convergence}, we saw that quantum Fisher efficiency depends on the iterates $\bm \Omega_j$ entering the locally quadratic convex region where

\begin{align}
\mathcal{I}(\bm \Omega_j) = \mathcal{I}(\bm \Omega^*) + \mathcal{O}\left(\frac{1}{j}\right)
.\end{align}

For this to be meaningful non-asymptotically, it is therefore  important that the constant factor of the $\mathcal{O}\left(\frac{1}{j}\right)$ is sufficiently small. So, this condition is a useful diagnosis for a sufficiently well-behaved region, and is derivative of the conditions \cref{eq:convex-bowl-expansion,eq:convex-bowl-other} which can be viewed as the primary approximations. 

Hence, when considering learning a sequences of states, we recall that chained initialization (\cref{sec:sequences-states}) means initializing the parameters for each state (after the first) at the optimal parameters corresponding to the previous target state in the sequence. Therefore, we may consider how strongly the information matrix varies in parameter space as we travel along the sequence. This allows one to evaluate whether we (approximately) remain within the quadratic convex bowl as we step forward in the sequence of states initialized at the previous optimum. This depends on both the physical process described by the sequence and the specific model parameterization. To this end, we work out the relevant expressions specific to real-time evolution, imaginary time-evolution, and generalizations of the latter to any generic path in parametric mixed state space in \Cref{sec:cts-sequence-limit}.

\subsection{Another perspective into estimator moments}

The following proposition allows one to approximate an arbitrary function of an online natural gradient estimator. Such a function may be any classical moment, for example.

\begin{proposition}[\cite{amari1967theory}]
\label{lemma-amari-1967}

Assume the update rule \eqref{apdx:eq:qcrb-gradient-update}. The expectation of a smooth function $f(\bm \Omega)$ at iteration ${j+1}$ is given by recursive equation
\begin{align}
\label{eq:amari-convergence}
      \mathbb{E}_{\bm \Omega_{j+1}}[f(\bm \Omega_{j+1})] &=  \begin{aligned}[t]&\mathbb{E}_{\bm \Omega_j}[f(\bm \Omega_j)] - \eta_j \mathbb{E}_{\bm \Omega_j}[ \langle \nabla f(\bm \Omega_j),  \mathcal{I}^{-1}(\bm \Omega_j) \nabla_{\bm \Omega_j} \Phi(\bm \Omega^* , {\bm{{\Omega}}_j }) \rangle] \\&+ \frac{\eta_j^2}{2} \tr[\mathbb{E}_{\bm \Omega_j}[ \mathcal{I}^{-1}(\bm \Omega_j) (\nabla^2 f(\bm \Omega_j)) \mathcal{I}^{-1}(\bm \Omega_j) Q(\bm \Omega_j) ] ] + \mathcal{O}(\eta_j^3)\end{aligned} \\
      \intertext{where,}
      Q(\bm \Omega) &:= \mathbb{E}_{\bm x}[\tilde{h}(\bm \Omega)\tilde{h}^\dag(\bm \Omega)]
,\end{align}
for sufficiently small $\eta_j$.
  
\begin{proof}
  For simplicity in notation, write that $\bm \Omega' \equiv \bm \Omega_{j+1}, \bm \Omega \equiv \bm \Omega_j, \eta \equiv \eta_j$. As in \eqref{apdx:eq:qcrb-gradient-update}, the observation of data $\ket{\bm x}$ results in the model parameters update $\bm \Omega' = \bm \Omega + \delta \bm \Omega(\ket{\bm x}, \bm \Omega)$ where $\delta \bm \Omega(\ket{\bm x}, \bm \Omega) = - \eta \mathcal{I}^{-1}(\bm \Omega) \tilde{h}(\bm \Omega)$.
  
  We have said that $\ket{\bm x}$ follows the spectral distribution of $\hat\rho_{\bm \Omega^*}$ which we write as $p(\bm x)$. The conditional probability density $q(\bm \Omega' \mid \bm \Omega)$ obeys
  \begin{align}
      q(\bm \Omega' \mid \bm \Omega) d\bm \Omega' &= p(\bm x(\bm \Omega, \bm \Omega')) d \bm x(\bm \Omega, \bm \Omega')
  \end{align}
  in that $d \bm x(\bm \Omega, \bm \Omega')$ describes the volume element such that example $\bm x(\bm \Omega, \bm \Omega')$ modifies the estimator from $\bm \Omega$ to a point in the rectangle $[\bm \Omega', \bm \Omega' + d \bm \Omega']$. Hence,
  \begin{align}
     q(\bm \Omega') &= \int_{\bm \Omega} q(\bm \Omega' \mid \bm \Omega) q(\bm \Omega) d \bm \Omega' d \bm \Omega  \\
     &= \int_{\bm \Omega} p(\bm x)  q(\bm \Omega) d \bm x d \bm \Omega 
  \end{align}
  And so, 
  \begin{align}
      \mathbb{E}_{\bm \Omega'}[f(\bm \Omega')] &= \int_{\bm \Omega'} f(\bm \Omega') q(\bm \Omega') d\bm \Omega' \\
      &= \int_{\bm \Omega}\int_{\bm x} f(\bm \Omega') p(\bm x)  q(\bm \Omega) d \bm x d \bm \Omega \\
      &= \mathbb{E}_{\bm \Omega}[ \mathbb{E}_{\bm x}[f(\bm \Omega')]]
  \end{align}
  Hence, we may expand $\bm \Omega'$ about $\bm \Omega$ and note that $\delta \bm \Omega = \mathcal{O}(\eta)$
  \begin{align}
      \mathbb{E}_{\bm x}[f(\bm \Omega')] &=  \mathbb{E}_{\bm x}[f(\bm \Omega + \delta \bm \Omega)] \\
      &=  f(\bm \Omega) + \langle \nabla f(\bm \Omega), \mathbb{E}_{\bm x}[\delta \bm \Omega ]\rangle + \frac{1}{2} \mathbb{E}_{\bm x}[ \langle \delta \bm \Omega , (\nabla^2 f(\bm \Omega)) \delta \bm \Omega \rangle ] + \mathcal{O}(\eta^3) \\
      &= f(\bm \Omega) - \eta \langle \nabla f(\bm \Omega), \mathcal{I}^{-1}(\bm \Omega) \mathbb{E}_{\bm x}[\tilde{h}(\bm \Omega) ]\rangle + \frac{\eta^2}{2} \tr[\mathcal{I}^{-1}(\bm \Omega) (\nabla^2 f(\bm \Omega)) \mathcal{I}^{-1}(\bm \Omega) \mathbb{E}_{\bm x}[\tilde{h}(\bm \Omega)\tilde{h}^\dag(\bm \Omega)] ] + \mathcal{O}(\eta^3)
  \end{align}
  Taking the expectation with respect to $\bm \Omega$ completes the proof.
\end{proof}
\end{proposition}

\subsection{Sample-efficient learning of many-body states}
\label{sec:many-body}

The authors of \cite{anshu2021sample} consider the particular exponential parameterization of target state $\hat \sigma_\beta \in \mathcal{M}^{(N)}$ with $\kappa$-local modular Hamiltonian $\hat K = \sum_{l = 1}^{N^2-1} \mu_l \hat E_l$ having $p$ nonzero coefficients $\mu_j$; $\lVert\bm{\mu}\rVert_0 = p$, $\lvert \mu_j \rvert \leq 1$ over a finite-dimensional lattice so that 
\begin{align}
    \hat \sigma_\beta &= \frac{\exp(- \beta \hat{K})}{\mathcal{Z}_\beta}
\end{align}
with $\beta$ as inverse-temperature. The interaction graph is assumed to be spatially local so that $p = \mathcal{O}(n)$ in terms of the number of qubits. The canonical representation is assumed so that $\hat E_j$ are known non-identity Pauli operators. It is then shown that there are constants $c, c' > 3$ depending on the geometric properties of the lattice such that the Hessian of the log-partition function (aka free energy) satisfies
\begin{align}
\label{eq:strongly-convex}
    \nabla^2 \log \mathcal{Z}(\mu) \succeq e^{- \mathcal{O}(\beta^c)}\tfrac{\beta^{c'}}{p} \mathbbm{1}
\end{align}
Now, \Cref{res:qpngd-achieves-qcrb} implies that, for quantum metric-aware descent in the $\bm \mu$ parameterization,

\begin{align}
\label{eq:quantum-fisher-efficient-many-body}
    \Cov(\hat{\bm{A}}_m ; \bm \mu) &= \tfrac{1}{m} \left(\mathcal{I}(\bm \mu)\right)^{-1} + \mathcal{O}\left(\tfrac{1}{m^2}\right),
\end{align}
where we may write the induced $\hat{\bm{A}}_m = \{ \hat A_m^1, \cdots , \hat A_m^p \}$ with that $\tr[\hat \sigma_{\beta} \hat A_m^j] = \mu_j$. In mixture coordinates, we have that $\sigma_{\beta} = \frac{1}{N}\mathbbm{1} + \sum_{j=1}^{N^2 - 1} \eta_j \hat E_j$ where $\eta_j = \tfrac{\partial \log \mathcal{Z}}{\partial \mu_j}$ as in \eqref{eq:coord-change-exp-to-mix}.

\begin{lemma}[\cite{hasegawa1997exponential}]
\label{lem:bkm-mle}
The $\hat{\bm{A}}$ which achieves the \gls{bkm} \gls{qcrb} scaling i.e. $\Cov(\hat{\bm{A}} ; \bm \mu) = \left(\mathcal{I}(\bm \mu)\right)^{-1}$ asymptotically is given by
\begin{align}
\label{eq:bkm-mle}
\hat{A}^j &= \mu_j \mathbbm{1}  + \int_0^\infty (\hat \sigma_{\beta} + s\mathbbm{1})^{-1} \hat E_j (\hat \sigma_{\beta} + s\mathbbm{1})^{-1} ds
\end{align}

\begin{proof}

We verify this explicitly.
\begin{align}
[ \Cov(\hat{\bm{A}} ; \bm \mu)]_{j, k} &= \int_0^1 \tr[ \hat \sigma_\beta^{s} \hat A^j \hat \sigma_\beta^{1-s} \hat A^k] ds \\
&= \mu_j \mu_k + \int_0^\infty \tr[\hat E_j (\hat \sigma_{\beta} + s\mathbbm{1})^{-1} E_k (\hat \sigma_{\beta} + s\mathbbm{1})^{-1}] ds \\
&= \mu_j\mu_k + \tr[(\partial_{\eta_j} \hat \sigma_\beta)(\partial_{\eta_k} \log \hat \sigma_\beta)]
,\end{align}
which verifies that $\hat{\bm{A}}$ achieves the \gls{qcrb} scaling because the metric tensor in mixture coordinates is inverse to the one in exponential coordinates given that mixture coordinates can be seen as covariant with exponential as contravariant 
 
(Crouzeix’s identity \cite{crouzeix1977relationship, nielsen2020elementary,hasegawa1997exponential}).
\end{proof}
\end{lemma}

From \Cref{lem:bkm-mle}, we may consider this choice of $\hat{\bm{A}}$ to be the efficient estimator from the perspective of the \gls{bkm} geometry. We note that $\int_0^\infty (\hat \sigma_{\beta} + s\mathbbm{1})^{-1} \hat E_j (\hat \sigma_{\beta} + s\mathbbm{1})^{-1} ds = \partial_{\eta_j} \log \hat \sigma_{\beta}$ and \eqref{eq:expected-derivative-is-zero} shows that $\tr[(\partial_{\eta_j} \log \hat \sigma_{\beta}) \hat \sigma_\beta] = 0$.

We will see in \eqref{eq:hessian-log-partition} that the \gls{bkm} information matrix is equivalent to the Hessian of the log-partition function in the exponential coordinates. In their work and subsequent works \cite{haah2021optimal}, the connection with information geometry and parameter estimation bounds (e.g. Fisher efficiency) is not identified. This is reasonable because their observable estimators (measurements are taken to be the $\{\hat E_1, \cdots, \hat E_p\}$) do not match the ones \eqref{eq:bkm-mle} which we termed the efficient estimators in the eyes of the \gls{bkm}, in general. However, clearly, in our case this identification is appropriate because quantum Fisher efficiency ties this quantity as a metric to our algorithm's convergence rate; this identification allows a straightforward sample-efficiency upper bound argument when paired with a strong convexity bound on the free energy.

With this in mind, we have that
\begin{align}
\left(\mathcal{I}(\bm \mu)\right)^{-1} &\preceq e^{ \mathcal{O}(\beta^c)}\tfrac{p}{\beta^{c'}} \mathbbm{1} \\
\label{eq:cov-strong-convexity-bound}
\implies \Cov(\hat{\bm{A}}_m ; \bm \mu)  + \mathcal{O}\left(\tfrac{1}{m^2}\right) &\preceq e^{ \mathcal{O}(\beta^c)}\tfrac{p}{m \beta^{c'}} \mathbbm{1}
,\end{align}
It remains to show how $\Cov(\hat{\bm{A}}_m ; \bm \mu)$ relates to expected variations in the estimator $\bar \mu$. To achieve this, we will consider a large deviations perspective \cite{nagaoka2005asymptotic}. Large deviations analysis characterizes the exponential rate of convergence in the probability of the estimated parameters being greater than some constant.  

\begin{lemma}

Define $\bar \mu_m^j$ to be the estimator of $\mu_j$ found by a finite averaging of outcomes from the observables $\{ \hat A_1^j, \hat A_2^j, \cdots, \hat A_m^j \}$. Suppose that $\bm {\bar \mu}_m \rightarrow \bm \mu$ converges asymptotically and satisfies quantum Fisher efficiency \eqref{eq:cov-strong-convexity-bound}. Then, there exists a sufficiently small neighborhood about $\bm \mu$ such that

\label{prop:deviation}
\begin{align}
\label{eq:deviation}
     \lim_{\eps \rightarrow 0}\lim_{m \rightarrow \infty}-\frac{2}{m \eps^2} \log \Pr[\lvert \bar \mu_m^j - \mu_j \rvert > \eps] &= (\tr[\hat \sigma_\beta (\partial_{\eta_j} \log \hat \sigma_\beta)^2])^{-1}
.\end{align}

\begin{proof}
  Because the efficient estimator \eqref{eq:bkm-mle} uniquely \cite{hasegawa1997exponential} satisfies \eqref{eq:cov-strong-convexity-bound}, it suffices to check that this estimator $\hat A^j = \mu_j \mathbbm{1} + (\partial_{\eta_j} \log \rho_{\eta})$ satisfies \eqref{eq:deviation}, which follows directly from \cite[Proof of Lemma 10]{hayashi2002two}.
 
\end{proof}
\end{lemma}

\begin{theorem}
\label{thm:sample-efficient}

With probability ${1-\delta}$, the algorithm \eqref{eq:online-gradient-estimator} learns the target Hamiltonian to error $\lVert \bar {\bm\mu}_m - \bm \mu \rVert_{\infty} \leq \eps$ with $\eps \rightarrow 0$ using
\begin{align}
    m &= \mathcal{O}\left(\frac{e^{ \mathcal{O}(\beta^c)}p^3}{\beta^{c'} \eps^2} \log(\frac{p}{\delta})\right)
\end{align}
data samples, asymptotically.

\begin{proof}
First, \Cref{eq:bh-info-mcp,eq:info-bkm} gives that
\begin{align}
    \tr[\hat \sigma_\beta (\partial_{\eta_j} \log \hat \sigma_\beta)^2] &= [\mathcal{I}({\bm \eta})]_{j,j} \times \mathcal{O}\left( p^2 \right)
.\end{align}
Hence, combining \Cref{prop:deviation} and \eqref{eq:cov-strong-convexity-bound} gives that the failure probability $\delta' = \Pr[\lvert \bar \mu_j - \mu_j \rvert > \eps]$ is met after
\begin{align}
    m &= \mathcal{O}\left(\frac{e^{ \mathcal{O}(\beta^c)}p^3}{\beta^{c'} \eps^2} \log(\frac{1}{\delta'})\right)
\end{align}
descent steps asymptotically. Since we want all $p$ estimates to fail with probability less than $\delta$, it suffices to set $\delta' = \delta / p$ and apply the union bound. Finally, each descent step, if implemented as \eqref{eq:online-gradient-estimator}, requires $\mathcal{O}(1)$ samples from the quantum data.
\end{proof}
\end{theorem}

This samples scaling is a factor of $p^3$ greater than the optimal in $p$ \cite{haah2021optimal} and so looser than the one found for the (differing) marginals matching strategy of \cite{anshu2021sample} by a factor of $p$. Our scaling does not involve a restriction on $\beta$. Note, however, that in the high-temperature regime it was shown that \eqref{eq:strongly-convex} can be improved to eliminate the dependence on $p$ \cite{haah2021optimal}.

Note that it is understood that achieving a stronger version of \eqref{eq:deviation} so that the RHS is given by $\mathcal{I}(\bm \mu)$ requires so-called super-efficient estimation \cite{hayashi2002two}. In particular, as a necessary condition, one must be able to measure multiple Gibbs states collectively i.e. leveraging quantum correlations between several states and a single measurement apparatus. We observe that the theoretically optimal samples scaling for the $l_\infty$-norm, for general $\beta$, may then be provable in such a scenario using similar techniques as above.

\section{Mirror descent equivalence}
\label{sec:md-duality}

In this section, we show that performing Natural Gradient Descent using the \gls{bkm} metric in the \textit{exponential family} coordinate representation is equivalent to performing so-called Mirror Descent in the \textit{mixture family} representation. This means that, under a particular Legendre transform, we can translate the proposed second-order method on our primal manifold to a first-order method on the dual manifold. In fact, under the \gls{bkm} metric, the primal and dual manifolds are equivalent \cite{hasegawa1997exponential}. Hence, we can interpret this strategy as a special change of coordinates for which differing Riemannian connections vanish (i.e. they are flat in differing senses). We will then interpret the implication in the context of other model parameterizations in \Cref{sec:duality-qhbm}.

Translating from a second-order to first-order method may be computationally beneficial (less parameter-shifted observables). Furthermore, compared to the method of minimizing over the non-truncated Lagrangian directly by gradient descent \eqref{sec:geometric-regularizers}, this decouples the loss function from the inner loop.

\subsection{Convex duality}
\label{sec:app:duality}

The first part of our description follows a known result that the \gls{bkm} metric is the unique monotone metric for which the mixture and exponential flat affine connections are mutually dual \cite{hasegawa1997exponential,grasselli2001uniqueness} (and so the Levi-Civita connection is the average of the two, tying these connections to the metric tensor \cite{amari2016information}).  We review in \Cref{apdx:mixture-exponential} the mixture coordinate decomposition of $\hat \rho_{\bm \theta \bm \phi}$ in a basis identifiable with $\mathfrak{su}(N)$: $\hat \rho_{\bm \theta \bm \phi} = \frac{1}{N}\mathbbm{1} + \sum_{j=1}^{N^2 - 1} \eta_j \hat X_j, \eta_j \in \mathbb{R}$ where $\{\hat X_j\}$ are traceless, Hermitian matrices and $\eta_j = \frac{1}{2}\tr[\hat X_j \hat \rho_{\bm \theta \bm \phi}]$. Similarly, we will recall the exponential coordinate decomposition of the modular Hamiltonian as $-\hat K_{\bm \theta \bm \phi} = \sum_{j=1}^{N^2 - 1} \varphi_j \hat X_j, \varphi_j \in \mathbb{R}$.  We now work within these two parameterizations and translate our results to arbitrary coordinate re-parameterizations in \Cref{sec:duality-qhbm}.

The (dual) Riemannian metric tensors for the mixture and exponential global coordinate charts can be expressed as the Hessians of dual convex potential functions, which are related by Legendre transform. This fact is guaranteed by the aforementioned dual flatness \cite{shima2007geometry}. The Legendre transform is a frequented device in thermodynamics and other physical study, for example describing the relationship between temperature and entropy as conjugate quantities. In this sense, we can think of $\varphi_i$, above, as inverse generalized temperature coordinates and $\eta_i$, below, as entropy coordinates.

Conversely, dual coordinates are guaranteed to exist when the metric can be given in terms of a scalar function termed the \textit{potential function} (this is in fact a necessary condition). In our consideration, we have this existence when we consider the \gls{bkm} metric and take the log-partition function (aka free energy) to be the potential, $\psi(\varphi) \equiv \log \mathcal{Z}_{\bm \theta}$, by observing that,
\begin{align}
\label{eq:entropy-coords}
\partial_{\varphi_j} \psi(\bm \varphi) &= \frac{1}{\mathcal{Z}_{\bm \theta}} \int_0^1 \tr[e^{-s \hat K_{\bm \Omega}} \hat X_j e^{-(1-s) \hat K_{\bm \Omega}} ]ds = \frac{1}{\mathcal{Z}_{\bm \theta}} \int_0^1 \tr[e^{-\hat K_{\bm \Omega}} \hat X_j  ]ds = \tr[\hat \rho_{\bm \Omega} \hat X_j ] \\
\label{eq:hessian-log-partition}
\partial^2_{\varphi_j\varphi_k} \psi(\bm \varphi) &= \int_0^1 \tr[\hat \rho_{\bm \Omega}^s X_j \hat \rho_{\bm \Omega}^{1-s}X_k ]ds
\end{align}
which matches the result of \eqref{eq:bkm-defining} since $$\int_0^1 \tr[\hat \rho_{\bm \Omega}^s \frac{\partial \log \hat \rho_{\bm \Omega}}{\partial \varphi_j} \hat \rho_{\bm \Omega}^{1-s} \frac{\partial \log \hat \rho_{\bm \Omega}}{\partial \varphi_k} ]ds = \tr[\frac{\partial \log \hat \rho_{\bm \Omega}}{\partial \varphi_j} \frac{\partial \hat \rho_{\bm \Omega}}{\partial \varphi_k}] = \int_0^\infty \tr[(\hat \rho_{\bm \Omega} + s\mathbbm{1})^{-1} \frac{\partial \hat \rho_{\bm \Omega}}{\partial \varphi_j} (\hat \rho_{\bm \Omega} + s\mathbbm{1})^{-1} \frac{\partial \hat \rho_{\bm \Omega}}{\partial \varphi_k} ]ds.$$

\begin{definition}[Legendre transform]
\label{def:legendre}

The Legendre transform of $\psi$ is the function $\Psi$ defined by 
\begin{align}
\label{eq:legendre}
    \Psi(\bm \eta)  &:= \max_{\bm \varphi} \left[\sum_j \eta_j \varphi_j - \psi(\bm \varphi)\right] \\
    \intertext{implying that}
    \label{eq:coord-change-exp-to-mix}
    \eta_j &= \frac{\partial \psi}{\partial \varphi_j}
\end{align}
\end{definition}

Hence, from \eqref{eq:entropy-coords}, $\eta_i = \tr[\hat \rho_{\bm \Omega} X_i]$. Of course, the above also implies that the metric tensor in $\varphi_i$ can be given by $\frac{\partial \eta_j}{\partial \varphi_i} = \frac{\partial \eta_i}{\partial \varphi_j}$.

\subsection{Mirror descent}

\begin{lemma}[Theorem 2 of \cite{petz2007bregman}]
\label{lem:bregman-petz}

Define,
\begin{align}
  \Gamma_{\hat \rho}(\hat{\rho}) &:=  \hat{\rho}\log {\hat{\rho}}
\end{align}
to be the quantum Shannon entropy, $S(\cdot)$, before tracing and with a sign flip. Then, the quantum Bregman divergence in $\tr[\Gamma_{\hat \rho}(\cdot)] = -S(\cdot)$ is equivalent to the quantum relative entropy,
\begin{align}
D_{\tr[\Gamma], \hat \rho}(\hat{\rho}_1 , \hat{\rho}_2) &= \tr[D_{\Gamma, \hat \rho}(\hat{\rho}_1 , \hat{\rho}_2)] = D(\hat{\rho}_1 \Vert \hat \rho_2)
\end{align}
where
\begin{align}
D_{\Gamma, \hat \rho}(\hat{\rho}_1 , \hat\rho_2) &:=  \Gamma_{\hat \rho}(\hat{\rho}_1) - \Gamma_{\hat \rho}(\hat\rho_2) - \lim_{t \rightarrow 0^+} \frac{1}{t} (\Gamma_{\hat \rho}(\hat{\rho}_2 + t (\hat{\rho}_1 - \hat{\rho}_2)) - \Gamma_{\hat \rho}(\hat\rho_2) )
\end{align}
\end{lemma}

Observe that the Bregman divergence in $\Gamma$ is simply the first-order expansion of $\Gamma$ around $\hat\rho_2$ evaluated at $\hat\rho_1$.

\begin{proposition}
\label{lem:bregman}

Using mixture coordinates $\bm \eta$, define,
\begin{align}
\label{eq:dual-potential}
  \Gamma_{\bm \eta}({\bm\eta'}) &:=  \hat{\rho}_{\bm\eta'}\log {\hat{\rho}}_{\bm\eta'}
.\end{align}
Then, the quantum Bregman divergence in $\tr[\Gamma(\cdot)]$ is equivalent to the quantum relative entropy,
\begin{align}
\label{eq:bregman-def}
D_{\tr[\Gamma], {\bm \eta}}({\bm\eta_1} , {\bm\eta_2,}) &= D(\hat{\rho}_{\bm\eta_1} \Vert\hat\rho_{\bm\eta_2})
\end{align}
where
\begin{align}
\label{eq:bregman-def-first-order}
D_{\Gamma, \bm \eta}({\bm\eta_1} , {\bm\eta_2}) &:=  \Gamma_{\bm \eta}({\bm\eta_1}) - \Gamma_{\bm \eta}({\bm\eta_2}) - \langle \nabla_{\bm \eta_2} \Gamma_{\bm \eta}(\bm \eta_2), \bm \eta_1 - \bm \eta_2 \rangle
\end{align}

\begin{proof}

Of course, 
\begin{align}
\langle \nabla_{{\bm \eta_2}} \Gamma_{\bm \eta}\left({\bm \eta_2}\right), {\bm \eta_1} - {\bm \eta_2} \rangle &= \lim_{t \rightarrow 0} \frac{1}{t} \left(\Gamma_{\bm \eta}\left(\bm \eta_2 + t(\bm\eta_2 - \bm\eta_1)\right) - \Gamma\left(\bm \eta_2\right)\right)
\end{align}
For this particular choice of coordinates, $\hat \rho_{\bm \eta}$ is linear in $\bm \eta$. Hence,
\begin{align}
\Gamma_{\bm \eta}(\bm \eta_2 + t(\bm\eta_2 - \bm\eta_1)) &= \hat\rho_{\bm \eta_2 + t(\bm\eta_2 - \bm\eta_1)} \log \hat\rho_{\bm \eta_2 + t(\bm\eta_2 - \bm\eta_1)} \\
&= \left(\hat \rho_{\bm \eta_2} + t(\hat \rho_{\bm\eta_2} - \hat \rho_{\bm\eta_1}) \right) \log \left(\hat \rho_{\bm \eta_2} + t(\hat \rho_{\bm\eta_2} - \hat \rho_{\bm\eta_1}) \right) \\
&= \Gamma_{\hat \rho} \left(\hat \rho_{\bm \eta_2} + t(\hat \rho_{\bm\eta_2} - \hat \rho_{\bm\eta_1}) \right)
\end{align}
Therefore, we may directly apply \Cref{lem:bregman-petz}.
\end{proof}
\end{proposition}

\begin{lemma}[Dual potential]
\label{lem:dual-potential}

The definition $\Psi := \tr[\Gamma]$ satisfies the Legendre transform dual to \eqref{eq:legendre} and so,
\begin{align}
\psi(\bm \varphi)  &:= \max_{\bm \varphi} \left[\sum_j \eta_j \varphi_j - \Psi(\bm \eta) \right]
.\end{align}
with that
\begin{align}
\frac{\partial \Psi}{\partial \eta_j} &= \varphi_j
.\end{align}

\begin{proof}
We check that
\begin{align}
\partial_{\eta_j} \Psi(\rho_{\bm\eta}) &=  \partial_{\eta_j} \tr[{\rho}_{\bm\eta}\log {\rho}_{\bm\eta}] \\
&= \tr[X_j \log \hat \rho_{\bm\eta}] + \int_0^\infty \tr[\rho_{\bm \eta}(\hat \rho_{\bm \eta} + s \mathbbm{1})^{-1} X_j (\hat \rho_{\bm \eta} + s \mathbbm{1})^{-1} ] ds \\
&= \varphi_j + 0
,\end{align}
where the second term is 0 as in \eqref{eq:first-order-relentropy-vanishes}.
\end{proof}
\end{lemma}

Hence, the metric tensor in $\eta$ can be given by $\tfrac{\partial^2 \Psi}{\partial \eta_j \partial \eta_k}$.

\begin{theorem}[Dual relationship between mirror descent and metric-aware descent, non-commutative case]
\label{thm:dual-md=ngd}

The mirror descent update rule in mixture coordinates $\bm \eta$,
\begin{align}
\label{apdx:eq:duality-md-update-rule}
    \bm\eta_{j+1} &= \arg\min_{\bm\eta} \left[ \langle \bm \eta,  \nabla_{\bm \eta_j} \mathcal{L}( {\bm\eta_j }) \rangle + \lambda {\Phi}^{\operatorname{BKM}}  ( {\bm \eta} , {\bm \eta_j} ) \right] \\
        \intertext{is equivalent to the Natural Gradient update rule in exponential coordinates $\bm\varphi$,}
\bm\varphi_{j+1} &= \underset{\bm{\varphi}}{\arg\min}\left[\langle \bm\varphi, \nabla_{\bm \varphi_j} \mathcal{L}( {\bm{{\varphi}}_j }) \rangle + \frac{\lambda}{2} \langle \bm\varphi - \bm\varphi_j , \mathcal{I}^{\operatorname{BKM}} (\bm \varphi_j) (\bm\varphi - \bm\varphi_j) \rangle\right].
\end{align}

\begin{proof}
  
Recall that $\Phi^{\operatorname{BKM}}( {\bm \Omega_1 }, \bm \Omega_2 ) = {D}  (\bm \Omega_1 \Vert \bm \Omega_2 )$. Taking \eqref{apdx:eq:duality-md-update-rule} with \Cref{lem:bregman} and differentiating in mixture coordinates $\bm \eta$ gives,
\begin{align}
0 &=  \partial_{\bm \eta^k_j} \mathcal{L}_\eta( {\bm{\eta}_j }) +  \lambda \partial_{\bm \eta^k} D_{\Psi, \bm \eta}({\bm \eta}, {\bm \eta_j})\\
\intertext{Hence, from \eqref{eq:bregman-def-first-order},}
\label{eq:md-other-repr}
 \partial_{\bm \eta^k}  \Psi_{\bm \eta}(\bm \eta)
 &= \partial_{\bm \eta^k_j} \Psi_{\bm \eta}(\bm \eta_j) - \frac{1}{\lambda} \partial_{\bm \eta^k_j} \mathcal{L}_\eta( {\bm{\eta}_j }) \\
 \intertext{By \Cref{lem:dual-potential} and \Cref{def:legendre},}
 \bm\varphi^k_{j+1} &= \bm\varphi^k_j - \frac{1}{\lambda} \partial_{\bm \eta^k_j} \mathcal{L}_\eta\left( \frac{\partial{\psi_j}}{\partial \bm \varphi} \right) \\
 \intertext{$\nabla_{\bm \varphi} \mathcal{L}_\eta\Big( \frac{\partial{\psi}}{\partial \bm \varphi}\Big) = \frac{\partial^2{\psi}}{\partial \bm \varphi \partial \bm \varphi^\dag} \nabla_{\bm \eta} \mathcal{L}_\eta\Big( \frac{\partial{\psi}}{\partial \bm \varphi}\Big)$ since $\bm \eta = \frac{\partial{\psi}}{\partial \bm \varphi}$ implies that}
 \label{eq:ngd}
  \bm\varphi_{j+1} &= \bm\varphi_j - \frac{1}{\lambda} \left(\tfrac{\partial^2{\psi_j}}{\partial \bm \varphi_j \partial \bm \varphi_j^\dag}\right)^{-1} \nabla_{\bm \varphi_j} \mathcal{L}_{\bm\varphi}\left(\bm \varphi_j \right) 
\end{align}
where the loss reparameterization $\mathcal{L}_{\bm \varphi}(\bm \varphi) := \mathcal{L}_{\bm \eta}(\bm \eta)$.
\end{proof}
\end{theorem}

A duality between mirror and natural gradient descent has been explored in the classical literature \cite{raskutti2015information} and so we have demonstrated a non-commutative analogue which required a specific choice of monotone metric. A parallel argument can be checked to hold if we were to begin with exponential coordinates and derive a natural gradient update rule in mixture coordinates. It is practically important to recall that exponential coordinates are unconstrained over the reals, whereas mixture coordinates are constrained.

Finally, note that in some cases the forward relative entropy may be computationally preferred to the reverse. In this light, it is worth noting that they are equivalent up to second order in the perturbation, $\bm \delta_{j+1}$ (\Cref{prop:forward-reverse-local-equiv}). Hence, we may consider the approximation, $\bm{\Omega}_{j+1} \approx \arg\min_{\bm\Omega} \left[ \langle \bm \Omega,  \nabla_{\bm {\Omega}_j} \mathcal{L}(\hat{\rho}_{\bm {\Omega}_j})\rangle + \lambda {D}  (\hat{\rho}_{\bm { \Omega}_j}  \Vert \hat{\rho}_{\bm\Omega} ) \right]$.

\subsection{Translating to general ansatzes}
\label{sec:duality-qhbm}

The result \Cref{thm:dual-md=ngd} is shown for a special choice of global coordinate charts, and we may seek to interpret the result for some other choice of coordinates. Relatedly, we have indicated in \Cref{sec:metric-aware-derivation} that metric-aware descent \eqref{eq:natural-gradient-update} is invariant under smooth, bijective re-parameterization in the limit of an infinitesimal learning rate, which we will now check explicitly. Assume exponential coordinates $\bm\varphi$, or any other global coordinate chart. Let us write this explicitly for \eqref{eq:discrete-diff-eq}, viewing $\bm \varphi$ dynamically and writing $\bm\delta_j \equiv d\bm \varphi(j)$ since iterations are taken to be infinitesimal $\lambda \rightarrow \infty$,
\begin{align}
\label{eq:rgf}
    \frac{d\bm \varphi(t)}{dt}  &= -\mathcal{I}^{-1}(\bm \varphi(t)) \nabla \mathcal{L}(\bm \varphi(t))
.\end{align}
When we specify the boundary condition $\bm \varphi(0)$, the evolved path has been referred to as the Riemannian Gradient Flow \cite{gunasekar2021mirrorless}, which we will see offers a coordinate-independent perspective into the relationship between metric-aware and mirror descent. Viewing \eqref{eq:rgf} as the primary object, we see that the metric-aware descent update is a forward Euler discretization with the learning rate $\frac{1}{\lambda}$ as the stepsize.

Let us consider what happens to the dynamics of \eqref{eq:rgf} when we perform a change of coordinate chart, for example from exponential coordinates $\bm \varphi$ to projected \gls{qhbm} coordinates\footnote{See \Cref{apdx:sec-suitable-projection}.} $\bm \Omega$. Let $\bm f = \bm\Omega\circ\bm\varphi^{-1}$ be the diffeomorphism that changes coordinates from exponential to another parameterization. Then, let $D \bm f$ designate the Jacobian matrix $[D \bm f]_{jk} = \tfrac{\partial \bm f_j}{\partial \bm \varphi_k}$. Hence, by the tensor transformation law (equivalently, thinking about how the second fundamental form transforms),
\begin{align}
    \mathcal{I}(\bm\Omega) &= (D \bm f^{-1}(\bm\Omega))^\dag \mathcal{I}(\bm f^{-1}(\bm\Omega))D \bm f^{-1}(\bm\Omega) 
,\end{align}
recalling that $(D \bm f)^{-1} = D \bm f^{-1}$. Therefore,
\begin{align}
\mathcal{I}^{-1}(\bm\Omega) \nabla \mathcal{L}(\bm\Omega) &= ((D \bm f^{-1}(\bm \Omega))^\dag \mathcal{I}(\bm f^{-1}(\bm \Omega))D \bm f^{-1}(\bm \Omega))^{-1} (D \bm f^{-1}(\bm \Omega))^\dag \nabla \mathcal{L}(f^{-1}(\bm \Omega)) \\ \label{eq:transformation-law-tangent-space}
&= D \bm f(\bm \varphi)\mathcal{I}^{-1}(\bm \varphi) \nabla \mathcal{L}(\bm \varphi)
\end{align}
which describes same flow. We check this explicitly by verifying that $\bm f(\varphi(t))$ is a solution to \eqref{eq:rgf} whenever $\varphi(t)$ is a solution. Indeed,
\begin{align}
  \frac{d \bm f(\bm\varphi(t))}{dt} &=  D \bm f(\bm\varphi) \frac{d \bm \varphi(t)}{dt}\\
  \label{eq:substitute-flow}
  &= - D \bm f(\bm\varphi) \mathcal{I}^{-1}(\bm\varphi) \nabla \mathcal{L}(\bm\varphi) =- D \bm f(\bm\varphi)( D \bm f^{-1}(\bm\varphi)) \mathcal{I}^{-1}(\bm f(\bm\varphi)) \nabla \mathcal{L}(f(\bm\varphi))\\
  &= - \mathcal{I}^{-1}(f(\bm\varphi)) \nabla \mathcal{L}(\bm f(\bm\varphi))
.\end{align}
Therefore, we have verified that the flow \eqref{eq:rgf} is invariant under re-parameterization, also termed \textit{intrinsic} in the sense that it is intrinsic to the manifold. As another interpretation, recall that \eqref{eq:transformation-law-tangent-space} is precisely the definition of the equivalence relation that defines the tangent space of a manifold when written in coordinates. Combined with \eqref{eq:rgf}, this observation implies that $\bm\varphi'(t) \in T_{\bm\varphi(t)} \mathcal{M}^{(N)}$ for all $t \in \mathbb{R}$, hence meaning that \eqref{eq:rgf} is a well-defined differential equation on $\mathcal{M}^{(N)}$.

We can view mirror descent as an alternative discretization of this flow when working in the \gls{bkm} geometry, similar to as we have commented for metric-aware descent. Explicitly, we have said that metric-aware descent \eqref{eq:natural-gradient-update} is a forward Euler discretization of \eqref{eq:rgf} meaning that it is evidently the linear interpolation of
\begin{align}
  \frac{d\bm \varphi(t)}{dt}  &= -\mathcal{I}^{-1}(\bm \varphi(\floor{t}_{\gamma})) \nabla \mathcal{L}(\bm \varphi(\floor{t}_{\gamma}))  
\end{align}
where $\floor{t}_\gamma := \gamma \floor{t / \gamma}$ and $\gamma := \frac{1}{\lambda}$. Consider if we instead were to only discretize the gradient, working now in mixture coordinates $\eta$,
\begin{align}
\label{eq:md-discretized-flow}
 \frac{d\bm \eta(t)}{dt}  &= -\mathcal{I}^{-1}(\bm \eta(t) ) \nabla \mathcal{L}(\bm \eta(\floor{t}_{\gamma}))     
\end{align}
We will check that this identifies with the mirror descent update \eqref{apdx:eq:duality-md-update-rule} as considered in \cite{gunasekar2021mirrorless}. So, consider the mirror descent update \eqref{eq:md-other-repr},
\begin{align}
\nabla  \Psi_{\bm \eta}(\bm \eta_{j+1})
 &= \nabla \Psi_{\bm \eta}(\bm \eta_j) - \gamma \nabla \mathcal{L}_\eta( {\bm{\eta}_j })
.\end{align}
From this relation, we can construct a path $\bm z(t)$ by linear interpolation, $\forall t \in [j \gamma, (j+1) \gamma)$,
\begin{align}
\nabla \Psi_{\bm \eta}(\bm z(t)) &= \nabla \Psi_{\bm \eta}(\bm \eta_j) - (t - \gamma j) \nabla \mathcal{L}_\eta( {\bm{\eta}_j })
\end{align}
so that $\bm z(j \gamma) \equiv \bm \eta_j$ at the interpolation points. Hence, we see that $\bm z$ is smooth between these points, obeying $\tfrac{d \nabla \Psi_{\bm \eta}(\bm z(t))}{dt} =  -\nabla \mathcal{L}_\eta( z(\floor{t}_{\gamma}) )$. In such a case, the chain rule $\tfrac{d \nabla \Psi_{\bm \eta}(d\bm z)}{dt} = \nabla^2 \Psi(\bm z) \tfrac{d\bm z}{dt}$ implies that
\begin{align}
\frac{d \bm z}{dt} &= - (\nabla^2 \Psi(\bm z))^{-1} \nabla \mathcal{L}(\bm z(\floor{t}_{\gamma})) \\
&= -\mathcal{I}^{-1}(\bm z(t) ) \nabla \mathcal{L}(\bm z(\floor{t}_{\gamma}))     
\end{align}
which matches \eqref{eq:md-discretized-flow}. Since we have only discretized the gradient, mirror descent \eqref{eq:md-relation} can be considered a "more accurate" discretization of \eqref{eq:rgf}, being more faithful to the geometry of the search space. This has been commented classically \cite{gunasekar2021mirrorless}.

Now, let us consider a solution $\bm \eta(t)$ to \eqref{eq:md-discretized-flow} and transform coordinates by smooth, bijective $f: \bm \eta \mapsto \bm \Omega$. As in \eqref{eq:substitute-flow},
\begin{align}
\frac{d\bm f(\eta)}{dt}  &= -D \bm f(\bm\eta) \mathcal{I}^{-1}(\bm \eta ) \nabla \mathcal{L}(\bm \eta(\floor{t}_{\gamma})) \\
\intertext{except now}
&= -\mathcal{I}^{-1}(\bm f(\bm \eta) ) ((D \bm f(\bm\eta(\floor{t}_{\gamma})))^{-1} D \bm f(\bm\eta) )^\dag \nabla \mathcal{L}(\bm \eta(\floor{t}_{\gamma})) 
.\end{align}
and so the discretized flow is not invariant under re-parameterization i.e. it is \textit{extrinsic} in general. This is true except when $(D \bm f(\bm\eta(\floor{t}_{\gamma})))^{-1} D \bm f(\bm\eta) \equiv \mathbbm{1}$ which would mean that $D \bm f(\bm\eta)$ is constant meaning that the coordinates are affinely related. A similar argument holds for the metric-aware descent discretization.

So, for arbitrary coordinate charts of $\mathcal{M}^{(N)}$ in the \gls{bkm} geometry, metric-aware descent \eqref{eq:natural-gradient-update} and mirror descent \eqref{eq:md-relation} are generally distinct and \textbf{extrinsic} discretizations of an \textbf{intrinsic} flow; however, they are equivalent discretizations for special parameterizations related by Legendre transform as in \Cref{thm:dual-md=ngd}.

\section{Quantum exponential and mixture family ansatzes}

\subsection{Mixture and exponential coordinates}
\label{apdx:mixture-exponential}

In the literature \cite{hasegawa1997exponential}, mixture or Bloch coordinates $\{\tau_j\}_{j=1}^{N^2 - 1}$ may refer to the decomposition of $\hat{\rho}_{\bm \Omega}$ in a basis identifiable with $\mathfrak{su}(N)$. In particular,
\begin{align}
   \hat{\rho}_{\bm \tau} = \frac{1}{N}\mathbbm{1} + \sum_{j=1}^{N^2 - 1} \tau_j \hat{\sigma}_j, \qquad \tau_j \in \mathbb{R} 
\end{align}
where $\hat{\sigma}_j$ are traceless, Hermitian matrices and $\tau_j = \frac{1}{2}\tr[\hat\sigma_j \hat{\rho}_{\bm \Omega}]$. The positivity of density matrices implies that the Bloch coordinates must be constrained; e.g., when $N=2$, $\sqrt{\tau_1^2 + \tau_2^2 + \tau_3^3} \leq 1$.

Similarly, exponential coordinates $\{ \mu_j\}_{j=1}^{N^2 - 1}$ may elsewhere refer to the $\mathfrak{su}(N)$-identifiable decomposition of the modular Hamiltonian as 
\begin{align}
\hat{\rho}_{\bm \mu} = \exp(-\hat{K}_{\bm \mu})/\mathcal{Z}_{\bm \mu}, \qquad\hat{K}_{\bm \mu} = \sum_{j=1}^{N^2 - 1} \mu_j \hat{\sigma}_j, \qquad \mathcal{Z}_{\bm \mu} = \tr[\exp(-\hat{K}_{\bm \mu})]
\end{align}
with $\mu_j \in \mathbb{R}$ unconstrained.

\subsection{Online loss gradients for exponential family}
\label{sec:online-loss-gradient}

We check here that the gradients of the loss for the mixed state ansatz \cite{amin2018quantum} (quantum exponential family, as above) which we used in \Cref{sec:many-body}
\begin{align}
    \hat \rho_{\bm \mu} = \exp(- \sum^p_{l} \mu_l \hat E_l) / Z_{\bm \mu}
,\end{align}
where $E_l$ are known non-identity Pauli operators, can be estimated using a single data sample (for learning) or Gibbs state preparation (for simulation). The motivation is so as to implement online natural gradient descent \eqref{apdx:eq:qcrb-gradient-update}. A similar calculation is feasible for \glspl{qhbm} (\Cref{sec:review-qhbm}).

For the reverse relative entropy loss \eqref{eq:RelativeEntropyQMHL},
\begin{align}
    \partial_{\mu_j} D(\hat \sigma_\beta \Vert \hat \rho_{\bm \mu}) &= -\tr[\hat \sigma_\beta (\partial_{\mu_j} \log \hat \rho_{\bm \mu})] \\
    &= \mathbb{E}_{p_{\hat \sigma_\beta}(\bm x)}[ \bra{\bm x} \hat E_j \ket{\bm x}] + \partial_{\mu_j} Z_{\bm \mu}
,\end{align}
where $p_{\hat \sigma_\beta}(\bm x)$ is the eigenvalue distribution over the eigenstates $\ket{\bm x}$ of $\hat \sigma_{\beta}$. Hence, a single sample from the \textit{data distribution} $p_{\hat \sigma_\beta}(\bm x)$ is sufficient to have an unbiased estimator of the above gradient which is the resource we count in determining quantum Fisher efficiency. 
Nevertheless, in general, many samples from the model distribution are required to estimate the quantum log-partition function gradient \cite{anschuetz2019realizing}, which involves complications due to the non-commutativity between the thermal state and its gradient \cite{amin2018quantum}. A key motivation for the \gls{qhbm} ansatz is to translate this sampling problem (via \eqref{eq:rgf-main}) -- while NP-hard to compute exactly -- to be classical by utilizing the spectral representation of density operators. In this way, the classical log-partition function gradient can be computed offline from the quantum computer, and diagonalization eliminates quantum-specific complications like the sign problem. Quantum algorithms which diagonalize density operators (thus splitting the quantum sampling problem into a classical sampling problem followed by a unitary transformation) are expected to be efficient in broad cases where classical sampling from the quantum thermal distribution is not~\cite{alhambra2022quantum} (for example, under the \gls{eth} assumption~\cite{chen2021fast}).

Similarly, for the forward relative entropy loss \eqref{eq:RelativeEntropyVQT} where we seek to simulate $\hat\sigma_\beta$ given its modular Hamiltonian,
\begin{align}
\partial_{\mu_j} D(\hat \rho_{\bm \mu} \Vert \hat \sigma_\beta) &= -\partial_{\mu_j} S(\hat \rho_{\bm \mu}) + \beta \tr[(\partial_{\mu_j} \hat \rho_{\bm \mu}) \hat K] \\
&= -\partial_{\mu_j} S(\hat \rho_{\bm \mu}) - \tr[(\partial_{\mu_j} \hat \rho_{\bm \mu}) \log \hat \sigma_\beta]
\intertext{since $\partial_{\mu_j} \hat \rho_{\bm \mu}$ is traceless and so $\log Z_{\beta } \tr[(\partial_{\mu_j} \hat \rho_{\bm \mu})] =0$,}
&= \mathbb{E}_{\log p_{\hat \sigma_\beta}(\bm x)} [ -\partial_{\mu_j} S(\hat \rho_{\bm \mu}) - \tr[\bra{\bm x}(\partial_{\mu_j} \hat \rho_{\bm \mu}) \ket{\bm x}] ]
,\end{align}
where $p_{\hat \sigma_\beta}$ is the data distribution as before. Hence, for the problem of simulation, the resource counted by Fisher efficiency is only the number of measurements of the known Hamiltonian against the variational Gibbs state.

\section{\texorpdfstring{Review of \acrfullpl*{qhbm}}{Review of Quantum Hamiltonian-Based Models}}
\label{sec:review-qhbm}

We refer the reader to Verdon et. al \cite{verdon2019qhbm} for a comprehensive description of \glspl{qhbm}. However, for completeness, we review the essential details to our work here. Now, generally, \Glspl{qhbm} are variational circuits intended to learn descriptions of mixed quantum systems. We can view these circuits under two equivalent representations, as subsequently defined.

\begin{definition}[``Mixture representation"]
\label{def:mixture-coords}

One may parameterize an arbitrary density operator $\hat\rho_{\bm \theta \bm\phi} \in \mathcal{M}^{(N)}$ as,
\begin{align}
\label{eq:ParameterizedDM}
    \hat\rho_{\bm \theta \bm\phi} & = \sum_x p_{\bm \theta}(x) \hat{U}_{\bm\phi}\ket{x}\bra{x} \hat{U}_{\bm\phi}^\dagger,
\end{align}
in terms of classical parameters $\bm\Omega := (\bm \theta, \bm \phi) \in \mathbb{R}^d$ which specify the classical probability distribution $p_{\bm \theta} : \{0, 1\}^n \rightarrow \mathbb{R}^+$ with $\sum_x p_{\bm \theta}(x) = 1$ and the unitary \gls{qnn} \cite{broughton2020tensorflow} $U_{\bm \phi}$.
\end{definition}

\begin{definition}[``Exponential representation"] \label{def:exponential-coords}

Equivalently, one may parameterize an arbitrary density operator as
\begin{align}
\label{eq:exponential-qhbm}
    \hat\rho_{\bm \theta \bm\phi} & = \frac{e^{-\hat{K}_{\theta\phi}}}{\mathcal{Z}_{\bm \theta}},                              \\
    \intertext{with the so-called \textit{modular Hamiltonian},}
    \label{eq:modular-hamiltonian}
    \hat{K}_{\bm \theta \bm \phi}       & = \hat{U}_{\bm \phi}\left(\sum_j E_{\bm \theta}(x) \ket{x}\bra{x}\right) \hat{U}^\dag_{\bm \phi}, \\
    \intertext{and its \textit{partition function},}
    \mathcal{Z}_{\bm \theta}            & = \tr[\exp(-\hat{K}_{\bm \theta \bm \phi})] = \sum_{x} \exp(-E_{\bm\theta}(x)),
\end{align}
where $E_{\bm \theta} : \{0, 1\}^n \rightarrow \mathbb{R}$ is referred to as the real-valued ``energy'' function.
\end{definition}

Note that our notions of exponential and mixture representations differ from similar terminology in the quantum thermodynamics literature \cite{hasegawa1997exponential}, though we adopt this naming due to a straightforward identification of the respective representations. In particular, referring to \Cref{apdx:mixture-exponential}, for mixture coordinates there exists a bijection between $(p(\bm \theta), \tilde {\bm\phi})$ and $\{\tau_j\}_j$ (with $\tilde{\bm \phi}$ describing the projected coordinates of \eqref{eq:tilde-omega}). And, notably, the important structure of $p(\bm \theta)$ being constrained over the reals is shared. For exponential coordinates, there is a bijection between $(E(\bm \theta), \tilde {\bm\phi})$ and $\{\mu_j\}_j$. Likewise, $E(\bm \theta)$ is unconstrained over the reals.

Evidently, $E_{\bm\theta}$ and $p_{\bm\theta}$ are related in this case by,
\begin{align}
  \label{eq:EBM}
  p_{\bm \theta}(x) = \frac{\exp({-E_{\bm\theta}(x)})}{\sum_y \exp({-E_{\bm\theta}(y)})},
\end{align}
and, in the classical statistical learning literature, a statistical model specified in $E_{\bm \theta}$ coordinates is termed an \gls{ebm} \cite{hinton2002training,dayan1995helmholtz} and so the apt naming of \acrfullpl{qhbm}. The \gls{ebm} serves as the main vehicle to add classical correlations in the \gls{qhbm} representation, while the \gls{qnn} serves to add multipartite quantum entanglement, the latter of which cannot be captured at scale by classical models \cite{nielsen2010quantum}, or sampled efficiently from using classical computers in general \cite{arute2019quantum}.

Optimizing graphical models in terms of $E_{\bm \theta}$ directly (which may, for example, be a neural network) is a popular choice because one effectively relaxes the constrained optimization problem in $p_{\bm \theta}$ to be unconstrained. One must however cope with the fact that estimating $\mathcal{Z}$ (required e.g. for gradient updates in $\bm \theta$) may be computationally intractable when $N$ is sufficiently large. So, \gls{mcmc} approximation algorithms are customarily classically invoked \cite{carlo2004markov, betancourt2017conceptual, welling2011bayesian, grathwohl2021oops}. One may proceed similarly in the quantum case (see the discussion in \cite{verdon2019qhbm}); preferred \gls{mcmc} algorithms may leverage the discreteness or possible differentiability of $E_{\bm \theta}$ for improved sample efficiency.

From the spectral decomposition theorem of Hermitian operators, for a sufficiently expressive energy function and \gls{qnn}, this hypothesis class can cover the whole space of density matrices, as the \gls{qnn} acts as a diagonalizing unitary. In our applications, we aim to have parameterizations for both the \gls{ebm} and the \gls{qnn} which possess an inductive bias for better trainability through metric-aware optimization, which will yield improved optimization behavior.

\subsection{Quantum Relative Entropy}

The \gls{qhbm} training loss function motivated in \cite{verdon2019qhbm} is the quantum relative entropy defined as \cite{wilde2017quantum}
\begin{align}
  \label{eq:RelativeEntropy}
  D(\hat\rho \Vert \hat\sigma) & = \tr[\hat{\rho}(\log \hat\rho - \log \hat\sigma)]     \\
                                                             & = -S(\hat\rho) - \tr[\hat\rho \log \hat\sigma] ,
\end{align}
in terms of the von Neumann entropy $S(\cdot)$ and with $\hat\rho, \hat\sigma \in \mathcal{M}^{(N)}$. Evidently, $D(\hat\rho \Vert \hat\sigma) \geq 0$ with equality if and only if $\hat\rho = \hat\sigma$.

The quantum relative entropy is a non-commutative generalization \cite{umegaki1962conditional} of the Kullback-Leibler divergence \cite{kullback1951information} which is a conventional loss function in classical probabilistic machine learning \cite{goodfellow2016deep}. Observe that the quantum relative entropy is asymmetric in its arguments as in the classical case. The two applications briefly described next highlight this asymmetry.

\subsubsection{Applications}
\label{sec:qhbm-applications}

One may learn an unknown mixed state $\hat \sigma$ in the \gls{qhbm} parameterization through gradient-based optimization in either the forward or reverse quantum relative entropies: $D({\hat \sigma} \Vert \hat{\rho}_{\bm \Omega}) $ and $D( \hat{\rho}_{\bm \Omega} \Vert {\hat \sigma}) $, respectively. As the true modular Hamiltonian of the data distribution is unknown, we usually minimize alternate but equivalent losses, being that of quantum cross-entropy and quantum free energy, respectively. The algorithms to do so via gradient-based optimizers or otherwise are called \gls{qmhl} and \gls{vqt}, respectively \cite{verdon2019qhbm}.

The preferred choice in terms of tractability depends on what is known \textit{a priori}, for example in simulation as compared to characterization scenarios. Gradients can be computed through special stochastic averages of parameter-shift rules \cite{mitarai2018quantum}. For notational convenience in some instances we might denote the quantum relative entropy between two \glspl{qhbm} as 
\(
D({\bm \Omega_1} \Vert {\bm \Omega_2}) := D(\hat{\rho}_{\bm \Omega_1} \Vert \hat\rho_{\bm \Omega_2}).
\)

\paragraph{\acrfull{vqt}}
\label{par:vqt}

Suppose the provided input is a Hamiltonian ${\hat H}$ and an inverse temperature $\beta$.  An important task using this information is to simulate the associated thermal state,
\begin{align}
  {\hat \sigma}_{\beta} = \frac{e^{-\beta\hat{H}}}{\mathcal{Z}_{\beta}}, \quad \text{where} \quad \mathcal{Z}_{\beta} = \tr\left[\exp(-\beta {\hat H})\right],
\end{align}
a task also known as \textit{quantum Gibbs sampling}.
Gradient-based optimization of this task is feasible using \glspl{qhbm}.  See the discussion at \cref{sec:numerics} for more on quantum Gibbs sampling.

Summarizing a result of \cite{verdon2019qhbm}, using \Cref{eq:RelativeEntropy} and the \gls{qhbm} ansatz of \Cref{eq:exponential-qhbm}, we may minimize the forward quantum relative entropy and so implicitly define the \gls{vqt} loss.

\begin{definition}[\gls{vqt} loss]

  The \gls{vqt} loss is given by,

  \begin{align}
    \label{eq:RelativeEntropyVQT}
    \min_{\bm \theta \bm \phi} D(\hat\rho_{\bm\theta\bm\phi} \Vert {\hat \sigma}_\beta) &= \min_{\bm \theta \bm \phi} \{ -S(\hat\rho_{\bm\theta}) + \beta\tr\left[\hat\rho_{\bm\theta\bm\phi}{\hat H}\right] + \ln \mathcal{Z}_\beta \}           \\
 & =  \min_{\bm \theta \bm \phi} \{-S(\hat\rho_{\bm\theta}) + \beta\tr\left[\hat\rho_{\bm\theta\bm\phi}{\hat H}\right] \} \\
& := \min_{\bm \theta \bm \phi} \mathcal{L}_{\operatorname{VQT}}(\theta, \phi)
.\end{align}
\end{definition}

In the following we assume the \gls{ebm} parameters $\bm{\theta}$ and \gls{qnn} parameters $\bm{\phi}$ are distinct. Then the derivative of the \gls{vqt} loss with respect to the classical model parameters is
\begin{align}
\label{eq:vqt-grad-ebm}
\nabla_{\bm{\theta}}\mathcal{L}_{\operatorname{VQT}}(\bm{\theta},\bm{\phi})  &= \begin{aligned}[t]&\mathbb{E}_{\bm{x}\sim p_{\bm{\theta}}(\bm{x})}\big[\beta H_{\bm{\phi}}(\bm{x}) -E_{\bm{\theta}}(\bm{x}) \big] \times \mathbb{E}_{\bm{y}\sim p_{\bm{\theta}}(\bm{y})}\big[\nabla_{\bm{\theta}}E_{\bm{\theta}}(\bm{y})\big] \\& - \mathbb{E}_{\bm{x}\sim p_{\bm{\theta}}(\bm{x})}\Big[(\beta H_{\bm{\phi}}(\bm{x}) -E_{\bm{\theta}}(\bm{x})) \nabla_{\bm{\theta}}E_{\bm{\theta}}(\bm{x}) \Big],\end{aligned}
\end{align}
where we defined the push-forwards Hamiltonian $\hat{H}_{\bm{\phi}} \equiv \hat{U}^\dagger({\bm{\phi}})\hat{H}\hat{U}({\bm{\phi}})$ and the push-forwards Hamiltonian expectation per basis state $H_{\bm{\phi}}(\bm{x}) \equiv \bra{\bm{x}}\hat{U}^\dagger({\bm{\phi}})\hat{H}\hat{U}({\bm{\phi}})\ket{\bm{x}}$. 

Similarly, the gradient of the \gls{vqt} loss with respect to the quantum model parameters is
\begin{align}
\label{eq:vqt-grad-qnn}
\nabla_{\bm{\phi}}\mathcal{L}_{\operatorname{VQT}}(\bm{\theta},\bm{\phi}) =\beta \nabla_{\bm{\phi}}\tr[\hat{U}({\bm{\phi}})\hat{\rho}_{\bm{\theta}}\hat{U}^\dagger({\bm{\phi}})\hat{H} ]
.\end{align}

\paragraph{\acrfull{qmhl}}

Dual to the task of \gls{vqt}, suppose we are given query access to prepared thermal state ${\hat \sigma}_D$ and seek to fit a parameterized modular Hamiltonian $\hat{K}_{\bm \theta \bm \phi}$ (as in \eqref{eq:modular-hamiltonian}) such that the corresponding model state $\hat\rho_{\bm{\theta\phi}}$ is as close as possible to ${\hat \sigma}_D$.  This is a modified version of the Hamiltonian learning task defined in section \eqref{sec:on-optimality}.  To derive the \gls{qmhl} loss, consider taking the relative entropy in the reverse order as compared to \gls{vqt}:

\begin{definition}[\gls{qmhl} loss]

The \gls{qmhl} loss is given by,
\begin{align}
    \label{eq:RelativeEntropyQMHL}
  \min_{\bm \theta \bm \phi} D({\hat \sigma}_D\Vert\hat{\rho}_{\theta\phi}) &= \min_{\bm \theta \bm \phi}\{-S({\hat \sigma}_D) + \tr\left[{\hat \sigma}_D \hat{K}_{\theta\phi}\right] + \ln \mathcal{Z}\} \\
  \label{eq:LossFunctionQMHL}
    &\equiv \min_{\bm \theta \bm \phi} \{D({\hat \sigma}_D \Vert \hat\rho_{\theta\phi}) + S({\hat \sigma}_D)\} \\
    &:= \min_{\bm \theta \bm \phi} \mathcal{L}_{\operatorname{QMHL}}(\theta, \phi).
\end{align}

\end{definition}

The gradient of the \gls{qmhl} loss with respect to the classical model parameters is
\begin{align}
    \label{eq:qmhl-grad-qnn}
\nabla_{\bm{\theta}}\mathcal{L}_{\operatorname{QMHL}}(\bm{\theta,\phi})=\mathbb{E}_{\bm{x}\sim \sigma_{\bm{\phi}}(\bm{x})}[\nabla_{\bm{\theta}} E_{\bm{\theta}}(\bm{x})] - \mathbb{E}_{\bm{y}\sim p_{\theta}(\bm{y})} 
[\nabla_{\bm{\theta}}E_{\bm{\theta}}(\bm{y})]
,\end{align}
where $\hat{\sigma}_{\bm{\phi}}(\bm{x}) \equiv\bra{\bm{x}}\hat{U}^\dagger(\bm{\phi})\hat{\sigma}_{\mathcal{D}}\hat{U}(\bm{\phi})\ket{\bm{x}}$. Using our notation for the \textit{pulled-back data state} 
\(\hat{\sigma}_{\mathcal{D},\bm{\phi}} \equiv \hat{U}^\dagger(\bm{\phi})\hat{\sigma}_{\mathcal{D}}\hat{U}(\bm{\phi})\),
the gradient with respect to unitary \gls{qnn} parameters $\bm{\phi}$ will be given by 
\begin{align}
    \label{eq:qmhl-grad-ebm}
    \nabla_{\bm\phi}\mathcal{L}_{\operatorname{QMHL}}(\bm{\theta,\phi}) = \nabla_{\bm\phi}\langle\hat{K}_{\bm{\theta}}\rangle_{\hat{\sigma}_{\mathcal{D},\bm{\phi}}}.
\end{align}

\subsubsection{Additional properties}
\label{sec:quantum-relative-entropy}

While the quantum relative entropy is asymmetric (as is the classical KL-divergence), it is nearly symmetric for nearby states, as we show in the following result. \Cref{prop:forward-reverse-local-equiv} is standard and well-known \cite{lashkari2016canonical}, but we give our own explicit proof, as it will be useful in other parts of this work.
\begin{proposition}
\label{prop:forward-reverse-local-equiv}
The relative entropy is symmetric up to third order. That is, for any (sufficiently smooth) parametrized density operator $\hat{\rho} : \mathbb{R} \to \mathcal{M}^{(N)}$, we have
\begin{align}
D(\hat{\rho}(\lambda) \|\hat{\rho}(0)) = D(\hat{\rho}(0)\|\hat{\rho}(\lambda)) + \mathcal{O}(\lambda^3).
\end{align}

\begin{proof}
We know that:
\begin{align*}
D(\hat{\rho}(\lambda)\|\hat{\rho}) = \lambda \frac{d}{d\lambda} D(\hat{\rho}(\lambda)\|\hat{\rho}(0))\bigg|_{\lambda=0} + \lambda^2 \frac{d^2}{d\lambda^2} D(\hat{\rho}(\lambda)\| \hat{\rho}(0))\bigg|_{\lambda=0} + \mathcal{O}(\lambda^3),
\end{align*}
where we used that $D(\hat{\rho}(0)\|\hat{\rho}(0)) = 0$. Now, a simple computation involving the definition of the relative entropy (\cref{eq:def-relative-entropy}) gives:
\begin{equation}
\label{eq:first-order-relentropy-vanishes}
\frac{d}{d\lambda} D(\hat{\rho}(\lambda)\|\hat{\rho}(0)) \bigg|_{\lambda=0} = \tr (\hat{\rho}(\lambda) \frac{d}{d\lambda} \log\hat{\rho}(\lambda))\bigg|_{\lambda =0} = -\frac{d}{d\lambda} D(\hat{\rho}(0)\|\hat{\rho}(\lambda)) \bigg|_{\lambda=0}
\end{equation}
Let $\hat{A}(\lambda) = \log\hat{\rho}(\lambda)$, so that $\hat{\rho}(\lambda) = \exp\hat{A}(\lambda)$. Then, we see from the infinitesimal form of the Baker-Campbell-Hausdorff formula -- and some algebraic manipulation -- that
\begin{align*}
\frac{d}{d\lambda} \hat{A}(\lambda) &= \hat{\rho}(\lambda)^{-1}\hat{\rho}'(\lambda) + \frac{1}{2!} [\hat{A}(\lambda), \hat{A}'(\lambda)] - \frac{1}{3!}[\hat{A}(\lambda), [\hat{A}(\lambda), \hat{A}'(\lambda)]] + \frac{1}{4!}[\hat{A}(\lambda), [\hat{A}(\lambda), [\hat{A}(\lambda), \hat{A}'(\lambda)]]] - \cdots \\
&= \hat{\rho}^{-1}\hat{\rho}' + \sum_{n=1}^{\infty} \frac{(-1)^{n+1}}{(n+1)!} \sum_{i=0}^n (-1)^i {\binom{n}{i}} \hat{A}^{n-i}\hat{A}' \hat{A}^i,
\end{align*}
where we've suppressed explicit $\lambda$ dependence in the last line for brevity. Substituting the above into \eqref{eq:first-order-relentropy-vanishes} and using the linearity of trace, we find
\begin{align*}
\tr(\hat{\rho}(\lambda)\frac{d}{d\lambda}\log\hat{\rho}(\lambda)) = \tr(\hat{\rho}(\lambda) \hat{A}'(\lambda)) &= \tr(\hat{\rho}'(\lambda)) + \sum_{n=1}^{\infty} \frac{(-1)^{n+1}}{(n+1)!} \sum_{i=0}^n (-1)^i {\binom{n}{i}} \tr(\hat{\rho}\hat{A}^{n-i} \hat{A}' \hat{A}^i) \\
&= \frac{d}{d\lambda}\tr(\hat{\rho}(\lambda)) + \sum_{n=1}^{\infty} \frac{(-1)^{n+1}}{(n+1)!} \sum_{i=0}^n (-1)^i {\binom{n}{i}} \tr(\hat{A}^{n-i} \hat{\rho} \hat{A}'\hat{A}^i) \\
&= \frac{d}{d\lambda}(1) + \sum_{n=1}^{\infty} \frac{(-1)^{n+1}}{(n+1)!} \sum_{i=0}^n (-1)^i {\binom{n}{i}} \tr(\hat{\rho} \hat{A}') \\
&= \tr(\hat{\rho} \hat{A}') \sum_{n=1}^{\infty} \frac{(-1)^{n+1}}{(n+1)!} \sum_{i=0}^n (-1)^i {\binom{n}{i}} = 0,
\end{align*}
where we used that $\tr(\hat{\rho}) = 1$, that $[\hat{\rho}, \hat{A}] = 0$, and finally the simple identity, $\sum_{i=0}^n (-1)^i {\binom{n}{i}} = 0$. Hence, combining the above with \eqref{eq:first-order-relentropy-vanishes}, we conclude that
\begin{align}
\label{eq:expected-derivative-is-zero}
\frac{d}{d\lambda} D(\hat{\rho}(\lambda) \|\hat{\rho}(0))\bigg|_{\lambda=0} = \frac{d}{d\lambda} D(\hat{\rho}(0) \|\hat{\rho}(\lambda))\bigg|_{\lambda=0} = 0.
\end{align}

Now, we consider the second order terms,
\begin{align*}
\frac{d^2}{d\lambda^2} D(\hat{\rho}(\lambda) \|\hat{\rho}(0)) \bigg|_{\lambda=0} &= \frac{d}{d\lambda}(\tr(\hat{\rho}'(\lambda)\log\hat{\rho}(\lambda)) + \tr(\hat{\rho}(\lambda) \frac{d}{d\lambda} \log\hat{\rho}(\lambda)) - \tr(\hat{\rho}'(\lambda)\log\hat{\rho}(0)))\big|_{\lambda=0} \\
&=\left[\tr(\hat{\rho}''(\lambda)\log\hat{\rho}(\lambda)) + 2 \tr(\hat{\rho}'(\lambda) \frac{d}{d\lambda}\log\hat{\rho}(\lambda)) + \tr(\hat{\rho}(\lambda) \frac{d^2}{d\lambda^2} \log\hat{\rho}(\lambda)) - \tr(\hat{\rho}''(\lambda)\log\hat{\rho}(0))\right]\bigg|_{\lambda=0} \\
&=\left[  2 \tr(\hat{\rho}'(\lambda) \frac{d}{d\lambda}\log\hat{\rho}(\lambda)) + \tr(\hat{\rho}(\lambda) \frac{d^2}{d\lambda^2} \log\hat{\rho}(\lambda)) \right]\bigg|_{\lambda=0}.
\end{align*}
But now, we observe that
\begin{align*}
\tr(\hat{\rho}'(\lambda) \frac{d}{d\lambda} \log\hat{\rho}(\lambda)) = \frac{d}{d\lambda} \tr(\hat{\rho}(\lambda) \frac{d}{d\lambda} \log\hat{\rho}(\lambda)) - \tr(\hat{\rho}(\lambda) \frac{d^2}{d\lambda^2} \log\hat{\rho}(\lambda)) = -\tr(\hat{\rho}(\lambda) \frac{d^2}{d\lambda^2}\log\hat{\rho}(\lambda)).
\end{align*}
Hence, combining the above calculations, we conclude that
\begin{align*}
\frac{d^2}{d\lambda^2} D(\hat{\rho}(\lambda)\|\hat{\rho}(0))\bigg|_{\lambda=0} = -\tr(\hat{\rho}(\lambda)\frac{d^2}{d\lambda^2}\log\hat{\rho}(\lambda))\bigg|_{\lambda=0}.
\end{align*}
Similarly, we obtain readily that
\begin{align*}
\frac{d^2}{d\lambda^2}D(\hat{\rho}(0) \|\hat{\rho}(\lambda)) \bigg|_{\lambda =0} = -\frac{d}{d\lambda} \tr(\hat{\rho}(0) \frac{d}{d\lambda}\log\hat{\rho}(\lambda)) = -\tr(\hat{\rho}(0) \frac{d^2}{d\lambda^2} \log\hat{\rho}(\lambda))\bigg|_{\lambda = 0},
\end{align*}
which immediately implies that $D(\hat{\rho}(0)\|\hat{\rho}(\lambda))$ and $D(\hat{\rho}(\lambda) \|\hat{\rho}(0))$ agree up to third order terms, i.e.,
\begin{align*}
D(\hat{\rho}(\lambda) \|\hat{\rho}(0)) = D(\hat{\rho}(0) \|\hat{\rho}(\lambda)) + \mathcal{O}(\lambda^3),
\end{align*}
as desired.
\end{proof}
\end{proposition}

\begin{corollary}
\label{cor:mixed-hessian-rear-hessian}

The above line of reasoning immediately yields the following equivalent forms of the \gls{bkm} metric:
\begin{align} \label{eq:equiv-hessians}
[\mathcal{I}^{\operatorname{BKM}}(\bm \Omega_0)]_{jk} &:= -\tr[(\partial_{\Omega_i} \hat{\rho}_{\bm\Omega}) (\partial_{\Omega_j} \log\hat{\rho}_{\bm\Omega})]\bigg|_{\bm\Omega=\bm\Omega_0} = \tr[\hat{\rho}_{\bm\Omega_0} \partial^2_{\Omega_i \Omega_j} \log \hat{\rho}_{\bm\Omega}]\bigg|_{\bm\Omega=\bm\Omega_0} 
\end{align}
where the starting definition \eqref{eq:bkm-simplest} stems from a more general scope of dual metrics \cite{hasagawa1995noncommutative}.

\begin{proof}
Recall from the proof of \Cref{prop:forward-reverse-local-equiv} that for any smooth parameterized density operator $\hat{\rho}(\lambda)$, we have
\begin{align*}
\tr[\hat{\rho}(\lambda) \tfrac{d}{d\lambda} \log \hat{\rho}(\lambda)] = 0.
\end{align*}
Now, for any $k \in \{1, \dots, d\}$, we can fix the other coordinates $\bm\Omega_{\setminus k} \in \mathbb{R}^{d-1}$ and view $\Omega_k \mapsto \hat{\rho}_{\bm\Omega}$ as a 1-dimensional family of density operators, which immediately gives
\begin{align}\label{eq:state-orthogonal-to-derivative-of-log}
\tr(\hat{\rho}_{\bm\Omega} \partial_{\Omega_k} \log\hat{\rho}_{\bm\Omega}) = 0.
\end{align}
Note that \eqref{eq:state-orthogonal-to-derivative-of-log} holds for \textit{any} parameterization $\bm\Omega \mapsto \hat{\rho}_{\bm\Omega}$. Now, differentiating \eqref{eq:state-orthogonal-to-derivative-of-log} with respect to $\Omega_j$ and applying the chain rule yields
\begin{align*}
\tr(\partial_{\Omega_j} \hat{\rho}_{\bm\Omega} \partial_{\Omega_k}\log\hat{\rho}_{\bm\Omega}) + \tr(\hat{\rho}_{\bm\Omega} \partial^2_{\Omega_j, \Omega_k} \log \hat{\rho}_{\bm\Omega}) = 0,
\end{align*}
which is exactly the the equality in \eqref{eq:equiv-hessians}.
\end{proof}
\end{corollary}

\subsection{On the over-parameterization of \texorpdfstring{\acrshortpl{qhbm}}{QHBMs}}
\label{apdx:over-param}

In the \gls{qhbm} (spectral) parameterization (\Cref{def:mixture-coords}), in general, $\bm \theta$ parameterizes an arbitrary categorical distribution (with full support as per \Cref{assum:full-rank}) in $\operatorname{Cat}(N)$ and $\dim \operatorname{Cat}(N) = N-1$. Similarly, in the absence of an inductive bias, $\bm \phi$ parameterizes an arbitrary element of $SU(N)$ and $\dim SU(N) = N^2  - 1$. Hence, there exists a straightforward projection map  $\pi_1 : \mathbb{R}^d \rightarrow \operatorname{Cat}(N) \times SU(N)$ which acts independently on the classical and quantum parameters of $\bm \Omega$.

In such a case, the parameterization map $\varphi : \operatorname{Cat}(N) \times SU(N) \cong \mathbb{R}^{N^2 + N - 2} \rightarrow \mathcal{M}^{(N)}$ which acts as $\varphi : \pi_1(\bm \Omega) \mapsto \hat{\rho}_{\bm\theta \bm\phi} $ cannot be injective given that $\dim \mathcal{M}^{(N)} = N^2 - 1$. So, to construct a manifold chart \cite{amari2016information} of $\mathcal{M}^{(N)}$ we must project out $N-1$ dimensions from the image of $\pi_1$. We describe a mechanism to do so in this section. Such a construction relies on the generalized Euler angle parameterization \cite{tilma2002generalized} and the fact that diagonal density operators commute with diagonal unitaries which leads to a characterization of the invariant $SU(N)$ actions.

With $\pi \equiv \pi_2 \circ \pi_1$ and $\pi : \mathbb{R}^d \rightarrow \operatorname{Cat}(N) \times SU(N) \rightarrow \mathbb{R}^{N^2 - 1}$, we say that,
\begin{align}
\pi: \bm \Omega \mapsto \tilde {\bm \Omega}
\end{align}
We note that $\pi_2$ acts only on the quantum parameters; hence, even with projection, the quantum and classical parameters of a \gls{qhbm} are decoupled from one another.

So, we have counted above that the classical parameters which encode a $\hat\rho_{\bm \theta \bm \phi}$ can be thought of as $\bm \theta\in \mathbb{R}^{N-1}$ and $\bm \phi \in \mathbb{R}^{N^2 - 1}$ after an implicit projection. We also know that the real dimension of $\mathcal{M}^{(N)}$ is $N^2 - 1$. Hence, it appears at first glance that \glspl{qhbm} are over-parameterized. This indicates that $\Omega$ cannot chart the Riemannian manifold $\mathcal{M}^{(N)}$ directly. Hence, we identify a (conveniently, global) projection map $\pi : \bm\Omega \rightarrow \tilde{\bm\Omega}$ so that its image gives appropriate (global) coordinates. Accordingly, we project out a dimensionality of ${N-1}$ through a map which acts precisely on the $\bm \phi$ parameters,
\begin{align}
\label{eq:tilde-omega}
\pi : \bm\Omega \rightarrow \tilde{\bm\Omega} = (\bm \theta, \bm {\tilde\phi})
\end{align}

\subsubsection{Isotropy of \texorpdfstring{$\mathcal{M}^{(N)}$}{density operator space} under action of \texorpdfstring{$U(N)$}{the unitary group}}
\label{apdx:isotropy}

A means to resolve the over-parameterization\footnote{Our discussion follows the one of \cite{bengtsson2017geometry} regarding the ``stratification" of $\mathcal{M}^{(N)}$ into orbits of the unitary group.} is to observe that the diagonal $\rho_{\bm \theta}$ commutes with certain unitary $\hat{V}$ and therefore is invariant under the conjugation $\hat{V}\hat\rho_{\bm \theta} \hat{V}^\dag = \hat\rho_{\bm \theta}$. And note, all such $V$ may be classified as the Cartan subgroup \cite{lang2002graduate} of $U(N)$ acting on $\hat\rho_{\bm \theta}$; the Cartan subgroup is an exponential of a set of simultaneously diagonalizable elements of the algebra $\mathfrak{su}(N)$ and so holds dimension ${N-1}$. We resolve the details for the full-rank, non-degenerate special case next. 

Consider a non-degenerate and full-rank $\hat\rho_{\bm \theta \bm \phi}$ transformed to its diagonal frame with image $\hat\rho_{\bm \theta}$. Evidently, $\rho_{\bm \theta}$ is invariant under precisely the conjugate action of an arbitrary diagonal unitary $V^{\operatorname{diag}}$. In this sense, $\hat\rho_{\bm \theta \bm \phi} = \hat{U}_{\bm \phi}\hat{V}^{\operatorname{diag}} \hat\rho_{\bm \theta}(\hat{V}^{\operatorname{diag}})^\dag U^\dag_{\bm \phi}$ and so $U_{\bm\phi}$ is determined up to the arbitrary phases entering $V^{\operatorname{diag}}$. Hence, we can identify the isotropy group at hand with the $({N-1})$-dimensional set of diagonal, unitary matrices. And in particular, we can identify this group with the $U(1) \times \cdots \times U(1)$ of $N{-}1$ factors (denoted $[U(1)]^{N-1}$ and thought of topologically as ${N-1}$ independent 1-tori).

Note that the isotropy group becomes more interesting when, for example, $\rho_{\bm \theta \bm \phi}$ has degenerate eigenvalues.

\subsubsection{Flag manifolds}

In \Cref{apdx:isotropy}, we have effectively described $\mathcal{M}^{(N)}$ as the \textit{flag manifold} which describes the space of all \textit{flags} by identifying $[U(1)]^{N-1}$ as an isotropy group of $U(N)$ acting on $\mathcal{M}^{(N)}$. In our case, a flag consists of sequences of subspaces $V_1 \subset V_2 \subset \cdots \subset V_{N-1}$ that reside in $U(1) \times U(2) \times \cdots \times U(N-1)$. 

\subsubsection{A suitable projection map}
\label{apdx:sec-suitable-projection}

Consider a generalized Euler angle parameterization \cite{tilma2002generalized} of $SU(N) \ni U_{\bm \phi}$. In light of \Cref{apdx:isotropy}, we can construct such a parameterization so that the isotropy group $[U(1)]^{N-1}$ projects out. We refer the reader to \cite{tilma2002generalized} for details in that our described algorithms ultimately utilize the unprojected \gls{qhbm} parameterization. However, we note and exemplify the essential points relevant whenever a manifold chart is required in our analysis.

In constructing a generalized Euler angle parameterization of $SU(N)$, one uses the so-called Cartan decomposition (a generalization of the singular value decomposition to a semisimple Lie group or Lie algebra) to decompose $\mathfrak{su}(N)$ into the semi-direct sum of two subspaces $\mathfrak{l}, \mathfrak{p}$. It can be shown that $\mathfrak{l}, \mathfrak{p}$ are orthogonal complements of each other with respect to the Killing form $B(\cdot, \cdot)$ on $\mathfrak{su}(N)$ (i.e. $B(X, Y) = 2N \tr[XY]$ and so under the Hilbert-Schmidt inner product). Any subalgebra of $\mathfrak{p}$ is commutative and in particular contains the $N-1$ diagonal elements of $\mathfrak{su}(N)$. In the generalize Euler angle parameterization, such elements act rightmost in the decomposition and so whose corresponding exponentials can be dropped. The following example demonstrates this for $N=2$.

\begin{example}[$N=2$]
Consider that we may write an arbitrary unitary when $N=2$ as,
\begin{align}
\hat{U}_\phi &= \exp(i \hat \sigma_z \alpha) \exp(i \hat \sigma_x \beta) \exp(i \hat \sigma_z \gamma),
\end{align}
and observe that the $\gamma$ term can be dropped since $\hat \sigma_z$ is diagonal and so $\exp(i\hat\sigma_z \gamma)$ always commutes with $\hat\rho_{\bm \theta}$.

\end{example}

\begin{proposition}
\label{apdx:lem:traceless-spanning}
The tangent vectors of $\hat{\rho}_{\bm \theta \bm \phi}$ in the \gls{qhbm} and projected (\Cref{apdx:sec-suitable-projection}) \gls{qhbm} parameterizations span the set of $N \times N$ traceless, Hermitian matrices. The unprojected \gls{qhbm} parameterization tangent basis is over-complete by a dimension of $N - 1$.

\begin{proof}

Let $\bm\Omega \in \mathbb{R}^d$. We can rewrite an arbitrary density operator in the affine (Bloch) parameterization (see \Cref{apdx:mixture-exponential}),
\begin{align}
\hat{\rho}_{\bm \Omega} &= \frac{1}{N} \mathbbm{1} + \sum_{j=1}^{N^2-1} \tau_j(\bm\Omega) \hat{\sigma}_j
\end{align}
where $\hat\sigma_j$ are traceless, Hermitian matrices i.e. Hermitian generators of $\mathfrak{su}(N)$. Hence,
\begin{align}
\label{eq:chain-rule-tangent}
\partial_{\Omega_k} \hat{\rho}_{\bm \Omega} &= \sum_{j=1}^{N^2-1} \frac{\partial \tau_j(\bm\Omega)}{\partial \Omega_k} \hat{\sigma}_j
\end{align}
and $\mathfrak{su}(N)$ is closed by definition of being an algebra\footnote{Another way to see the tracelessness is that $\tr[(\partial_{\Omega_k} \rho_{\bm \Omega})] = \partial_{\Omega_k} \tr[\rho_{\bm \Omega}] = \partial_{\Omega_k} 1 = 0$.}. From \eqref{eq:chain-rule-tangent}, we see that the projected \gls{qhbm} parameterization admits tangent vectors which span $\mathfrak{su}(N)$ so long as the vectors
\begin{align*}
v_k := \left(\frac{\partial\tau_1}{\partial\Omega_k}, \dots, \frac{\partial\tau_{N^2-1}}{\partial\Omega_k}\right) \in \mathbb{R}^{N^2-1}  
\end{align*}
are linearly independent. Since $\{\tau_j\}$ charts $\mathcal{M}^{(N)}$, these coordinates are locally diffeomorphic to $\mathbb{R}^{N^2 - 1}$. The differential of a diffeomorphism is injective and therefore surjective, and so we have the desired spanning property. Since the unprojected \gls{qhbm} parameterization includes redundant degrees of freedom, the same holds although its tangent vectors specify an over-complete basis.
\end{proof}
\end{proposition}

\section{\texorpdfstring{Estimating the \gls{bkm} metric tensor}{Estimating the BKM metric tensor}}
\label{sec:estimating-bkm-tensor}

We will now cover how to obtain the \gls{bkm} metric tensor of the parameters in terms of the \gls{qhbm} parameterization. In particular, we provide analytical expressions for sampling-based techniques to obtain unbiased estimates of the information matrix elements. We split up our information matrix calculation into three types of blocks of this matrix; the cases where the tangent vectors are both of $\bm{\theta}$ parameters, the cases where they are both $\bm{\phi}$ parameters, and the cases where they are a mixture of both types of parameters. The fact that we can compute analytic expressions for the metric tensor for which we can sample the values using a mixture of the quantum and classical computers is unique to the \gls{qhbm} class of models.

In particular, when resolving the metric to a basis we may use that of the $\Omega_j$ tangent vectors so as to assume the parameter space dynamics induced by the \gls{qhbm} parameterization \cite{toth2017lower},
\begin{align}
\label{eq:bkm-defining}
    [\mathcal{I}^{\operatorname{BKM}}(\bm \Omega)]_{j, k} &= \int_0^\infty \tr [ (\partial_{\Omega_j} \hat \rho_{\bm \Omega})  (\hat \rho_{\bm \Omega}+s\mathbbm{1})^{-1} (\partial_{\Omega_k} \hat \rho_{\bm \Omega}) (\hat \rho_{\bm \Omega}+s\mathbbm{1})^{-1}  ] ds \\
    \label{eq:bkm-simplest}
    &= \tr[(\partial_{\Omega_j} \hat \rho_{\bm \Omega}) (\partial_{\Omega_k} \log \hat{\rho}_{\bm\Omega}) ]
,\end{align}
using \cref{eq:monotone-info-matrix,eq:bkm-lowering}. We have also used the following identity, which will be very helpful throughout the course of this work:
\begin{align}
\label{eq:deriv-of-log}
\frac{d}{dt} \log(\hat{A} + t\hat{B}) \bigg|_{t=0} = \int_{0}^{\infty} (\hat{A} + s\mathbbm{1})^{-1} \hat{B} (\hat{A} + s\mathbbm{1})^{-1} \, ds.
\end{align}
Note as well that
\begin{align}
\tr[(\partial_{\Omega_j} \hat \rho_{\bm \Omega}) (\partial_{\Omega_k} \log \hat{\rho}_{\bm\Omega}) ] &= - \tr[(\partial_{\Omega_j} \hat \rho_{\bm \Omega}) (\partial_{\Omega_k} \hat{K}_{\bm\Omega}) ] - \tr[\partial_{\Omega_j} \hat \rho_{\bm \Omega}] (\partial_{\Omega_k} \log Z_{\bm \theta})\\
&= - \tr[(\partial_{\Omega_j} \hat \rho_{\bm \Omega}) (\partial_{\Omega_k} \hat{K}_{\bm\Omega}) ]
\end{align}
using  \Cref{apdx:lem:traceless-spanning}. We have termed \eqref{eq:bkm-defining} the \textit{\gls{bkm} information matrix}. We now discuss its relationship to the relative entropy; recall that the quantum relative entropy can be written in the following form:
\begin{align}
    \label{eq:def-relative-entropy}
D(\hat\rho \| \hat \sigma ) = \tr[\hat \rho (\log \hat \rho - \log \hat \sigma)] = \int_{0}^\infty \tr[\hat \rho (\hat \sigma + s\mathbbm{1} )^{-1} (\hat \rho - \hat \sigma) (\hat \rho + s\mathbbm{1} )^{-1}] ds.
\end{align}

In \Cref{sec:qxe-bkm}, we claimed that the \gls{bkm} inner product is related to the Hessian of the relative entropy. To establish this claim, write:
\begin{align}
 \partial^2_{\alpha \beta} D(\hat\rho + \alpha \hat{H} \| \hat \rho + \beta \hat{H}' ) &= \frac{\partial}{\partial \beta}\int_{0}^\infty \Big(
 \label{eq:first-derivative-relative-entropy}
 \begin{aligned}[t]&\tr[ \hat{H} (\hat \rho + \beta \hat{H}' + s\mathbbm{1} )^{-1} (\alpha \hat{H}- \beta \hat{H}') (\hat \rho + \alpha \hat{H} + s\mathbbm{1} )^{-1}] \\
&+ \tr[ (\hat \rho + \alpha \hat{H}) (\hat \rho + \beta \hat{H}' + s\mathbbm{1} )^{-1}\hat{H}(\hat \rho + \alpha \hat{H} + s\mathbbm{1} )^{-1}] \\
&+ \tr[ (\hat \rho + \alpha \hat{H}) (\hat \rho + \beta \hat{H}' + s\mathbbm{1} )^{-1} \alpha \hat{H} (\hat \rho + \alpha \hat{H} + s\mathbbm{1} )^{-1}H(\hat \rho + \alpha \hat{H} + s\mathbbm{1} )^{-1}] \\
&- \tr[ (\hat \rho + \alpha \hat{H}) (\hat \rho + \beta \hat{H}' + s\mathbbm{1} )^{-1} \beta \hat{H}' (\hat \rho + \alpha \hat{H} + s\mathbbm{1} )^{-1}\hat{H}(\hat \rho + \alpha \hat{H} + s\mathbbm{1} )^{-1}] \Big) ds \end{aligned} 
\end{align}
Evaluating the above at $\alpha = \beta = 0$, we find:
\begin{align}
\partial^2_{\alpha\beta} D(\hat{\rho} + \alpha \hat{H} \| \hat{\rho} + \beta \hat{H}')|_{\alpha,\beta = 0} = -\int_{0}^\infty \tr[\hat{H} (\hat \rho + s\mathbbm{1} )^{-1} \hat{H}' (\hat \rho + s\mathbbm{1} )^{-1}] ds
.\end{align}
Therefore,
\begin{align}
\label{eq:bkm-hessian-canonical}
  [\mathcal{I}^{\operatorname{BKM}}(\bm \Omega)]_{j, k} &= -\partial^2_{\alpha\beta} D(\hat\rho_{\bm \Omega} + \alpha (\partial_{\Omega_j} \hat \rho_{\bm \Omega})  \| \hat \rho_{\bm \Omega} + \beta (\partial_{\Omega_k} \hat \rho_{\bm \Omega}))\mid_{\alpha = \beta = 0} 
.\end{align}

We will use \eqref{eq:bkm-simplest} in what follows. Recall from \Cref{sec:qxe-bkm} that our choice of lowering and raising operators implies we are working in mixture coordinates (\Cref{def:mixture-coords}) meaning that the partition function is parameterized directly by $\bm \Omega$. The metric tensor resolves equivalently in exponential coordinates, where the raising and lowering operators are exchanged. In exponential coordinates, tangent vectors $H$ enter as $\partial_t \exp(-K_{\bm \Omega} - \mathbbm{1}\mathcal{Z}_{\bm \theta} + tH)$.

\subsection{EBM block}

We first compute the \gls{bkm} logarithmic derivative,
\begin{align}
\label{eq:ebm-grad}
   \partial_{\theta_k} \log\hat\rho_{\bm\Omega} &= -\partial_{\theta_k} (\hat K_{\bm \Omega} + \mathbbm{1}\log \mathcal{Z}_{\bm\theta}) \\
   &= -U_{\bm \phi} (\partial_{\theta_k}\hat{K}_{\bm \theta})\hat{U}_{\bm\phi}^\dag  + \frac{1}{\mathcal{Z}_{\bm\theta}} \tr[(\partial_{\theta_k} \hat{K}_{\bm \theta}) e^{-\hat{K}_{\bm \theta}}]
\end{align}
and tangent vector,
\begin{align}
\partial_{\theta_k} \hat\rho_{\bm\Omega} &= \hat{U}_{\bm \phi }\left(\partial_{\theta_k} \frac{e^{-\hat{K}_{\bm \theta}}}{\mathcal{Z}_{\bm \theta}}\right)\hat{U}^\dag_{\bm \phi} \\
&= \hat U_{\bm \phi }\frac{-\mathcal{Z}_{\bm \theta}(\partial_{\theta_k}\hat{K}_{\bm \theta}) e^{-\hat{K}_{\bm \theta}} + e^{-\hat{K}_{\bm \theta}} \tr[(\partial_{\theta_k} \hat{K}_{\bm \theta}) e^{-\hat{K}_{\bm \theta}}] }{\mathcal{Z}^2_{\bm \theta}}\hat U^\dag_{\bm \phi} \\
&= \hat U_{\bm \phi } \left(-(\partial_{\theta_k}\hat{K}_{\bm \theta}) \hat{\rho}_{\bm \theta} + \hat{\rho}_{\bm \theta} \tr[(\partial_{\theta_k} \hat{K}_{\bm \theta}) \rho_{\bm \theta}]\right) \hat U^\dag_{\bm \phi} \\
\label{eq:rho-ebm-derivative}
&= \mathbb{E}_{x \sim p_{\bm\theta}(x)}\left[ \partial_{\theta_j} E_{\bm \theta}(x) \right] \hat\rho_{\bm \Omega} - \sum_x (\partial_{\theta_k} E_\theta) p_{\bm\theta}(x) \hat{U}_\phi \ket{x}\bra{x} \hat{U}_\phi^\dag
.\end{align}
Critically, $[\partial_{\theta_j} \hat{\rho}_{\bm \Omega}, \hat{\rho}_{\bm \Omega}] = 0 = [\partial_{\theta_j} \log \hat{\rho}_{\bm \Omega}, \hat{\rho}_{\bm \Omega}] $ given that the eigenbasis of $\hat{\rho}_{\bm \Omega}$ is independent of $\bm\theta$. Therefore,
\begin{align}
 [\mathcal{I}^{\operatorname{BKM}}(\bm \Omega)]_{\theta_j, \theta_k} &= \tr[\frac{(\partial_{\theta_j}\hat{K}_{\bm \theta})(\partial_{\theta_k}\hat{K}_{\bm \theta}) e^{-\hat{K}_{\bm \theta}} }{\mathcal{Z}_{\bm \theta}}  ] - \frac{\tr[(\partial_{\theta_j} \hat{K}_{\bm \theta}) e^{-\hat{K}_{\bm \theta}}]\tr[(\partial_{\theta_k} \hat{K}_{\bm \theta}) e^{-\hat{K}_{\bm \theta}}]}{\mathcal{Z}^2_{\bm \theta}} \\
&= \sum_{x} p_{\bm \theta}(x) \partial_{\theta_j} E_{\bm \theta}(x) \partial_{\theta_k} E_{\bm \theta}(x) - \sum_{x} p_{\bm \theta}(x) \partial_{\theta_j} E_{\bm \theta}(x) \sum_y p_{\bm \theta}(y) \partial_{\theta_k} E_{\bm \theta}(y) \\
\label{Hessianblock:classical-model-parameters}
 &= \mathbb{E}_{x \sim p_{\bm\theta}(x)}\left[ \partial_{\theta_j}  E_{\bm\theta}( x) \partial_{\theta_k} E_{\bm\theta}( x) \right]
    - \mathbb{E}_{ x \sim p_{\bm\theta}( x)}\left[\partial_{\theta_j}  E_{\bm\theta}( x) \right] \mathbb{E}_{y \sim p_{\bm\theta}( y)}\left[ \partial_{\theta_k} E_{\bm\theta}(y) \right]
\end{align}
The result reads as the covariance matrix of the gradient vector of the energy function subject to the sampled \gls{ebm} distribution. Note that this quantity does not require a quantum computer to be evaluated.

\subsection{QNN block}
\label{apdx:sec:qnn-block}

For the \gls{bkm} metric tensor elements which only depend on the gradients with respect to \gls{qnn} parameters, we can use an intuitive double parameter shift rule. A gradient technique for unitary \glspl{qnn} was recently pointed out in \cite{cerezo2020impact}; here we can apply it to the gradients of the \gls{qhbm} \gls{qnn} parameters. For a hardware efficient ansatz (i.e. a \gls{qnn} whose parameterized operations are independently parameterized and are of the form of simple exponentials of single Pauli operators, $\{\hat P_j\}_j$, e.g. $U_{\bm \phi} = \prod_j e^{i\phi_j \hat{P}_j}$), we have the parameter shift rules,
\begin{align}
\label{eq:qnn-grad}
\partial_{\phi_k} \log \hat{\rho}_{\bm \Omega} &=  -\hat{K}_{\bm{\theta} (\bm{\phi}+\bm{\Delta}^k)} + \hat{K}_{\bm{\theta} (\bm{\phi}-\bm{\Delta}^k)} \\
\label{eq:qnn-grad-rho}
\partial_{\phi_k} \hat{\rho}_{\bm{\Omega}} &=  \hat{\rho}_{\bm{\theta} (\bm{\phi}+\bm{\Delta}^k)} - \hat{\rho}_{\bm{\theta} (\bm{\phi}-\bm{\Delta}^k)} \\
\label{Hessianblock:QNN-parameters}
[\mathcal{I}^{\operatorname{BKM}}(\bm \Omega)]_{\phi_j, \phi_k} &= \begin{aligned}[t]&-\tr[\hat{\rho}_{\bm{\theta} (\bm{\phi}+\bm{\Delta}^j)} \hat{K}_{\bm{\theta} (\bm{\phi}+\bm{\Delta}^k)}] - \tr[\hat{\rho}_{\bm{\theta} (\bm{\phi}-\bm{\Delta}^j)} \hat{K}_{\bm{\theta} (\bm{\phi}-\bm{\Delta}^k)}] \\
&+ \tr[\hat{\rho}_{\bm{\theta} (\bm{\phi}+\bm{\Delta}^j)} \hat{K}_{\bm{\theta} (\bm{\phi}-\bm{\Delta}^k)}] + \tr[\hat{\rho}_{\bm{\theta} (\bm{\phi}-\bm{\Delta}^j)} \hat{K}_{\bm{\theta} (\bm{\phi}+\bm{\Delta}^k)}]
\end{aligned}
\end{align}
with $\bm{\Delta}^j = \tfrac{\pi}{4} \bm{\hat{e}_j}$ where standard basis vector has entries $(\bm{\hat{e}_j})_k = \delta_{j,k}$.

\subsection{Coupled block}

Finally, let us compute the terms of the \gls{bkm} metric tensor which include the coupling of \gls{qnn} and \gls{ebm} parameters,
\begin{align}
\label{Hessianblock:cross-terms}
[\mathcal{I}^{\operatorname{BKM}}(\bm \Omega)]_{\phi_j, \theta_k} &= -\tr[(\partial_{\theta_k}\hat{K}_{\bm \theta})\hat{U}_{\bm\phi}^\dag \hat{U}_{\bm{\phi}+\bm{\Delta}^j}\hat{\rho}_{\bm{\theta}}\hat{U}^\dag_{\bm{\phi}+\bm{\Delta}^j} \hat{U}_{\bm \phi} ] + \tr[(\partial_{\theta_k}\hat{K}_{\bm \theta})\hat{U}_{\bm\phi}^\dag \hat{U}_{\bm{\phi}-\bm{\Delta}^j}\hat{\rho}_{\bm{\theta}}\hat{U}_{\bm{\phi}-\bm{\Delta}^j}^\dag \hat{U}_{\bm \phi} ]
,\end{align}
where we have essentially combined \cref{eq:ebm-grad,eq:qnn-grad}.

\section{Estimating the \texorpdfstring{\gls{bh}}{BH} metric tensor}
\label{sec:estimating-bh-tensor}

\subsection{EBM block}

We can find the same \gls{ebm} block result as \eqref{Hessianblock:classical-model-parameters} in terms of the \gls{qfi} which corresponds to the \gls{bh} metric. As we will see from \eqref{eq:logarithmic-derivative}, the defining relation of the \gls{sld} $\hat{L}^{\operatorname{BH}}_{\bm \Omega, \theta_j}$ is given by
\begin{align}
\label{eq:sld-theta}
   \partial_{\theta_j} \hat\rho_{\bm \Omega} &= \frac{1}{2}\{ \hat{L}^{\operatorname{BH}}_{\bm \Omega, \theta_j}, \hat\rho_{\bm \Omega}\}.
\end{align}

\Cref{eq:sld-theta} can be viewed as a special case of the Lyapunov equation which admits solution $L^{\operatorname{BH}}_{\bm \Omega, \theta_j} = 2 \int_{0}^\infty e^{-z \hat\rho_{\bm \Omega}} ( \partial_{\theta_j} \hat\rho_{\bm \Omega}) e^{-z \hat\rho_{\bm \Omega}} dz = \mathcal{L}^{\operatorname{BH}}_{\bm \Omega}(\partial_{\theta_j}\hat\rho_{\bm \Omega})$. Since $[\partial_{\theta_j} \hat\rho_{\bm \Omega}, \hat\rho_{\bm \Omega}] = 0$,
\begin{align}
    L^{\operatorname{BH}}_{\bm \Omega, \theta_j} &= (\hat\rho_{\bm \Omega})^{-1} \partial_{\theta_j} \hat\rho_{\bm \Omega}, \\
    \intertext{noting that,} \\
    \label{eq:rho-inverse}
    \hat\rho_{\bm \Omega}^{-1} &= \sum_x \frac{1}{p_{\bm\theta}(x)} \hat{U}_{\bm\phi} \ket{x} \bra{x} \hat{U}_{\bm\phi}^\dag
.\end{align}
Furthermore, noting \eqref{eq:cov-information} and using \cref{eq:rho-ebm-derivative,eq:rho-inverse},
\begin{align}
    [\mathcal{I}^{\operatorname{BH}}(\bm \Omega)]_{\theta_i,\theta_j}  &= \frac{1}{2}\tr[\hat{\rho}_{\bm \Omega} \{\hat{L}^{\operatorname{BH}}_{\bm \Omega, \theta_j},\hat{L}^{\operatorname{BH}}_{\bm \Omega, \theta_j}\}] \\
    &= \tr[(\hat\rho_{\bm \Omega}^{-1}) (\partial_{\theta_i} \hat\rho_{\bm \Omega})(\partial_{\theta_j} \rho_{\bm \Omega})] \\
    &= \begin{aligned}[t]& \tr[\langle \partial_{\theta_i} E_{\bm\theta} \rangle \langle \partial_{\theta_j} E_{\bm\theta} \rangle  \hat\rho_{\bm \Omega} - \langle \partial_{\theta_i} E_\theta \rangle \sum_x (\partial_{\theta_j} E_\theta) p_{\theta}(x) \ket{\phi(x)} \bra{\phi(x)}) - \langle \partial_{\theta_j} E_\theta \rangle \sum_x (\partial_{\theta_i} E_{\bm\theta}) p_{\theta}(x) \ket{\phi(x)} \bra{\phi(x)})] \\
    &+ \tr[\sum_x (\partial_{\theta_i} E_{\bm\theta})(\partial_{\theta_j} E_\theta)p_{\theta}(x) \ket{\phi(x)} \bra{\phi(x)}] \end{aligned}\\
    &= \mathbb{E}_{x \sim p_{\bm\theta}(x)}\left[ \partial_{\theta_i}  E_{\bm\theta}( x) \partial_{\theta_j} E_{\bm\theta}( x) \right]
    - \mathbb{E}_{ x \sim p_{\bm\theta}( x)}\left[\partial_{\theta_i}  E_{\bm\theta}( x) \right] \mathbb{E}_{y \sim p_{\bm\theta}( y)}\left[ \partial_{\theta_j} E_{\bm\theta}(y) \right]
,\end{align}
as expected, classically. 

\subsection{QNN block}

We now proceed similarly for $\bm\phi$, 
\begin{align}
    \hat{\rho}_{\theta (\bm\phi +\bm{\Delta}^j)} - \hat{\rho}_{\bm\theta (\bm\phi - \bm{\Delta}^j)} &= \frac{1}{2}\{ \hat{ L}^{\operatorname{BH}}_{\bm \Omega, \phi_j}, \hat{\rho}_{\bm \Omega}\},
    \intertext{referring to \eqref{eq:qnn-grad-rho}. Taking matrix elements in the $\hat\rho_{\bm \Omega}$-basis with $\hat\rho_{\bm \Omega} = \sum_x p_{\bm \theta}(x) \ket{\phi(x)}\bra{\phi(x)}$,}
    \label{SLD-unitary-params}
    \hat{L}^{\operatorname{BH}}_{\bm \Omega, \phi_j} &= \sum_{x,y} \frac{2}{p_{\bm\theta}(x) + p_{\bm\theta}(y)} (\bra{\phi(x)}\hat\rho_{\bm\theta (\bm\phi +\bm{\Delta}^j)}\ket{\phi(y)} - \bra{\phi(x)}\hat\rho_{\bm\theta (\bm\phi - \bm{\Delta}^j)}\ket{\phi(y)})\ket{\phi(x)}\bra{\phi(y)} \\
    \intertext{Hence,}
    [\mathcal{I}^{\operatorname{BH}}(\bm \Omega)]_{\phi_j,\phi_k} &= 2\sum_{x,y} \frac{\Re[\bra{\phi(x)}(\hat\rho_{\bm\theta (\bm\phi +\bm{\Delta}^j)} - \hat\rho_{\bm\theta (\bm\phi - \bm{\Delta}^j)}) \ket{\phi(y)}\bra{\phi(y)}(\hat\rho_{\bm\theta (\bm\phi +\bm{\Delta}^k)} - \hat\rho_{\bm\theta (\bm\phi -\bm{\Delta}^k)}) \ket{\phi(x)} ]}{p_{\bm\theta}(x) + p_{\bm\theta}(y)}
,\end{align}
which matches \eqref{eq:mc-info-matrix}. In this (non-commutative) case, we see that the result does not match the \gls{bkm} metric result of \Cref{apdx:sec:qnn-block}.

\subsection{Coupled block}

Using \eqref{eq:mc-info-matrix} directly and again taking matrix elements in the $\hat{\rho}_{\bm \Omega}$-basis,
\begin{align}
    [\mathcal{I}^{\operatorname{BH}}(\bm \Omega)]_{\theta_j,\phi_k} &= 2\sum_{x,y} \frac{\Re[\bra{\phi(x)} \partial_{\theta_j}\hat\rho_{\bm\Omega} \ket{\phi(y)}\bra{\phi(y)}(\hat\rho_{\bm\theta (\bm\phi +\bm{\Delta}^k)} - \hat\rho_{\bm\theta (\bm\phi -\bm{\Delta}^k)}) \ket{\phi(x)} ]}{p_{\bm\theta}(x) + p_{\bm\theta}(y)} \\
    \intertext{From \eqref{eq:rho-ebm-derivative},}
    \bra{\phi(x)} \partial_{\theta_j}\hat\rho_{\bm\Omega}\ket{\phi(y)} &= \delta_{x, y} (\langle \partial_{\theta_j} E_{\theta} \rangle p_{\bm \theta}(x) - \partial_{\theta_j} p_{\theta}(x)) \\
    \intertext{Hence,}
    [\mathcal{I}^{\operatorname{BH}}(\bm \Omega)]_{\theta_j,\phi_k} &= \sum_{x} \frac{\left(\mathbb{E}_{x \sim p_{\bm\theta}(x)}\left[ \partial_{\theta_j} E_{\bm\theta}( x)\right] p_{\bm \theta}(x) - \partial_{\theta_j} p_{\theta}(x)\right)\bra{\phi(x)}(\hat\rho_{\bm\theta (\bm\phi +\bm{\Delta}^k)} - \hat\rho_{\bm\theta (\bm\phi -\bm{\Delta}^k)}) \ket{\phi(x)} ]}{p_{\bm\theta}(x)}
.\end{align}

\section{Metric-aware optimization of sequences of tasks}
\label{sec:sequences}

It is useful to classify sequence models as either recurrent or regressive. In the case of a recurrent (or recursive) task, the sequence is specified in terms of a quantum map which links each modeled density operator to the next one in the sequence. On the other hand, in a regression task, query access to the entire sequence of states is supplied directly. For our numerics, we focused on sequential conditional optimization (\Cref{alg:seq}), and recursive conditional optimization for \gls{qvartz}.

\subsection{Quantum-stochastic processes description of quantum dynamics}\label{app:qsto_dyn}

First, let us review the general mathematical theory of quantum open system dynamics \cite{gardiner2004quantum}. We can define a \textit{superoperator} which takes a density operator as argument and outputs a different density operator in the same space of operators, 
$\mathcal{L} : \mathcal{M}^{(N)} \rightarrow \mathcal{M}^{(N)}$.  The output density matrix is then denoted as $\mathcal{L}[\hat{\rho}]$.  One general form of this operator is the \textit{Liouvillian superoperator} written as
\begin{align}
\label{eq:quantum-master}
    \tfrac{d}{dt}\hat{\rho}(t) &= \mathcal{L}[\hat \rho(t)] = -i [\hat H(t), \hat \rho(t) ] + \mathcal{N}[\hat \rho(t)] + \int_s^t \mathcal{K}(t, t')[\hat \rho(t)] dt' \\
    \intertext{where,}
\mathcal{N}[\hat \rho(t)] &= \textstyle\sum_{a=1}^{N^2}(\hat{L}_a\hat{\rho}\hat{L}_a^\dagger -\tfrac{1}{2}\hat{L}_a^\dagger\hat{L}_a\hat{\rho} - \tfrac{1}{2}\hat{\rho}\hat{L}_a^\dagger\hat{L}_a).
\end{align}
such that the Hamiltonian $\hat{H}$ is a Hermitian operator, $\hat{L}_a$ are termed the jump operators (both of these can generally be time-dependent), and $\mathcal{K}(t, t')$ is a general superoperator known as the \textit{quantum memory kernel}.  The standard \textit{Liouville-Von-Neumann equation} describes a closed quantum system and is the special case where the latter two terms on the RHS vanish.

The solution to \Cref{eq:quantum-master} can then be expressed as a superoperator-valued time-ordered exponential of the Liouvillian,
\begin{align}
  \hat{\rho}(t) = \mathcal{T}e^{\int_0^t \mathcal{L}_{t'}dt'}[\hat{\rho}(0)] = \mathcal{V}_{t,0}[\hat{\rho}(0)], \quad \mathcal{V}_{t,s}[\cdot] \equiv \mathcal{T}e^{\int_s^t \mathcal{L}_{t'}dt'}[\cdot],  
\end{align}
where the effective channel superoperator $\mathcal{V}_{t,s}(\hat{\rho}_s)$ which takes the state at time $s$ and outputs the state at time $t$ is called the \textit{propagator}.

The differential equation describing the time evolution of an open quantum system under the Born-Markov \cite{accardi2013quantum} approximation assumes that the memory-carrying term of \Cref{eq:quantum-master} vanishes,
\begin{align}\label{eq:lindblad}
    \mathcal{L}_t[\hat{\rho}(t)] \equiv -i[\hat{H}(t),\hat{\rho}(t)] + \mathcal{D}(\hat \rho(t)).
\end{align}
In such a case, we may refer to the Liouvillian as a \textit{Lindblad superoperator}. For the scope of this paper, we consider how to learn quantum states coming from Markovian time evolutions where the above approximation holds, though all of our described techniques should generalize to non-Markovian sequences of completely-positive trace-preserving maps.

\subsection{Generic Sequential Optimization for Sequences of Tasks}\label{sec:gen_seq}

\begin{figure}
  \begin{algorithm}[H]
    \begin{algorithmic}[1]
    \Require target state sequence $\{\hat \sigma_{\bm \Lambda(\tau_k)}\}_{k=1}^M$, loss $\mathcal{L}$ 
    \Require $\textsc{Optimizer}(\texttt{init}, \texttt{lr\_sched},  \texttt{loss}, \texttt{target\_state})$
    \State initialize first model parameters $\bm \Omega^*(\tau_0)$
      \For{$k=1,2,\ldots,M$}
      \State choose learning rate schedule $\bm \eta_k$
      \State $\bm \Omega^*(\tau_k) \gets \textsc{Optimizer}(\bm \Omega^*(\tau_{k-1}), \bm \eta_k, \mathcal{L}, \hat \sigma_{\bm \Lambda(\tau_k)})$
      \EndFor
    \end{algorithmic}
    \caption{Sequentially chained optimization}
    \label{alg:seq}
  \end{algorithm}
  \caption{The simple initialization strategy for fitting sequences of density operators. Using metric-aware optimizers leverages the information-geometric proximity of neighboring states in the sequence.
  }
\end{figure}

Suppose we have a sequence of target states $\{\hat{\sigma}_{\bm{\Lambda}({\tau_{k}})}\}_{k=1}^M$ given by a partition $\{\tau_{k}\}_{k=1}^M\subset [0,T]$ of a discretized parametric path  $ \bm{\Lambda}(\tau)$ in task parameter space $\bm{\Lambda}\in \mathbb{R}^R$. We wish to learn a sequence of optimal \gls{qhbm} parameters so that each $\hat{\rho}_{\bm{\Omega}^{*}(\tau_k)} \approx \hat{\sigma}_{\bm{\Lambda}({\tau_{k}})}$ for each $k$. Hence, we can have a collection of loss functions
\begin{align}
\label{eq:chained-objective}
  \bm{\Omega}^*(\tau_k)  = \underset{\bm{\Omega}(\tau_k))}{\arg\min} \  \mathcal{L}(\bm{\Omega}(\tau_k)),\qquad\forall k \in \{1, \cdots, M\}.
\end{align}
For sequential optimization, we can apply \Cref{alg:seq} to the above sequence of losses in order to find these optimal parameters. We then obtain a full approximation of the quantum-stochastic process \cite{milz2021quantum} formed by the sequence of states along the chosen task path,
\begin{equation}
    \bigotimes_{k=1}^M  \hat{\rho}_{\bm{\Omega}^{*}(\tau_k)} \approx \bigotimes_{k=1}^M \hat{\sigma}_{\bm{\Lambda}({\tau_{k}})}.
\end{equation}
This is the approach for both \acrshort{meta-vqt} (\Cref{sec:meta-vqt}) and \acrshort{qspl} (\Cref{sec:meta-qmhl}). At an abstract level, the principle difference between both these algorithms is that for \acrshort{meta-vqt}, we minimize sequentially the forwards quantum relative entropy between model and target state; $\mathcal{L}(\bm{\Omega}(\tau_k)) = D(\hat{\rho}_{\bm{\Omega}(\tau_k)}\Vert \hat{\sigma}(\bm{\Lambda}(\tau_{k})))$, while in \acrshort{qspl}, we minimize the backwards quantum relative entropy between model and target state; $\mathcal{L}(\bm{\Omega}(\tau_k)) = D(\hat{\sigma}(\bm{\Lambda}(\tau_{k}))\Vert \hat{\rho}_{\bm{\Omega}(\tau_k)})$. Let us flesh out below how the physical contexts of the target datasets differ.

\subsection{\texorpdfstring{\acrfull*{meta-vqt}}{Meta-Variational Quantum Thermalization}}
\label{sec:meta-vqt}

Consider the case where we are given a sequence in the space of target modular Hamiltonians \(\{\hat{H}_{\bm{\Lambda}(\tau_{k})}\}_{k=1}^M\) rather than the states themselves.  Then our sequence of target states is a sequence of thermal (Gibbs) states \begin{equation}\hat{\sigma}(\bm{\Lambda}(\tau_{k})) = e^{-\hat{H}_{\bm{\Lambda}(\tau_{k})}}/\mathcal{Z}_{\bm{\Lambda}(\tau_{k})},\end{equation} where
\(\bm{\Lambda}(\tau): \mathbb{R}\rightarrow \mathbb{R}^R\) is a path function in the task parameterization space of dimension $R$. Note that, this path parameter vector $\bm{\Lambda}$ can include both coldness (inverse temperature; $\beta(\tau_k)$) and Hamiltonian parameters.

In order to find sequence of optimal QHBM parameters $\bm{\Omega}^*(\tau_k)$ such that the relative entropy between our models and the target states are minimized throughout the sequence, we minimize each of the free energies:
\begin{align}\label{eq:free_en_equiv}
\bm{\Omega}^*(\tau_k) = \underset{\bm{\Omega}(\tau_k)}{\arg\min} \  D(\hat{\rho}_{\bm{\Omega}(\tau_k)}\Vert \hat{\sigma}(\bm{\Lambda}(\tau_{k}))) =  \underset{\bm{\Omega}(\tau_k)}{\arg\min}
  \mathcal{F}_{\hat{H}_{\bm{\Lambda}(\tau_{k})}}(\bm{\Omega}(\tau_k))
,\end{align}
$\forall k \in \{1, \cdots, M\}$ and where we denote the free energy  $\mathcal{F}_{\hat{H}_{\bm{\Lambda}}}(\bm{\Omega})$ of the state $\hat{\rho}_{\bm{\Omega}}$ with respect to the Hamiltonian $\hat{H}_{\bm{\Lambda}}$, also known as the \gls{vqt} loss, 

\begin{equation}\mathcal{F}_{\hat{H}_{\bm{\Lambda}}}(\bm{\Omega})\equiv D(\hat{\rho}_{\bm{\Omega}}\Vert \hat{\sigma}(\bm{\Lambda}))  -  \ln \mathcal{Z}_{\bm{\Lambda}}   =    \tr\left[\hat\rho_{\bm\Omega}{\hat{H}_{\bm{\Lambda}}}\right]  -S(\hat\rho_{\bm\Omega}).     \end{equation} 
Results from \Cref{sec:sequences-states} came from choosing this \gls{vqt} free energy loss  for $\mathcal{L}$
as in \eqref{eq:chained-objective} and \Cref{alg:seq}.
Note that the gradients of this loss can be straightforwardly estimated, see \eqref{eq:RelativeEntropyVQT} for the explicit gradient estimators. We term this Gibbs-sequence-generating problem \gls{meta-vqt},\footnote{One can understand this choice of name as akin to a Meta-learned VQT optimization algorithm.} and we can apply our chained metric-aware optimizer (\Cref{alg:seq}) to sequentially optimize these free energy losses. In our results section, we mainly focused on chained \gls{qpmd}.

A special case of \gls{meta-vqt} is for \textit{imaginary time evolution}, where the path in coldness-Hamiltonian space is variable only in coldness (inverse temperature); $\hat{H}_{\bm{\Lambda}(\tau)}= \beta_\tau\hat{H}$. In this case of application of \acrshort{meta-vqt} to imaginary time evolution, the \textit{Meta-} prefix can also be understood to stand for \textit{Multi-Euclidean-Time-Annealing}. This imaginary time evolution special case is what we focus on for the results of \Cref{sec:sequences-states}.

\subsection{\texorpdfstring{\acrfull*{qspl}}{QSPL}}
\label{sec:meta-qmhl}

For \gls{qspl}, the task is to generatively model the sequence of states which arise throughout a quantum-stochastic process \cite{milz2021quantum}, given direct quantum data access to each state in the sequence. We mainly focus on time evolution processes, but any sequence of states with sufficient geometric locality between subsequent states would suffice as a dataset. 

Consider a sequence dataset \(\{\hat{\sigma}_{t_j} \}_{j=1}^M\) consisting of states along a Markovian time evolution \footnote{Open or closed.  Our techniques generalize straightforwardly to non-Markovian quantum-stochastic processes, see \Cref{app:qsto_dyn} for background.}
\begin{align}
\label{eq:discretized-dynamical-maps}
    \hat{\sigma}_{t_j} = \mathcal{V}_{t_j,0}(\hat{\sigma}_0) = \mathcal{V}_{t_j,t_{j-1}}\circ \mathcal{V}_{t_{j-1},t_{j-2}}\circ\ldots \circ \mathcal{V}_{t_1,0} (\hat{\sigma}_0),
\end{align}
where, for simplicity, we can assume \(t_j - t_{j-1} = \delta\) for all $j$. For this dataset, the goal of \gls{qspl} is to learn a sequence of optimal \gls{qhbm} generative model parameters \(\{ \bm{\Omega}^{*}_{t_j} \}_{j=1}^M\) such that \(\hat{\sigma}_{t_j} \approx \hat{\rho}_{\bm{\Omega}^{*}_{t_j}}\) for all \(j\). These optimal parameters can be learned by minimizing the cross entropy between the target state and the corresponding generative model,
\begin{align}\label{eq:meta-qmhl}
  \bm{\Omega}^*_{t_k}
  = \underset{\bm{\Omega}_{t_k}}{\arg\min} D( \hat{\sigma}_{t_k}\Vert \hat{\rho}_{\bm{\Omega}_{t_k}})
  = \underset{\bm{\Omega}_{t_k}}{\arg\min}   \mathcal{X}_t(\bm{\Omega}),
  \end{align}
$\forall k \in \{1, \cdots, M\}$, where for compactness of notation, we denoted the quantum cross entropy, also known as the \gls{qmhl} loss (see \eqref{eq:RelativeEntropyQMHL}) between our model and the target state at time $t$ as 
\begin{equation}\label{eq:time_xent_def}
    \mathcal{X}_t(\bm{\Omega}) \equiv -\tr[\hat{\sigma}_{t}\log \hat{\rho}_{\bm{\Omega}}].
\end{equation}
By leveraging our chained metric-aware optimization described in \Cref{alg:seq} for the above cross-entropy loss, we can learn the sequence of states representing the quantum-stochastic process. When using metric-aware optimizers, this process leverages the inherent geometric locality between subsequent states. See \Cref{sec:cont_qspl} for an analysis of when this this assumption holds approximatively.

\subsection{\texorpdfstring{\acrfull*{qvartz}}{Quantum Variational Recursive Time Evolution AnsatZ}}
\label{sec:qvartz}

Suppose we are given access to copies of an initial quantum state $\hat{\sigma}_0$ and the ability to apply a \gls{cptp} dynamical map to an arbitrary density operator. Such a map may encode unitary (Schrodinger), Markovian (Lindbladian), or non-Markovian (Nakajima-Zwanzig) dynamics. Our goal is then to simulate \cite{kliesch2011dissipative} the evolution of the initial quantum state under the action of the dynamical map over some time interval $[0, T]$.

Assume we can discretize the dynamical map over the time interval such that we can apply $\mathcal{V}_{t_{k+1}, t_{k}}$ for $\{t_k\}_{k=1}^{M}$, where, for simplicity, we have $t_{k+1} = t_k + \Delta t$, with $\Delta t = T / M$. The corresponding sequence of states we seek to learn are the evolved quantum states at each time step,
\begin{align}
\label{eq:discretized-dynamical-maps-qvartz}
    \hat{\sigma}(t_k) = \mathcal{V}_{t_k,0}(\hat{\sigma}_0) = \mathcal{V}_{t_k,t_{k-1}}\circ \mathcal{V}_{t_{k-1},t_{k-2}}\circ\ldots \circ \mathcal{V}_{t_1,0} (\hat{\sigma}_0).
\end{align}
The naive approach would be to simply identify each target state as $\hat{\sigma}(t_k)$ and formulate the problem as a specific instantiation of \gls{qspl} \eqref{eq:meta-qmhl}. However, we note that to construct each $\hat{\sigma}(t_k)$, the quantum circuit depth grows linearly with $k$. \Gls{qvartz} aims to circumvent this scaling by recursively learning our \gls{qhbm} representations. Given the optimal \gls{qhbm} at the previous time step $\hat \rho_{\bm \Omega^*(t_{k-1})}$, we apply the single channel for the current time step $\mathcal{V}_{t_{k}, t_{k-1}}$ and learn the current model $\hat \rho_{\bm \Omega(t_{k})}$ against the resulting evolved state, which serves as approximation of the true evolved state,
\begin{align}
    \hat{\sigma}(t_k) \approx \mathcal{V}_{t_{k}, t_{k-1}}(\hat \rho_{\bm \Omega^*(t_{k-1})}).
\end{align}
Formally, we replace $\hat{\sigma}(\tau_{k}) \leftrightarrow \mathcal{V}_{t_{k}, t_{k-1}}(\hat \rho_{\bm \Omega^*(t_{k-1})})$ in the \gls{qspl} objective \eqref{eq:meta-qmhl}. We may intuitively view this approach as a variational recursive checkpointing the quantum dynamics of a system in the classical parameters of a \gls{qhbm}, or alternatively, a variational form of temporal integration of quantum-stochastic processes. To contrast with our sequential optimizations, we can define the optimal parameters recursively:

\begin{align}\label{eq:qvartz_rec}
  \bm{\Omega}^*_{t_k}
   \equiv \underset{\bm{\Omega}_{t_k}}{\arg\min} \  D( \mathcal{V}_{t_{k}, t_{k-1}}(\hat \rho_{\bm \Omega^*(t_{k-1})})\Vert \hat{\rho}_{\bm{\Omega}_{t_k}}) \approx \underset{\bm{\Omega}_{t_k}}{\arg\min} \  D( \hat{\sigma}_{t_k}\Vert \hat{\rho}_{\bm{\Omega}_{t_k}}),
  \end{align}
 $\forall k \in \{1, \cdots, M\}$, with exact equality in the case where each model converges to the true state. Due to this recursive definition of the optima, the algorithm to find them differs from \Cref{alg:seq}; instead, we use \Cref{alg:recurs} with $\mathcal{W}_{k}\equiv \mathcal{V}_{t_{k}, t_{k-1}}$, and the loss being quantum relative entropy. Note that once again, quantum relative entropy minimization is equivalent to quantum cross-entropy minimization:
\begin{equation}\label{eq:qvartz-estimators}
    \bm{\Omega}^*_{t_k}
   \equiv \underset{\bm{\Omega}_{t_k}}{\arg\min} \  D( \mathcal{V}_{t_{k}, t_{k-1}}(\hat \rho_{\bm \Omega^*(t_{k-1})})\Vert \hat{\rho}_{\bm{\Omega}_{t_k}})= \underset{\bm{\Omega}_{t_k}}{\arg\min} \left(  -\tr[\mathcal{V}_{t_{k}, t_{k-1}}(\hat \rho_{\bm \Omega^*(t_{k-1})})\log \hat{\rho}_{\bm{\Omega}_{t_k}}]\right),
\end{equation}
$\forall k \in \{1, \cdots, M\}$. We can evaluate the gradients of this cross-entropy as it is simply the \gls{qmhl} loss (see \eqref{eq:RelativeEntropyQMHL} for gradient estimator) between our model and the single-step propagated state of the previous time step's optimized model.

\subsection{Generalized Quantum Variational Recursive Propagation}

Note that we can write down a generalization of this \gls{qvartz} algorithm to the recursive variational propagation through any sequence of \gls{cptp} maps, 

\begin{equation}
     \hat{\sigma}_k = \mathcal{W}_{k}\circ \mathcal{W}_{k-1}\circ\ldots \circ \mathcal{W}_{1} (\hat{\sigma}_0),
     \end{equation}
$\forall k \in \{1, \cdots, M\}$. We aim to learn a set of optimal parameters $\{\bm\Omega^*_k\}_k$ such that we approximate the resulting quantum-stochastic process
\begin{equation}
      \bigotimes_{k=1}^M  \hat{\rho}_{\bm{\Omega}^{*}_k} \approx \bigotimes_{k=1}^M \hat{\sigma}_{k}.
\end{equation}
We can do so by recursively minimizing some choice of loss $\mathcal{L}$ for each element of the sequence forming the process. Typically, this loss is chosen to contrastive between the model and the target state, e.g. quantum relative entropy. We describe in pseudocode this generalization in \cref{alg:recurs}.
    
\subsection{Discussion and outlook for \texorpdfstring{\gls{qvartz}}{QVARTZ}}

As a result of this recursive construction, our quantum circuit depth requirements remain constant with respect to $k$, assuming an upper bound to the quantum complexity of the state over its evolution during the quantum-stochastic process.  This is because we initialize the evolution at each time step $t$ from our latest \gls{qhbm} representation at $t-1$ instead of propagating our initial state through all timesteps. The buildup of complexity over time evolution remains an open question, though some recent works have begun to tackle this question \cite{haferkamp2022linear}. We leave exploration of the fixed-depth representability of quantum states to future work. Additionally, our methods could readily be used for explorations of quantum mechanics foundations, such as implementing a Bayesian calculus for quantum theory~\cite{NIPS2005_4191ef5f,Leifer_2008}.

Note that in order to guarantee convergence of this approach for both \gls{qspl} and \gls{qvartz}, we assumed geometric locality between steps. In \Cref{sec:cont_qspl}, we explore bounds on the Lindbladian superoperator and how one can choose an appropriate temporal step size in order to ensure that each optimization loop is within an approximately convex quadratic region. 

A big remaining open question with \gls{qvartz} is the rate of error buildup for finite numbers of samples or training iterations. Even assuming that the hypothesis submanifold spanned by our ansatz contains the entire task path of the time evolution, for finite training iterations we can expect some amount of drift from the true optimal parameters in parameter space. As recursive time evolution methods build up drift additively after each iteration, and this parameter space drift gets amplified through several maps, we can expect an exponential buildup of errors, as one would expect from composing noisy channels \cite{muller2016relative}.

In future work, we plan to use geometric methods to quantify the rate of error buildup over time. As we will show in future work, the semigroup of Markovian time evolution can be represented as a semiflow when lifted to model parameter space. As part of that work, in \Cref{sec:cont_qspl} we derived the expression for tangent vector along the parameter space representation of the task path. As a single time evolution is the integral of this tangent vector field, one could imagine having many initial states and forming a (semi)flow from the collection of tangent vector fields from this collection of paths. This is how one obtains a semi-flow representation of Markovian dynamics. From this semi-flow, one can model the error in convergence at each step as approximately Gaussian with a covariance dependent on the local metric along the path. Mapping this Gaussian noise through the semiflow can yield a Lyapunov growth  of the distribution that could be quantified, and is often studied in classical numerical integration \cite{moir2010reconsidering}. We plan to describe this in further detail in future work.

\begin{figure}
  \begin{algorithm}[H]
    \begin{algorithmic}
    \Require initial target state $\hat \sigma_{0}$, map sequence $\{\mathcal{W}_{k}\}_{k=1}^M$, loss $\mathcal{L}$
    \Require $\textsc{Optimizer}(\texttt{init}, \texttt{lr\_sched},  \texttt{loss}, \texttt{target\_state})$
    \State initialize first model parameters $\bm \Omega_0$
    \State choose learning rate schedule $\bm \eta_0$
    \State $\bm \Omega^*_0 \gets \textsc{Optimizer}(\bm \Omega_0, \bm \eta_0, \mathcal{L},\hat{\sigma}_0)$
      \For{$k=1,2,\ldots,M$}
      \State choose learning rate schedule $\bm \eta_k$
      \State $\bm \Omega^*_k \gets \textsc{Optimizer}(\bm \Omega^*_{k-1}, \bm \eta_k, \mathcal{L},\mathcal{W}_{k}(\hat{\rho}_{\bm{\Omega}^*_{k-1}})))$
      \EndFor
    \end{algorithmic}
    \caption{Quantum Variational Recursive Propagation}
    \label{alg:recurs}
  \end{algorithm}
  \caption{Generic algorithm for variationally recursively propagating density operators through sequences of maps. The initialization strategy works best when each of the maps retains some degree of information geometric locality between its input and output.
  }
\end{figure}

\section{Bayesian interpretation of algorithms}

\subsection{Tuning learning rates}
\label{sec:information-length}

The metric-aware update rule \Cref{eq:natural-gradient-update} produces a collection of \gls{qhbm} parameters $\{ \bm \Omega_j \}_j$. We derived this update rule from \eqref{eq:discrete-diff-eq} where we see that $\bm \delta = \mathcal{O}(\tfrac{1}{\lambda})$. Hence, when $\lambda \gg 1$, we can view $\{ \bm \Omega_j \}_j$ as the discretization of a curve over $\mathcal{M}^{(N)}$. The length-squared of a segment $\{\Omega_j\}_{j=A}^B$ of this curve in terms of the chosen metric is found by integrating the second fundamental form which (again with $\lambda \gg 1$) looks as,
\begin{align}
    ds^2 &\approx \frac{1}{2}\langle \bm \delta, \mathcal{I}({\bm{{\Omega}}_j}) \bm \delta \rangle \\
    \int_{\Omega_A}^{\Omega_B} ds^2 &\approx \frac{1}{2}\sum_{j = A}^{B - 1} \langle \bm \delta, \mathcal{I}({\bm{{\Omega}}_j}) \bm \delta \rangle
\end{align}
We see that $ds^2 = \mathcal{O}\left(\tfrac{1}{\lambda^2}\right)$. As we do not know the preferred length between successive descent parameters a priori, we generally treat $\lambda$ as drawn from a learnable hyper-prior.

Alternatively, one may interpret the relation \eqref{eq:ngd-inner-product} as a Gaussian approximation to the conditional prior over $\bm \Omega_{j+1}$ given $\bm \Omega_j$ whenever $\nabla_{\bm{{\Omega}}_j}\mathcal{L}(\hat{\rho}_{\bm{{\Omega}}_j})$ vanishes (i.e. about the so-called \gls{map} estimate) and the parameterization induces diffeomorphism. This is known elsewhere as the Laplace approximation \cite{murphy2012machine,rasmussen2003gaussian}. Consider that a $d$-dimensional Gaussian parameterized by mean vector and covariance matrix, $\bm x \sim \mathcal{N}(\bm \mu, \hat \Sigma)$, has a log probability density, $\log p(\bm x) = -\tfrac{1}{2} \langle (\bm x-\bm \mu), \hat\Sigma^{-1} (\bm x - \bm \mu)\rangle + \text{const}$ where the constant is the normalizing factor. Accordingly, we can view the second-order expansion \eqref{eq:ngd-inner-product} as,
\begin{align}
\label{eq:laplace}
\bm \Omega_{j+1} \mid \bm \Omega_j \sim \mathcal{N} 
\left(\bm \Omega_j , \frac{1}{\lambda} \mathcal{I}^{-1}({\bm{{\Omega}}_j})\right).
\end{align}
This probability density acts as a prior in the metric-aware descent rule \eqref{eq:ngd-inner-product} whereas the loss-gradient term acts as the likelihood. The prior density of $\bm \Omega_{j+1}$ decays quadratically with the distance from $\bm \Omega_j$ at a decay rate modulated by the $\lambda$ hyperparameter. By tuning this hyperparameter suitably, our prior should have good overlap with the neighborhood of ideal step size.

\subsubsection{Interpretation for sequences}
\label{sec:tuning-length-sequences}

A similar interpretation as above applies for learning sequences. In particular, for chained initialization (\Cref{sec:sequences-states}), we can collect $\bm \lambda = (\lambda_1, \lambda_2, \cdots, \lambda_M)$. We may tune $\bm \lambda$ as in \Cref{sec:information-length} thinking again in terms of either information length or posterior width (now corresponding to a Hidden Markov Model of Gaussians as below). In the case of $\bm \lambda$, we are now making claims about the believed distance between optimal parameters in the sequence.

\paragraph{Quantum-Probabilistic Hidden Markov Models Interpretation}
\label{sec:qphmm}

An alternative interpretation of our approach for quantum sequence modelling with geodesic priors is that we are leveraging a \gls{qphmm} as a conditional prior. Our metric-based regularizer can be understood\footnote{This interpretation requires that the classical parameterization induces a diffeomorphism between classical parameter space and $\mathcal{M}^{(N)}$ or a submanifold thereof. See \Cref{sec:tuning-length-sequences} for details on this conditional prior interpretation.} as a Gaussian conditional prior: 
\begin{align}
\begin{aligned}[t]p&(\bm \Omega_{j+1}(\tau_k) \mid \bm \Omega_j(\tau_k)) \\
&= p_{\mathcal{N}}\left( \bm \Omega_{j+1}(\tau_k); \bm\mu = \bm \Omega_j(\tau_k) , \hat\Sigma = \frac{1}{\lambda_{k}} \mathcal{I}^{-1}(\bm{{\Omega}}_j(\tau_k)\right).
\end{aligned}
\end{align}
Thus, our quantum sequence model corresponding to the first optimization step for each $\tau_k, k >0$ using chained initialization (\Cref{sec:sequences-states}) can be interpreted as giving \textit{prior predictive} density matrix:
\begin{align}
\label{eq:posterior-dm}
\begin{aligned}[t] \bigotimes_{k=1}^{M} &\hat \rho_{\bm{\Omega}_0(\tau_{k}) \mid \bm \Omega^*(\tau_{k-1}) }  \\ &=  \bigotimes_{k=0}^{M-1} \int  \hat \rho_{\bm{\Omega}}  \cdot  p_{\mathcal{N}}\left(\bm \Omega ; \bm \Omega^*(\tau_k), \frac{1}{\lambda_{k}} \mathcal{I}^{-1}(\bm \Omega^*)\right) d\bm \Omega  \end{aligned}
\end{align}
In this interpretation, when executing our chained mirror descent for modelling sequences, for the first optimization step, we are simply using \gls{map} \cite{murphy2012machine} inference on the parameter nodes of the \gls{qphmm} under this particular prior and a likelihood which depends on the loss gradient.

\subsection{Bayesian network generalization for learning sequences}
\label{sec:chain-bayesian-network}

We may generalize the chained initialization strategy of \Cref{sec:sequences-states} by always minimizing about $\bm \Omega_j$ while yet including contribution(s) from $\{ \bm{ \Omega}^*(\tau_{k'})\}_{k' \leq k}$ within the metric constraint:
\begin{align}
\label{eq:md-update-rule-chain-history}
  \bm{\Omega}_{j+1}(\tau_{k+1}) & = \arg\min_{\bm\Omega} \left[ \left\langle \bm \Omega,  \nabla_{\bm { \Omega}_j(\tau_{k+1})} \mathcal{L}\left({\bm { \Omega}_j(\tau_{k+1})}\right)\right \rangle
               + \lambda_{k+1} D \left( \hat{\rho}_{\bm { \Omega}_j(\tau_{k+1})} \Vert \hat{\rho}_{\bm\Omega} \right)
               + \sum_{k' \leq k} \gamma_{k'} D  \left( \hat{\rho}_{\bm{{\Omega}}^*(\tau_{k'})} \Vert \hat{\rho}_{\bm\Omega}  \right)
    \right],
\end{align}
We may choose, for example, $\gamma_j = (\zeta)^j$ for a fixed $\zeta$ so as to utilize an exponential decay scheduling. This is akin to including a conditional prior between subsequent generative models in the sequence, where the prior is an exponential decay with rate $\gamma_j$ with respect to squared information-geometric distance. In general, decay schedules may influence that less local optimal parameters in the sequence contribute decreasingly in the optimization. Observe that the chained initialization procedure discussed in \Cref{sec:sequences-states} is a special case of \Cref{eq:md-update-rule-chain-history} given the $\gamma_j$ degree of freedom.

\section{Continuous Limit: Flows and Paths in Task Space Geometry}
\label{sec:cts-sequence-limit}

In order to understand the dependency of our sequential optimization algorithms with respect to step size, it will be illuminating to consider the continuum limit of these protocols.

\subsection{Task Space Geometry Basics}\label{sec:task_geo_basic}

Before we introduce the continuous limits of our sequence optimization algorithms, it is worth briefly fleshing out the basics of task space geometry and curves within it.
\paragraph{Paths and Sequences of target states}

Consider a one-parameter continuous path $\bm{\Lambda}(\tau)\equiv \bm\Lambda_\tau$, where  $\bm{\Lambda}:\mathbb{R}\rightarrow \mathbb{R}^R$ is a function from the path parameter $\tau$ to the $R$-dimensional task parameter space of a  manifold $\hat{\sigma}_{\bm{\lambda}}$. We can define the tangent vector and the path as the integral of this tangent vector field,
 \begin{equation}\label{eq:flowing}
   \bm{\Lambda}_\tau \equiv \int_0^\tau d\tau'\, \partial_{\tau'} \bm\Lambda_{\tau'},
 \end{equation}
 this is a simple equation but it will later allow us to flow along this path in model parameter space.
 
 Using the \gls{bkm} metric for this parameter space, we can define the length of any segment $[\tau_A,\tau_B]$ of the path as
\begin{equation}\int ds = \int_{\tau_A}^{\tau_B}\!\!\!\! d\tau \sqrt{  \partial_\tau\bm{\Lambda}^\top_\tau \cdot \mathcal{I}(\bm{\Lambda}_\tau)\cdot\partial_\tau\bm{\Lambda}_\tau}, \quad  \mathcal{I}_{jk} (\bm{\lambda})= \tfrac{1}{2}\tr[ \hat{\sigma}_{\bm{\lambda}}\partial_{\lambda_j}\partial_{\lambda_k}\log \hat{\sigma}_{\bm{\lambda}}]. \end{equation}
Alternatively, one can write down the squared line element as
\(ds^2  = (d\tau)^2 \partial_\tau\bm{\Lambda}^\top_\tau \cdot \mathcal{I}(\bm{\Lambda}_\tau)\cdot\partial_\tau\bm{\Lambda}_\tau \).

We can consider a $M$-point sequence of states $\{\hat{\sigma}_{\bm{\Lambda}{\tau_j}}\}_{j=1}^M$ as the states along the one-parameter path in the multi-parameter manifold, with $\tau_1\leq \tau_2\leq \ldots \leq \tau_M$. Let $\bm{\varepsilon}^{(j)} \equiv \bm{\Lambda}_{\tau_{j+1}}-\bm{\Lambda}_{\tau_{j}}$ be the vector of difference in parameter space between the points. For $\tau_{j+1} - \tau_j =\delta_j$, we have $\bm{\varepsilon}^{(j)} \approx \delta_j\left. \partial_\tau\bm{\Lambda}_{\tau}\right|_{\tau_j}$.

Let us examine the sum of relative entropies between adjacent states in the sequence and relate it to path length, 

\spliteq{\label{eq:target_man_length}
\sum_{j=1}^M& D(\hat{\sigma}_{\bm{\Lambda}(\tau_{j+1})} \Vert \hat{\sigma}_{\bm{\Lambda}(\tau_j)} )  =\sum_{j=1}^M \bm{\varepsilon}^{(j)\top} \cdot \bm{g}(\bm{\Lambda}_\tau)\cdot\bm{\varepsilon}^{(j)} + \mathcal{O}(\delta_j^3)
=\sum_{j=1}^M \delta^2_j\left. \partial_\tau\bm{\Lambda}_{\tau}\right|_{\tau_j}^\top \cdot \bm{g}(\bm{\Lambda}_\tau)\cdot \left. \partial_\tau\bm{\Lambda}_{\tau}\right|_{\tau_j} + \mathcal{O}(\delta_j^3)
}

and so we see that if the sequence of $\{{\tau_j}\}_{j=1}^M$ is an infinitesimal partition of the path between $\tau_1$ and $\tau_M$, then
\begin{equation}\sum_{j=1}^M D(\hat{\sigma}_{\bm{\Lambda}(\tau_{j+1})} \Vert \hat{\sigma}_{\bm{\Lambda}(\tau_j)} )\approx \sum_{j=1}^M \int_{\tau_j}^{\tau_{j+1}} ds^2 \approx \int_{\tau_1}^{\tau_{M}} ds^2. \end{equation}

Most importantly, up to third order, the value of path length versus computing relative entropies between states are equivalent, and we can understand total path length as the sum of relative entropies between subsequent points in the sequence. Each segment's length is a local metric-dependent norm of the path tangent vector (tangent vector contracted with itself, with the contraction modulated by the metric). 

Thus, to understand the relevant limits for our practical implementations of chained metric-aware optimization, it is thus important to understand the tangent vector to the target path $\partial_\tau\bm{\Lambda}_{\tau}$, as it represents the instantaneous rate at which the target state changes. We will do so below and derive how to change the representation of this tangent vector from target parameter space to model parameter space. This way, one could in principle estimate the model parameter space representation of the tangent vector and update the parameters directly in order to flow along the path. Readers can refer to \cref{fig:md-sequence} to recall the intuitive picture of task space versus model space embeddings.

\subsubsection{Future Directions for Task Space Geometry Analysis}

Before we advance to examining how the continuous analogue of our chained optimizers could apply to representing task parameter space paths, let us briefly mention a few potential directions of future inquiry.

In terms of further extensions to our work in terms of quantum information geometry, a few possible extensions come to mind. First, a study of the total path length of task space paths would be of interest to evaluate the total thermodynamic length of quantum evolutions \cite{scandi2019thermodynamic}. This has various applications in the study of quantum thermodynamics \cite{alhambra2016fluctuating} and has a strong classical analog in the theory of out-of-equilibrium thermodynamics \cite{crooks2007}.

Additionally, exploring the computations of the Riemann curvature tensors and of Christoffel symbols in the Riemannian geometry \cite{dittmann2000curvature} of the space of quantum-probabilistic models for our choice of \gls{bkm} metric may be of interest to the subcommunity of quantum machine learning theorists interested in trainability of models \cite{arrasmith2021equivalence,holmes2022connecting}, as rapid metric variation would be detrimental to the possible step sizes one can take with any sort of \acrshort{ngd}-based optimization scheme.

\subsection{Continuous limits of Sequential Metric-aware optimization}

In a previous \Cref{sec:duality-qhbm}, we derived a gradient flow vector field which is the continuum analogue of our mirror descent algorithm update. In this section, we show how this flow can be leveraged to traverse the geometry of a given task space, for both \gls{meta-vqt} and \gls{qspl}/\gls{qvartz}.

In the previous subsection \ref{sec:task_geo_basic}, we have covered the basics of task-space geometry paths, we now delve into how one could leverage this picture to find how to continuously traverse such a geometry using variational models, without necessarily resorting to sequential optimization (see \eqref{eq:flowing_model_space}). As the path is simply the integral curve of the tangent vector, we seek to derive the pulled back representation of the task parameter space tangent vector in model space coordinates.

To achieve this, as we will show below, we can take a continuum limit of our sequence of losses between our models and target states along a path to create a loss functional over a continuum of parameter spaces along the path. We can then derive a functional variational principle to find the function minimizing this functional, this function will be the model parameter space representation of the target path. From this curve, we can derive the path tangent vector representation which we are after.

Let us starts with a generic derivation for a semi-generic choice of loss function, then specialize our expression to the relative entropic losses that appear in \gls{qspl} and for \gls{meta-vqt} for a Hamiltonian parameter sweep. 

As \gls{qvartz} is simply a recursive rather than sequential optimization of time evolution representation compared to \gls{qspl}, they inherently both share the same target path tangent vector representation, we will thus focus on deriving it for \gls{qspl}, as it is a sequential optimization task like \gls{meta-vqt} and shares many common elements to its optimal path representation.

\subsubsection{Continuous Variational Principle for Sequential Metric-aware optimization}

Let us derive the model parameter space representation of the optimal path for a generic contrast function loss. We can start from our definitions in section \ref{sec:gen_seq} and take the continuum limit.

Suppose we have a sequence of target states $\{\hat{\sigma}_{\bm{\Lambda}({\tau_{k}})}\}_{k=1}^M$ given by a partition $\{\tau_{k}\}_{k=1}^M\subset [0,T]$ of a discretized parametric path  $ \bm{\Lambda}(\tau)$ in task parameter space $\bm{\Lambda}\in \mathbb{R}^R$. For discrete sequence learning, we wish to obtain a sequence of optimal \gls{qhbm} parameters so that each $\hat{\rho}_{\bm{\Omega}^{*}(\tau_k)} \approx \hat{\sigma}_{\bm{\Lambda}({\tau_{k}})}$ for each $k$. Hence, we can have a collection of loss functions  
\begin{align}\label{eq:chained_path}
  \bm{\Omega}^*(\tau_k)  = \underset{\{\bm{\Omega}(\tau_k))\}}{\arg\min} \  \mathcal{L}_{\bm{\Lambda}(\tau_k)}(\bm{\Omega}(\tau_k)),\qquad\forall k \in \{1, \cdots, M\},
\end{align}
given a choice of loss $\mathcal{L}_{\bm{\Lambda}(\tau_k)}(\bm{\Omega}(\tau_k))$ between  $\hat{\rho}_{\bm{\Omega}^{*}(\tau_k)}$ and $\hat{\sigma}_{\bm{\Lambda}({\tau_{k}})}$. In general, this loss can be any sort of contrastive function $\Phi$ which compares the two density operators, as we have seen throughout this text. For our purposes in this section, we focus on the choices of loss being relative entropy, as is used for \acrshort{meta-vqt} (\Cref{sec:meta-vqt}) and \acrshort{qspl} (\Cref{sec:meta-qmhl}). Recall that for \acrshort{meta-vqt}, we minimize sequentially the forwards quantum relative entropy between model and target state; $\mathcal{L}_{\bm{\Lambda}(\tau_k)}(\bm{\Omega}(\tau_k)) = D(\hat{\rho}_{\bm{\Omega}(\tau_k)}\Vert \hat{\sigma}_{\bm{\Lambda}(\tau_{k})})$, while in \acrshort{qspl}, we minimize the backwards quantum relative entropy between model and target state; $\mathcal{L}_{\bm{\Lambda}(\tau_k)}(\bm{\Omega}(\tau_k)) = D(\hat{\sigma}_{\bm{\Lambda}(\tau_{k})}\Vert \hat{\rho}_{\bm{\Omega}(\tau_k)})$. 

Notice we made the loss $\mathcal{L}_{\bm{\Lambda}(\tau_k)}(\bm{\Omega}(\tau_k))$ dependence on the target path parameter space position $\bm{\Lambda}(\tau_k)$ explicit in \eqref{eq:chained_path}. This will come in useful for the extension to the continuum that follows.

First, notice that minimizing each of the losses sequentially is equivalent to minimizing the sum of the losses along the sequence:
\begin{align}
  \{\bm{\Omega}^*(\tau_k)\}_{k=1}^M =\underset{\{\bm{\Omega}(\tau_k)\}_{k=1}^M}{\arg\min} \left\{  \mathcal{L}_{\bm{\Lambda}(\tau_k)}(\bm{\Omega}(\tau_k))\right\}_{k=1}^{M}=  \underset{\{\bm{\Omega}(\tau_k)\}_{k=1}^M}{\arg\min} \sum_{k=1}^{M} 
  \mathcal{L}_{\bm{\Lambda}(\tau_k)}(\bm{\Omega}(\tau_k))
.\end{align}

To go from sequential optimization for discrete sequences to continuous-path sequential optimization, we will assume infinitesimal step sizes $\Delta \tau_k = \tau_k-\tau_{k-1}$ for the path time $\tau$ between the steps in the sequence. 
We can take the path total loss summed over sequence elements, equivalently minimize this sum with each summand multiplied by $\Delta \tau_k>0$ as each summand is nonnegative\footnote{this follows from the fact that our intended choice of losses are relative entropies, and these are nonnegative.} and take the continuum (infinite partition; $M\rightarrow \infty$) limit:

\begin{align}
  \{\bm{\Omega}^*(\tau_k)\}_{k=1}^M = \underset{\{\bm{\Omega}(\tau_k)\}_{k=1}^M}{\arg\min} \sum_{k=1}^{M} 
  \mathcal{L}_{\bm{\Lambda}(\tau_k)}(\bm{\Omega}(\tau_k))= \underset{\{\bm{\Omega}(\tau_k)\}_k}{\arg\min} \sum_{k=1}^{M} \Delta\tau_k\,
  \mathcal{L}_{\bm{\Lambda}(\tau_k)}(\bm{\Omega}(\tau_k))\overset{M\rightarrow\infty}{\longrightarrow} \underset{\{\bm{\Omega}(\tau)\}}{\arg\min} \int_0^T \mathrm{d}\tau  \mathcal{L}_{\bm{\Lambda}(\tau)}(\bm{\Omega}(\tau))
,\end{align}
which we can recognize as a line integral of the loss density functional of a continuous family of models along the target manifold path. We are thus looking for an optimal path parameter function $\bm{\Omega}^*:[0,T]\rightarrow\mathbb{R}$, which minimizes the integrated path loss functional:

\begin{equation}
\bm{\Omega}_{\bm{\Lambda}}^*(\tau) \equiv \underset{\bm{\Omega}(\tau)}{\arg\min\, } \bar{\mathcal{L}}[\bm{\Omega},\bm{\Lambda}] , \quad \bar{\mathcal{L}}[\bm{\Omega},\bm{\Lambda}] = \int_0^T \mathrm{d}\tau  \mathcal{L}_{\bm{\Lambda}(\tau)}(\bm{\Omega}(\tau)).    
\end{equation}

Now that we have phrased the problem as functional minimization, we must use the calculus of variations to define our optimal condition. At the optimum, the path loss functional is extremized when the functional derivative vanishes, 
\begin{equation}\label{eq:var_free_eng-2}\frac{\delta \bar{\mathcal{L}}[\bm{\Omega},\bm{\Lambda}]}{\delta \bm{\Omega}} \overset{!}{=} 0 \implies \left[\partial_{\Omega_k}\mathcal{L}_{\bm{\Lambda}(\tau)}(\bm{\Omega})\right]_{\bm{\Omega} =\bm{\Omega}^*(\tau)} \overset{!}{=} 0\quad  \forall \, k,\tau \end{equation}
the left hand side denotes a functional derivative, and the right hand side can be considered to be a set of Euler-Lagrange equations from this variational principle, and simply state that the gradient of loss function of our model with respect to the target state must be minimized at all points along the trajectory. 

Let us write the Euler-Lagrange condition as a multi-dimensional equilibrium condition, consider the following optimum constraint function define in terms of the target manifold space and variational coordinates, $\bm{F}:\mathbb{R}^D\times \mathbb{R}^T \rightarrow \mathbb{R}^D$ where 
\begin{equation}\bm{F}(\bm{\Omega}^*_{\bm{\Lambda}},\bm{\Lambda}) \equiv  \left[\partial_{\bm{\Omega}}\mathcal{L}_{\bm{\Lambda}}(\bm{\Omega})\right]_{\bm{\Omega} =\bm{\Omega}^*_{\bm{\Lambda}}} \overset{!}{=}\bm{0},\end{equation}
where \(  \bm{\Omega}^*_{\bm{\Lambda}}\) is the optimum of the loss \(\mathcal{L}_{\bm{\Lambda}}(\bm{\Omega})\) and $\partial_{\bm{\Omega}}$ is a slight abuse of notation to denote the Jacobian with respect to model (variational) coordinates; in this case as this operator is applied on a scalar it is a gradient vector. For any target path in the target manifold space $\bm{\Lambda}(\tau)$, then by the chain rule, 
\begin{equation}\partial_\tau \bm{F}(\bm{\Omega}^*_{\bm{\Lambda}(\tau)},\bm{\Lambda}(\tau))  = [\partial_{\bm{\Omega}}\bm{F}(\bm{\Omega},\bm{\Lambda}(\tau))]_{\bm{\Omega} = \bm{\Omega}^*_{\bm{\Lambda}(\tau)}}\cdot[\partial_{\bm{\Lambda}}\bm{\Omega}^*_{\bm{\Lambda}}]_{\bm{\Lambda} = \bm{\Lambda}(\tau)} \cdot \partial_\tau\bm{\Lambda}(\tau) +   [\partial_{\bm{\Lambda}}\bm{F}(\bm{\Omega}^*_{\bm{\Lambda}},\bm{\Lambda}(\tau))]_{\bm{\Lambda} = \bm{\Lambda}(\tau)} \cdot \partial_\tau\bm{\Lambda}(\tau) \overset{!}{=} 0\end{equation}
given that we have an expression multiplying the tangent vector to the path, we can turn this into a vector-valued equation again:
\begin{equation}[\partial_{\bm{\Omega}}\bm{F}(\bm{\Omega},\bm{\Lambda}(\tau))]_{\bm{\Omega} = \bm{\Omega}^*_{\bm{\Lambda}(\tau)}}\cdot[\partial_{\bm{\Lambda}}\bm{\Omega}^*_{\bm{\Lambda}}]_{\bm{\Lambda} = \bm{\Lambda}(\tau)} +   [\partial_{\bm{\Lambda}}\bm{F}(\bm{\Omega}^*_{\bm{\Lambda}},\bm{\Lambda}(\tau))]_{\bm{\Lambda} = \bm{\Lambda}(\tau)}\overset{!}{=}\bm{0} \end{equation}

we recover the equations of implicit differentiation for bi-level optimization problems \cite{blondel2021efficient}. We can rewrite this equation in order to get the Jacobian of the optimal variational coordinates in terms of the target manifold coordinates:
\begin{equation}[\partial_{\bm{\Lambda}}\bm{\Omega}^*_{\bm{\Lambda}}]_{\bm{\Lambda} = \bm{\Lambda}(\tau)} = -([\partial_{\bm{\Omega}}\bm{F}(\bm{\Omega},\bm{\Lambda}(\tau))]_{\bm{\Omega} = \bm{\Omega}^*_{\bm{\Lambda}(\tau)}})^{-1}\cdot[\partial_{\bm{\Lambda}}\bm{F}(\bm{\Omega}^*_{\bm{\Lambda}},\bm{\Lambda}(\tau))]_{\bm{\Lambda} = \bm{\Lambda}(\tau)} \end{equation}
we can then plug in our variational principle condition, i.e. the fact that the $\bm{F}$ function is the gradient of the loss: 
\begin{equation}\label{eq:tan_vec_rep}
\partial_{\bm{\Lambda}}\bm{\Omega}^*_{\bm{\Lambda}} = -([\partial_{\bm{\Omega}}\partial_{\bm{\Omega}}\mathcal{L}_{\bm{\Lambda}}(\bm{\Omega})]_{\bm{\Omega} = \bm{\Omega}^*_{\bm{\Lambda}}})^{-1}\cdot[\partial_{\bm{\Lambda}}\partial_{\bm{\Omega}}\mathcal{L}_{\bm{\Lambda}}(\bm{\Omega})]_{\bm{\Omega} = \bm{\Omega}^*_{\bm{\Lambda}}}    
\end{equation}
we see we get the inverse Hessian of the loss with respect to model parameters contracted with the Jacobian of the loss gradient. 

Let us now establish a relation of the above expression to the notion of change of coordinates in differential geometry. For both \acrshort{meta-vqt}; $\mathcal{L}_{\bm{\Lambda}}(\bm{\Omega}) = D(\hat{\rho}_{\bm{\Omega}}\Vert \hat{\sigma}_{\bm{\Lambda}})$, and \acrshort{qspl}; $\mathcal{L}_{\bm{\Lambda}}(\bm{\Omega}) = D(\hat{\sigma}_{\bm{\Lambda}}\Vert \hat{\rho}_{\bm{\Omega}})$, the loss is a relative entropy. As such, upon convergence of the optimization, assuming that at a given path time $\tau$ of interest, we truly have $\hat{\sigma}_{\bm{\Lambda}(\tau)}\approx \hat{\rho}_{\bm{\Omega}_{\bm{\Lambda}(\tau)}^*}$ and a convergence to the parameter space optimum, then the inverse Hessian of the loss should be exactly equal to the inverse \gls{bkm} metric\footnote{In practice, this only holds if the optimal parameters found by the optimizer have converged exactly onto the local optimum. Otherwise, it is a close approximation.}:
\begin{equation}[\partial_{\bm{\Omega}}\partial_{\bm{\Omega}}\mathcal{L}_{\bm{\Lambda}}(\bm{\Omega})]_{\bm{\Omega} = \bm{\Omega}^*_{\bm{\Lambda}}} = [\{\partial_{\bm{\Omega}}\partial_{\bm{\Omega}}D(\hat{\rho}_{\bm{\Omega}}\Vert \hat{\sigma}_{\bm{\Lambda}}) \ \lor \ \partial_{\bm{\Omega}}\partial_{\bm{\Omega}}D(\hat{\sigma}_{\bm{\Lambda}}\Vert\hat{\rho}_{\bm{\Omega}}) \}]_{\bm{\Omega} = \bm{\Omega}^*_{\bm{\Lambda}}}= [\partial_{\bm{\Omega}}\partial_{\bm{\Omega}}D(\hat{\rho}_{\bm{\Omega}}\Vert \hat{\rho}_{\bm{\Omega}^*_{\bm{\Lambda}}})]_{\bm{\Omega} = \bm{\Omega}^*_{\bm{\Lambda}}} = \mathcal{I}(\bm{\Omega}^*_{\bm{\Lambda}}),\end{equation}
where $\lor$ here is a logical or to encompass both \acrshort{meta-vqt} and for \acrshort{qspl} loss cases, and $\mathcal{I}(\bm{\Omega})$ is the \acrshort{bkm} metric for model coordinates used throughout this paper. Note we leveraged the symmetry of the metric near the optimum. The above yields our final expression for the Jacobian given by:
\begin{equation}\label{eq:Jacobian}\partial_{\bm{\Lambda}}\bm{\Omega}^*_{\bm{\Lambda}} = -([\partial_{\bm{\Omega}}\partial_{\bm{\Omega}}\mathcal{L}_{\bm{\Lambda}}(\bm{\Omega})]_{\bm{\Omega} = \bm{\Omega}^*_{\bm{\Lambda}}})^{-1}\cdot[\partial_{\bm{\Lambda}}\partial_{\bm{\Omega}}\mathcal{L}_{\bm{\Lambda}}(\bm{\Omega})]_{\bm{\Omega} = \bm{\Omega}^*_{\bm{\Lambda}}}  = -(\mathcal{I}(\bm{\Omega}^*_{\bm{\Lambda}}))^{-1}\cdot[\partial_{\bm{\Lambda}}\partial_{\bm{\Omega}}\mathcal{L}_{\bm{\Lambda}}(\bm{\Omega})]_{\bm{\Omega} = \bm{\Omega}^*_{\bm{\Lambda}}} .\end{equation}

We see that in terms of differential geometry, this can be interpreted as a change of coordinate representation of the loss gradient vector from the task space geometry coordinate chart to the model coordinate chart. The inverse metric appears as the above is the one-form rather than tangent vector in this model coordinate dual tangent space.

This provides us our generic expression for the model coordinate representation of the task path tangent vector. Estimating this tangent vector representation in principle allows one to circumvent the need to perform sequential optimization, one could in principle obtain the path representation in model parameter space by lifting equation \ref{eq:flowing}

 \begin{equation}\label{eq:flowing_model_space}
   \bm{\Omega}^*_{\bm{\Lambda}}(\tau) \equiv \int_0^\tau d\tau'\, \partial_{\tau'} \bm{\Omega}^*_{\bm{\Lambda}}(\tau') = \int_0^\tau d\tau'\,   [\partial_{\bm{\Lambda}}\bm{\Omega}^*_{\bm{\Lambda}}]_{\bm{\Lambda} = \bm{\Lambda}(\tau')} \cdot \partial_{\tau'}\bm{\Lambda}(\tau')
 \end{equation}

We can now examine how this generic form can be specialized for \acrshort{meta-vqt} and for \acrshort{qspl} and see how this change of coordinate Jacobian $\partial_{\bm{\Lambda}}\bm{\Omega}^*_{\bm{\Lambda}}$ can be directly estimated in some cases. We do so in the following subsections.

\subsubsection{Parametric Hamiltonian flow and Continuous \texorpdfstring{\gls{meta-vqt}}{Meta-VQT} variational principle}

Let us focus first on the case of \gls{meta-vqt} for a general parametric path in the space of Hamiltonians for which we would like to compute the thermal states.

To go from \gls{meta-vqt} for sequences to continuous-path \gls{meta-vqt}, we consider a scenario where we have a countinuous task path is in the space of parameterized Hamiltonians $\hat{H}_{\bm{\Lambda}}$ and our sequence of target states is a continuous family of thermal (Gibbs) states \begin{equation}\hat{\sigma}_{\bm{\Lambda}(\tau)} = e^{-\hat{H}_{\bm{\Lambda}(\tau)}}/\mathcal{Z}_{\bm{\Lambda}(\tau)},\end{equation} where
\(\bm{\Lambda}(\tau): \mathbb{R}\rightarrow \mathbb{R}^R\) is a our path function in the task parameterization space of dimension $R$, which can include both coldness (inverse temperature; $\beta(\tau)$, as is the case for imaginary time evolution) and/or Hamiltonian parameters. We seek to find a set of optimal parameters $\bm{\Omega}^*_{\bm{\Lambda}(\tau)}$ such that 
\begin{equation}\hat{\rho}_{\bm{\Omega}^*_{\bm{\Lambda}(\tau)}} \approx \hat{\sigma}_{\bm{\Lambda}(\tau)}, \quad \forall \tau\in[0,T].\end{equation}

Do do so, we can simply follow the integral curve to the task space tangent vector. To find its representation in model parameter space, we can simply put the quantum relative entropy as our loss in  \eqref{eq:Jacobian}, \begin{equation}\mathcal{L}_{\bm{\Lambda}}(\bm{\Omega}) = D(\hat{\rho}_{\bm{\Omega}}\Vert \hat{\sigma}_{\bm{\Lambda}}).\end{equation}

Now, inputting this into equation \eqref{eq:Jacobian}, we obtain, 
\begin{equation}\label{eq:jacobian_meta-vqt1}\partial_{\bm{\Lambda}}\bm{\Omega}^*_{\bm{\Lambda}}  = -(\mathcal{I}(\bm{\Omega}^*_{\bm{\Lambda}}))^{-1}\cdot[\partial_{\bm{\Lambda}}\partial_{\bm{\Omega}}D(\hat{\rho}_{\bm{\Omega}}\Vert \hat{\sigma}_{\bm{\Lambda}})]_{\bm{\Omega} = \bm{\Omega}^*_{\bm{\Lambda}}} = -(\mathcal{I}(\bm{\Omega}^*_{\bm{\Lambda}}))^{-1}\cdot[\partial_{\bm{\Lambda}}\partial_{\bm{\Omega}}\mathcal{F}_{\hat{H}_{\bm{\Lambda}}}(\bm{\Omega})]_{\bm{\Omega} = \bm{\Omega}^*_{\bm{\Lambda}}}  ,\end{equation}
where we used the fact that any gradient of relative entropy with respect to the model is also equal to the gradient of the free energy, as such, we can insert the variational free energy   $\mathcal{F}_{\hat{H}_{\bm{\Lambda}}}(\bm{\Omega})$ of the state $\hat{\rho}_{\bm{\Omega}}$ with respect to the Hamiltonian $\hat{H}_{\bm{\Lambda}}$ as instead of the relative entropy loss, also known as the \gls{vqt} loss. 

We can expand the expression corresponding to this gradient more explicitly, 
\spliteq{\partial_{\bm{\Lambda}}\partial_{\bm{\Omega}}\mathcal{F}_{\hat{H}_{\bm{\Lambda}}}(\bm{\Omega}) &= \partial_{\bm{\Lambda}}\partial_{\bm{\Omega}}\tr\left[\hat\rho_{\bm\Omega}{\hat{H}_{\bm{\Lambda}}}\right]  -\cancel{\partial_{\bm{\Lambda}}\partial_{\bm{\Omega}}S(\hat\rho_{\bm\Omega})} = \tr\left[(\partial_{\bm{\Omega}}\hat\rho_{\bm\Omega})(\partial_{\bm{\Lambda}}{\hat{H}_{\bm{\Lambda}}})\right]. \\ 
}
we see that it is simply the contraction of the model gradient of the state contracted with the Hamiltonian gradient.

In order to estimate these terms directly, for each component of the gradient observable $(\partial_{\bm{\Lambda}}{\hat{H}_{\bm{\Lambda}}})_j$, this is simply standard gradient of a quantum state observable expectation which we encounter in VQT calculations. Refer to \Cref{sec:review-qhbm} for unbiased estimators of the gradient of an expectation value with respect to a \gls{qhbm}, or to the \href{https://github.com/google/qhbm-library}{QHBM library} for open-source implementations of these gradient estimators.

\subsubsection{Time evolution flow and Continuous meta-temporal variational principle for \texorpdfstring{\gls{qspl} and \gls{qvartz}}{QSPL and QVARTZ}}\label{sec:cont_qspl}

Consider the scenario where our continuous-limit dataset of quantum states is given by the time evolution of a quantum state under open quantum system time evolution, i.e. the problem of \gls{qspl}. That is, a dataset consisting of continuous family of states of time-evolved quantum states according to some form of open or closed Markovian\footnote{Again, our techniques generalize straightforwardly to non-Markovian quantum-stochastic processes, see \Cref{app:qsto_dyn} for background} quantum system evolution:
\(    \hat{\sigma}_{t} = \mathcal{V}_{t,0}(\hat{\sigma}_0).\)
We can learn a continuous family of models which approximate each state along the continuous rollout sequence, i.e., we aim to a function of optimal parameters \(\{ \bm{\Omega}^{*}_{t} \}_{t\in[0,T]}\) such that \(\hat{\sigma}_{t} \approx \hat{\rho}_{\bm{\Omega}^{*}_{t}}\) for all \({t\in[0,T]}\).

To relate continuous \gls{qspl} to our generic path parameter space is quite simple: the task path parameter space is single-dimensional; $R=1$, and the path is simply the linear evolution of time;
\begin{equation}\label{eq:time_path_simple}
    \bm{\Lambda}(\tau) = \tau \equiv t,
\end{equation}
we identify our path time $\tau$ as real time $t$ of the time evolution to avoid any possible confusions with imaginary time evolution, which is a different application.

To go from \gls{qspl} for discrete sequences to continuous \acrshort{qspl}, we can simply put the backwards quantum relative entropy as our loss in  \eqref{eq:Jacobian}, \begin{equation}\mathcal{L}_{t}(\bm{\Omega}) = \mathcal{L}_{\bm{\Lambda}(\tau)}(\bm{\Omega}) =   D( \hat{\sigma}_{\bm{\Lambda}(\tau)}\Vert \hat{\rho}_{\bm{\Omega}}) = D( \hat{\sigma}_{t}\Vert \hat{\rho}_{\bm{\Omega}}).\end{equation}

This yields an expression for the Jacobian \eqref{eq:Jacobian} given by:
\begin{equation}\partial_t\bm{\Omega}^*_{t} =  - 
\mathcal{I}(\bm{\Omega}^*_{t})^{-1}\cdot[\partial_t\partial_{\bm{\Omega}}D( \hat{\sigma}_{t}\Vert \hat{\rho}_{\bm{\Omega}}) ]_{\bm{\Omega} = \bm{\Omega}^*_{t}} =
 - \mathcal{I}(\bm{\Omega}^*_{t})^{-1}\cdot[\partial_t\partial_{\bm{\Omega}}\mathcal{X}_t(\bm{\Omega}) ]_{\bm{\Omega} = \bm{\Omega}^*_{t}}.
 \end{equation}

Here, we use the definition of cross entropy from \eqref{eq:time_xent_def}, as the gradient of the backwards relative entropy with respect to model parameters is equal to the gradient of the cross entropy loss, also known as the \gls{qmhl} loss.

We can expand the expression corresponding to this gradient more explicitly, using our expression of the time derivative of states as Lindblad superoperators from \eqref{eq:lindblad}:
\spliteq{\partial_t\partial_{\bm{\Omega}}\mathcal{X}_t(\bm{\Omega})  = - \tr[\partial_t\hat{\sigma}_{t}\partial_{\bm{\Omega}}\log \hat{\rho}_{\bm{\Omega}}] =  - \tr[\mathcal{L}_t(\hat{\sigma}_{t}) \partial_{\bm{\Omega}}\log \hat{\rho}_{\bm{\Omega}}],
}
yielding a total time derivative of the optimal parameters given by:
\begin{equation}\partial_t\bm{\Omega}^*_{t}  =
  \mathcal{I}(\bm{\Omega}^*_{t})^{-1}\cdot \tr[\mathcal{L}_t(\hat{\sigma}_{t}) \partial_{\bm{\Omega}}\log \hat{\rho}_{\bm{\Omega}}].
 \end{equation}
we can interpret this expression as a model coordinate space pulled back representation of the generator of the time evolution path, i.e. the Lindbladian applied on the current state at a give time contracted the operator-valued score vector basis of our model. Note that in practice, applying the Lindbladian directly on a state and evaluating its operator contraction with other Hermitian operators can be difficult to do directly. If we were able his is simply apply the Linbladian superoperator directly onto the state $\hat{\sigma}_{t}$, then we could use this state as data for a standard \gls{qmhl} gradient estimator\footnote{
Refer to \Cref{sec:review-qhbm} for unbiased estimators of the gradient of an expectation value with respect to a \gls{qhbm}, or to the \href{https://github.com/google/qhbm-library}{QHBM library} for open-source implementations of these gradient estimators.
} in order to obtain this estimate of the time derivative of the optimum. In practice, it is best to use finite differences to estimate this derivative, which amounts to our scheme of infinitesimal steps along the time evolution path (which is what is used for both \gls{qspl} and \gls{qvartz} for discrete sequences of time evolutions).

\subsection{Metric Variation Bounds}

We begin the work of quantifying the required fineness of discretization for the sequence tasks.  We will provide a bound on the step size in terms of the difference in quantum Fisher information; we also derive the model parameter space representation of the path tangent vector. This is related to the rate of entropy production~\cite{deffner2011nonequilibrium,spohn1978entropy,callens2004quantum} for open system time evolutions, while singularities in this tangent vector for Gibbs state paths indicate the presence of phase transitions~\cite{PhysRevA.80.062326,PhysRevLett.104.020502}. Both of these areas are primed for follow-up theoretical and numerical investigation.

To begin, we can use our expression of the directional derivative of the optimal parameters in order to bound the third order term in the metric potential function, or interpreted alternatively, bound the metric variation from equation \eqref{eq:metric_Var}. We want to show that

\begin{equation}
  \lVert \mathcal{I}(\bm{\Omega}^*(\tau_{k+1})) - \mathcal{I}(\bm{ \Omega}^*(\tau_{k})) \rVert_F \leq L
\end{equation}
can be adjusted to be small (with tight control of the constant $L$) for appropriate choice of path step size $\Delta \tau_k \equiv \tau_{k+1}-\tau_{k}$.

We begin by looking back at our metric potential Taylor expansion, letting \(\bm{\delta} \equiv \bm \Omega - \bm\Omega^*\),

\spliteq{
   \Phi( \bm \Omega,\bm \Omega^*) &= \frac{1}{2}  \bm\delta^\perp\cdot\mathcal{I}({\bm\Omega})\cdot\bm\delta+ \mathcal{O}(\lVert\bm\delta\rVert^3)  = \frac{1}{2} \sum_{jk} \left[ \delta_j\delta_k\mathcal{I}_{jk}({\bm\Omega})+ \tfrac{1}{3}\sum_l \delta_j\delta_k\delta_l\partial_{\Omega_l}\mathcal{I}_{jk}({\bm\Omega}) \right]+ \mathcal{O}(\lVert\bm\delta\rVert^4)\\
   &=  \frac{1}{2}  \bm\delta^\perp\cdot[\mathcal{I}({\bm\Omega})+\tfrac{1}{3}\bm{\delta}^\perp\cdot \nabla_{\bm{\Omega}}\mathcal{I}({\bm\Omega})]\cdot\bm\delta+ \mathcal{O}(\lVert\bm\delta\rVert^4)
.}
Taking two derivatives with respect to $\bm{\delta}$ in the above, using the three-way symmetry of the gradient of the metric, we see the information matrix with its leading order term (ignoring the $ \mathcal{O}(\lVert\bm\delta\rVert^4)$ term) is given by
\begin{equation}
   \mathcal{I}({\bm\Omega})+ \mathcal{E}(\bm{\Omega}) \equiv  \mathcal{I}({\bm\Omega})+\bm{\delta}^\perp\cdot \nabla_{\bm{\Omega}}\mathcal{I}({\bm\Omega}) 
\end{equation}
where $\mathcal{E}(\bm{\Omega})$ is our parameter-dependent error term which we would like to bound. By the Cauchy-Schwarz inequality, we can have the element-wise inequality:
\begin{equation}
    \lvert \mathcal{E}_{jk}(\bm{\Omega})\rvert = \lVert\bm{\delta}^\perp\cdot \nabla_{\bm{\Omega}}\mathcal{I}_{jk}({\bm\Omega}) \rVert_2 \leq \lVert\bm{\delta}\rVert_2 \lVert \nabla_{\bm{\Omega}}\mathcal{I}_{jk}({\bm\Omega}) \rVert_2 = \lVert\bm\Omega - \bm\Omega^*\rVert_2 \lVert \nabla_{\bm{\Omega}}\mathcal{I}_{jk}({\bm\Omega}) \rVert_2 
\end{equation}
Let us instead use a more sophisticated norm. Considering the Jacobian of the metric as a 3-tensor, let us define the following norm
\begin{equation}\lVert \partial_{\bm{\Omega}}\mathcal{I}({\bm\Omega}) \rVert_2 \equiv  \sup_{\bm{\epsilon}: \lVert\bm\epsilon\rVert =1}\lvert \textstyle\sum_{jkl}\epsilon_j\epsilon_k\epsilon_l \partial_{\Omega_l}\mathcal{I}_{jk}({\bm\Omega}) \rvert \end{equation}
as the tensor generalization of the 2-norm, this is given by the largest singular value of the 3-tensor's canonical polyadic decomposition \cite{hitchcock}. Note that since this is the Jacobian of a Hessian, this tensor is symmetric across permutations of its indices.

Clearly,
\begin{equation}
    \lVert \mathcal{E}(\bm{\Omega})\rVert_F \leq P\lVert\bm\Omega - \bm\Omega^*\rVert_2\lVert \partial_{\bm{\Omega}}\mathcal{I}({\bm\Omega}) \rVert_2
\end{equation}
where $P$ is the dimension of our parameter space.

So 

\spliteq{
 \lVert \mathcal{I}(\bm{\Omega}) - \mathcal{I}(\bm{ \Omega}^*) \rVert_F \leq \lVert \mathcal{E}(\bm{\Omega}) \rVert_F  +\mathcal{O}(\lVert\bm\delta\rVert^2) \leq P\lVert\bm\Omega - \bm\Omega^*\rVert_2\lVert \partial_{\bm{\Omega}}\mathcal{I}({\bm\Omega}) \rVert_2+\mathcal{O}(\lVert\bm\delta\rVert^2) 
}
ignoring the $\mathcal{O}(\lVert\bm\delta\rVert^2)$ error term, we have
\begin{equation}
    \lVert \mathcal{I}(\bm{\Omega}) - \mathcal{I}(\bm{ \Omega}^*) \rVert_F \leq P\lVert\bm\Omega - \bm\Omega^*\rVert_2\lVert \partial_{\bm{\Omega}}\mathcal{I}({\bm\Omega}) \rVert_2.
\end{equation}
Plugging this metric variation expression, we have
\begin{equation}
           \lVert \mathcal{I}(\bm{\Omega}^*(\tau_{k+1})) - \mathcal{I}(\bm{ \Omega}^*(\tau_k)) \rVert_F  \leq P\lVert\bm{\Omega}^*(\tau_{k+1}) - P\tau_k)\rVert_2\lVert\left. \partial_{\bm{\Omega}}\mathcal{I}({\bm\Omega})\right|_{\bm{\Omega}=\bm{\Omega}^*(\tau_k)} \rVert_2.
\end{equation}
this leaves us with a metric variation bound 
\begin{equation}
    \lVert \mathcal{I}(\bm{\Omega}^*(\tau_{k+1})) - \mathcal{I}(\bm{ \Omega}^*(\tau_k)) \rVert_F \leq P\lVert\bm\Omega^*(\tau_{k+1}) - \bm\Omega^*(\tau_k)\rVert_2\lVert\left. \partial_{\bm{\Omega}}\mathcal{I}({\bm\Omega})\right|_{\bm{\Omega}=\bm{\Omega}^*(\tau_k)} \rVert_2.
\end{equation}
We can bound this distance between optima using our expressions for the tangent vectors of the paths in task space,
\begin{equation}
    \bm\Omega^*(\tau_{k+1}) - \bm\Omega^*(\tau_k) =\Delta\tau_k [\partial_{\bm{\Lambda}}\bm{\Omega}^*_{\bm{\Lambda}}]_{\bm{\Lambda} =  \bm{\Lambda}(\tau_k)} \cdot \partial_\tau\bm{\Lambda}(\tau_k) + \mathcal{O}(\Delta\tau_k^2)  
\end{equation}
where $\Delta\tau_k \equiv \tau_{k+1} - \tau_k$ is the path time difference for the outer loop index $k$. Note that the second order term here would depend on the path curvature. We can bound the two-norm of the above, 
\begin{equation}
    \lVert\bm\Omega^*(\tau_{k+1}) - \bm\Omega^*(\tau_k)\rVert_2 \leq\Delta\tau_k \lVert[\partial_{\bm{\Lambda}}\bm{\Omega}^*_{\bm{\Lambda}}]_{\bm{\Lambda} =  \bm{\Lambda}(\tau_k)}\rVert_2  \lVert \partial_\tau\bm{\Lambda}(\tau_k)\rVert_2  + \mathcal{O}(\Delta\tau_k^2)  
\end{equation}
putting it all together, we have
\begin{equation}
   \lVert \mathcal{I}(\bm{\Omega}^*(\tau_{k+1})) - \mathcal{I}(\bm{ \Omega}^*(\tau_k)) \rVert_F  \leq P\Delta\tau_k \lVert[\partial_{\bm{\Lambda}}\bm{\Omega}^*_{\bm{\Lambda}}]_{\bm{\Lambda} =  \bm{\Lambda}(\tau_k)}\rVert_2  \lVert \partial_\tau\bm{\Lambda}(\tau_k)\rVert_2\lVert\left. \partial_{\bm{\Omega}}\mathcal{I}({\bm\Omega})\right|_{\bm{\Omega}=\bm{\Omega}^*(\tau_k)} \rVert_2  + \mathcal{O}(\Delta\tau_k^2) ,
\end{equation}
it is thus clear that, assuming the gradient terms are non-divergent, for any $\varepsilon>0$, there exists a choice of step size $\Delta \tau_k$ such that $\lVert \mathcal{I}(\bm{\Omega}^*(\tau_{k+1})) - \mathcal{I}(\bm{ \Omega}^*(\tau_k)) \rVert_F < \varepsilon$.

It is interesting to consider how the terms could diverge, let us bound the tangent vector norm multiplicative term and see when the latter could be the case. For \acrshort{meta-vqt}, 
\begin{equation}
    \lVert[\partial_{\bm{\Lambda}}\bm{\Omega}^*_{\bm{\Lambda}}]_{\bm{\Lambda} =  \bm{\Lambda}(\tau_k)}\rVert_2  \leq  \lVert  [\partial_{\bm{\Omega}}\partial_{\bm{\Omega}}D(\hat{\rho}_{\bm{\Omega}}\Vert \hat{\sigma}_{\bm{\Lambda}})  ]^{-1}_{\bm{\Omega} = \bm{\Omega}^*_{\bm{\Lambda}}}\rVert_2\lVert\partial_{\bm{\Lambda}}\partial_{\bm{\Omega}}\mathcal{F}_{\hat{H}_{\bm{\Lambda}}}(\bm{\Omega})\rVert_2
\end{equation}
one way this could diverge is if the minimum singular value of the Hessian of the relative entropy is null, or if the gradient of the Jacobian of the free energy diverges. The latter could be a sign of phase transition \cite{PhysRevLett.104.020502}.

As for time evolution paths such as \acrshort{qspl} and \acrshort{qvartz},
\begin{equation}\label{eq:time_evo_bound1}
\lVert[\partial_{\bm{\Lambda}}\bm{\Omega}^*_{\bm{\Lambda}}]_{\bm{\Lambda} =  \bm{\Lambda}(\tau_k)}\rVert_2  \leq \lVert \left[\partial_{\bm{\Omega}}\partial_{\bm{\Omega}}\mathcal{X}_t(\bm{\Omega}) \right]_{\bm{\Omega} =\bm{\Omega}^*_{t}}^{-1}\rVert_2\lVert[\partial_t\partial_{\bm{\Omega}}\mathcal{X}_t(\bm{\Omega}) ]_{\bm{\Omega} = \bm{\Omega}^*_{t}}\rVert_2 =\lVert\left[\partial_{\bm{\Omega}}\partial_{\bm{\Omega}}\mathcal{X}_t(\bm{\Omega}) \right]_{\bm{\Omega} =\bm{\Omega}^*_{t}}^{-1}\rVert_2 \lVert\tr[\mathcal{L}_t(\hat{\sigma}_{t}) \partial_{\bm{\Omega}}\log \hat{\rho}_{\bm{\Omega}}]\rVert_2
\end{equation}
the first factor is approximately equal to the two-norm of the metric, while the second factor is more interesting, it is the two-norm of the gradient of the logarithm of the model, each element contracted with the target state mapped through the Liouvillian superoperator at time $t$. We can bound each element of the second factor further:
\spliteq{
    |\tr[\mathcal{L}_t(\hat{\sigma}_{t}) \partial_{\Omega_j}\log \hat{\rho}_{\bm\Omega}]|\leq \lVert\mathcal{L}_t(\hat{\sigma}_{t})  \rVert_{*} \lVert \partial_{\Omega_j}\log \hat{\rho}_{\bm\Omega}]\rVert_{*} &\leq \left( \lVert[\hat{H}(t),\hat{\sigma}_{t}(t)]\rVert_{*} + \lVert\mathcal{D}(\hat{\sigma}_{t})\rVert_{*}\right)\lVert \partial_{\Omega_j}\log \hat{\rho}_{\bm{\Omega}}\rVert_{*} 
}
where $\lVert \cdot\rVert_{*} $ denotes the norm induced by Hilbert-Schmidt operator inner product; \(\smash{\lVert \hat{A}\rVert_{*} \equiv \sqrt{\langle\!\langle \hat{A} \vert \hat{A} \rangle\!\rangle}}\), where \(\smash{\langle\!\langle \hat{A} \vert \hat{B} \rangle\!\rangle \equiv \text{tr}(\hat{A}^\dagger \hat{B})}\). This is also technically known as the Schatten 2-norm for quantum operator space \cite{watrous_2018}.\footnote{We use $\lVert \cdot\rVert_{*}$ here for norms on the space of operators (dimension $N^2$) instead of simply $\lVert \cdot\rVert_{2}$ in order to avoid ambiguity with norms defined in the parameter space (dimension $P$).} \footnote{Using this norm on the space of operators induced by the Hilbert-Schmidt inner product, we could define an induced norm on the space of \textit{superoperators}, given a superoperator $\mathcal{S}$, we have its norm given by
\begin{equation}\lVert\mathcal{S}\rVert_* \equiv \sup_{\hat{O}} \frac{\lVert \mathcal{S}[\hat{O}]\rVert_{*} }{\lVert \hat{O}\rVert_{*} },
\end{equation}
this is in fact the Schatten 2-norm for superoperators \cite{watrous_2018}.} Let us use the explicit form of the Master equation \eqref{eq:quantum-master} expression to create a more explicit bound in terms of jump operators, 
\begin{equation}
\Vert\mathcal{D}[\hat{\sigma}_{t}]\rVert_* \leq \textstyle\sum_{a=1}^{D^2}\lVert\hat{L}_a\hat{\sigma}_{t}\hat{L}_a^\dagger\rVert_*  +\lVert\hat{L}_a^\dagger\hat{L}_a\hat{\sigma}_{t}\rVert_* \leq \textstyle\sum_{a=1}^{D^2}2\lVert\hat{L}_a\rVert_*^2
\end{equation}
and we get a similar bound for the Hamiltonian commutator term:
\(
     \lVert[\hat{H}(t),\hat{\sigma}_{t}(t)] \rVert_*\leq 2\lVert\hat{H}(t) \rVert_*.
\)

This brings our total bound to
\begin{equation}
    \lVert\tr[\mathcal{L}_t(\hat{\sigma}_{t}) \partial_{\bm\Omega}\log \hat{\rho}_{\bm{\Omega}}]\rVert_2 \leq 2\left(\lVert\hat{H}(t) \rVert_*+\textstyle\sum_{a=1}^{D^2}\lVert\hat{L}_a\rVert_*^2\right) \lVert\lVert \partial_{\bm{\Omega}}\log \hat{\rho}_{\bm{\Omega}}\rVert_*\rVert_2
\end{equation}
we see \footnote{Note that here we did a slight abuse of notation with nested norms. Let us clarify here the notation, for a operator-valued vector $\bm{\hat{v}} = \{\hat{v}_j\}_j$, the 2-norm of the operator norm is given by $\lVert\lVert \bm{\hat{v}}\rVert_*\rVert_2 = \sqrt{\sum_j \lVert\hat{v}_j\rVert_*^2}.$} that this corresponds to the vector norm of the score vector's element-wise operator norm times the sum of the operator norms of the Hamiltonian and jump operator at time $t$. The above information may be useful in choosing step sizes $\Delta t$ in a way that is inversely proportional to the jump operator and Hamiltonian norms in order to keep the same guarantee of small metric variation. The above bound may also be related to the rate of entropy production over time in the system\footnote{Looking back at \Cref{eq:time_evo_bound1}, we also see that this term bounded above corresponds to the time derivative of the model-parameter gradient of the cross entropy at the optimum. If one were to drop the gradient $\partial_{\bm{\Omega}}$ in the above, one could obtain a bound on entropy production (as cross-entropy at the optimum is approximating the target state entropy) over time.}.

\section{Lagrange descent}

\begin{figure}
\begin{algorithm}[H]
	\begin{algorithmic}[1]
		\For {$j=1,2,\ldots$}
		    \State pick $\lambda_j > 0$
		    \State // inner loop to approximately solve \eqref{eq:ngd_fund}
		    \For {$k=1,2,\ldots,N$}
		    \State choose learning rate $\eta_k > 0$
		    \State evaluate $\nabla_{\delta}\mathcal{L}(\hat{\rho}_{\Omega+\delta})|_{\delta=\delta_k}$ (e.g., via methods in Appendix C)
		    \State evaluate $\nabla_{\delta}D(\hat{\rho}_{\Omega}\|\hat{\rho}_{\Omega+\delta})|_{\delta = \delta_k}$ via methods in Appendix E
		    \State update $\delta_{k+1} \gets  \delta_k-\eta_k\left(\nabla_{\delta}\mathcal{L}(\hat{\rho}_{\Omega+\delta})|_{\delta=\delta_k} + \lambda_j \nabla_{\delta}D(\hat{\rho}_{\Omega}\|\hat{\rho}_{\Omega+\delta})|_{\delta = \delta_k}\right)$
		    \EndFor
			\State update parameters: $\Omega_{j+1} \gets \Omega_j + \delta_{N+1}$
		\EndFor
	\end{algorithmic} 
	\caption{``Lagrange" descent as an alternative to mirror descent.} 
	\label{qplgd}
\end{algorithm}
\end{figure}

Let us consider an alternative way to get a similar descent behaviour to \gls{qpngd} without having to compute the actual metric itself (\Cref{qplgd}). Looking back at equation \eqref{eq:ngd_fund}, and assuming the quantum relative entropy as the constrast functional, we could simply minimize the objective in \eqref{eq:ngd_fund} with respect to $\bm\delta$ by a gradient descent update:

\begin{equation}\nabla_{\bm{\delta}}\mathcal{L}(\hat{\rho}_{\bm{\Omega}+\bm{\delta}}) + \lambda  \nabla_{\bm{\delta}} D(\hat{\rho}_{\bm{\Omega}}\Vert \hat{\rho}_{\bm{\Omega}+\bm{\delta}}).\end{equation}

The first term is just the gradient of our loss whereas the second term is just like a reverse relative entropy gradient where the data state is the model at the anchored value of $\bm{\Omega}$. By nesting a few iterations of gradient descent according to the above gradient and iteratively resetting the anchor point according to the optimum parameter space direction found from these iterations, $\bm{\Omega} \mapsto \bm{\Omega} + \bm{\Omega}^*$ we can get a form of natural gradient descent where we did not have to compute a matrix Hessian nor invert a matrix, two steps that can be computationally costly, as we have described.

In our numerical investigations, we found mirror descent to be superior to this approach.

\subsection{Convexity of Lagrange descent optimization}

\begin{proposition}
\label{apdx:prop:convexity-convergence}
Let $\mathcal{L} \in C^2(\mathbb{R}^n)$ and consider the Lagrange descent minimization problem \eqref{eq:ngd_fund}. Then, for all $\Omega\in\mathbb{R}^n$ such that $\hat\rho_{\bm\Omega}$ is positive definite, there exist $c, \lambda > 0$ such that the optimization problem is convex.

\begin{proof}

Let $\Omega\in\mathbb{R}^n$. Then, taking the Hessian of the objective \eqref{eq:ngd_fund} gives
\begin{align}
    \nabla^2_{\delta} f(\delta) &= \nabla^2_{\delta} \mathcal{L}(\hat{\rho}_{\Omega+\delta}) + \lambda \nabla^2_{\delta} \Phi(\hat{\rho}_{\Omega}, \hat{\rho}_{\Omega+\delta})
\end{align}
The second term on the RHS relates to the metric corresponding to $\Phi$ as in \eqref{eq:expand-potential}. Now, let $H := \nabla^2_{\delta} \mathcal{L}(\hat{\rho}_{\Omega+\delta})|_{\delta=0}$ denote the Hessian matrix of the loss at $\delta=0$. Then, we choose
\begin{align}
\lambda > \frac{|\lambda_{\min}(H)|}{\min_{\|v\| = 1}\left(v^\dag \nabla^2 \Phi(\hat{\rho}_{\Omega}\|\hat{\rho}_{\Omega+\delta} )v|_{\delta=0}\right)}.
\end{align}
Note that positive definiteness of monotone metrics over $\mathcal{M}^{(N)}$ for all positive definite $\rho_{\bm \Omega}$ (\Cref{cor:positive-definite}) guarantees
\begin{align}
\min_{\|v\| = 1}\left(v^\dag \nabla^2 \Phi(\hat{\rho}_{\Omega}, \hat{\rho}_{\Omega+\delta} )v|_{\delta=0}\right) > 0,
\end{align}
which makes our choice of $\lambda$ well defined and strictly greater than $0$. This yields, for any $w \in \mathbb{R}^n$,
\begin{align}
w^\dag (\nabla_{\delta}^2 f(\delta)|_{\delta = 0}) w &= w^\dag H w + \lambda w^\dag \Phi(\hat{\rho}_{\Omega} , \hat{\rho}_{\Omega+\delta})w|_{\delta=0} \\
&\geq \|w\|^2 \lambda_{\min}(H) + \lambda \|w\|^2 \left(\min_{\|v\| = 1}\left(v^\dag \nabla^2 \Phi(\hat{\rho}_{\Omega}, \hat{\rho}_{\Omega+\delta} )v|_{\delta=0}\right)\right) \\
&> \|w\|^2\left(\lambda_{\min}(H) + |\lambda_{\min}(H)| \right) \geq 0.
\end{align}
Hence we have that $\nabla_{\delta}^2 f(\delta)|_{\delta=0}$ is positive definite, which implies that $f$ is convex in some neighborhood of $0$, say $B_{\varepsilon}(0) \subset \mathbb{R}^n.$ To see why this is the case, note that the map defined by
\begin{align}
\delta_0 \mapsto \lambda_{\min}( \nabla^2 f(\delta)|_{\delta=\delta_0}) = \min_{i=1, \dots, n} \lambda_i(\nabla^2 f(\delta))|_{\delta=\delta_0})
\end{align}
is continuous, since the eigenvalue maps $\delta_0\mapsto\lambda_i(\nabla^2 f(\delta))|_{\delta=\delta_0}$ are continuous, and the minimum of $n$ continuous functions is continuous. Thus, since its value at $\delta_0 = 0$ is (strictly!) positive, there exists a ball around zero $B_{\varepsilon}(0)$ where it is also strictly positive, i.e., the Hessian $\nabla^2 f(\delta)|_{\delta=\delta_0}$ is positive definite for all $\delta_0 \in B_{\varepsilon}(0).$ Explicitly restricting our step $\delta$ to have $\|\delta\| \leq\varepsilon$ (which is a convex constraint) makes $f$ convex on the feasible set, hence making \eqref{eq:ngd_fund} into a constrained convex optimization problem.
\end{proof}
\end{proposition}

This is a useful result, as \eqref{eq:ngd_fund} being a (locally) convex problem gives us a guarantee that gradient descent will converge in polynomial time, hence efficiently solving the inner-loop of the optimization problem.

\clearpage 

\printglossaries


\end{document}